\documentclass[12pt]{article}
\usepackage[slantedGreek]{mathptmx}
\usepackage{fullpage}
\usepackage{hyperref}

\usepackage{graphicx}        % standard LaTeX graphics tool 
\usepackage{amssymb, amsmath,bm}
\usepackage{stmaryrd}
\usepackage{enumitem}
\usepackage{comment}
\usepackage{caption, rotating}
\usepackage[small]{subfigure}
\usepackage{amsfonts}
\usepackage{pdflscape}
\usepackage{xcolor}
\usepackage{upgreek}
\renewcommand{\d}{\text{d}}

\DeclareSymbolFont{newfont}{OML}{cmm}{m}{it}                     
\DeclareMathSymbol{\Varrho}{3}{newfont}{37}

\numberwithin{equation}{section}
\usepackage[numbers,square,comma,sort&compress]{natbib}
\graphicspath{{./figures/}} 

\title{Constitutive theory for mechanics of amorphous thermoplastic polymers under extreme dynamic loading} 

\author{J.D. Clayton$^{1}$\footnote{Email: john.d.clayton1.civ@army.mil} \\ \\
$^1$Terminal Effects Division, Army Research Directorate
\\
DEVCOM ARL, Aberdeen, MD 21005-5066, USA 
}

\date{}
\begin{document}
\graphicspath{{figures/}}
\maketitle

\begin{abstract}
A geometrically nonlinear continuum mechanical theory is formulated for deformation and failure behaviors of amorphous polymers. The model seeks to capture material response over a range of loading rates, temperatures, and stress states encompassing shock compression, inelasticity, melting, decomposition, and spallation.
Thermoelasticity, viscoelasticity, viscoplasticity, ductile failure with localized shear yielding, and brittle fracture with crazing 
can all emerge under this ensemble of intense loading conditions. 
Known prior theories have considered one or more, but not all, such physical mechanisms.
The present coherent formulation invokes thermodynamics with internal state variables
for dynamic molecular and network configurational changes affecting viscoelasticity and plastic deformation, and it uses
order parameters for more abrupt structural changes across state-dependent glass-transition and shock-decomposition thresholds.
A phase-field order parameter captures material degradation from ductile or brittle fracture, including evolving porosity from crazing.
The theory is applied toward polymethyl methacrylate (PMMA) under intense dynamic loading.
The high-pressure equilibrium response, with shear strength and temperature over known ranges, is well represented along the principal Hugoniot to pressures far exceeding shock decomposition. Predicted release wave velocities agree with experiment.
A semi-analytical solution for steady waves describes the relatively lower-pressure viscoelastic setting, providing insight into relaxation times. One-dimensional calculations assess suitability of the model for representing spall fracture strengths seen in experiments over a range of initial temperatures and loading rates. 
 \end{abstract}

%\noindent

\tableofcontents

\section{Introduction}
\label{sec1}
Thermoplastic polymers are widely used in commercial and structural applications.
Examples include polymethyl methacrylate (PMMA), polyethylene, polystyrene, polyvinylchloride, and polycarbonate (PC).
Under the glass transition temperature $\theta_G$,
behaviors can be brittle or ductile depending on loading mode, whereas rubbery or viscous fluid-like response   manifests above $\theta_G$.
Many physical properties vary in the vicinity of $\theta_G$, where the latter itself, for a polymer of specific composition and processing history, is unique only in an ambient equilibrium state. For example, the glass transition in PMMA can
be interpreted to vary with pressure and loading rate \cite{saeki1992,richeton2007,srivastava2010,clements2012a}. 
The term ``melting'' is used interchangeably to denote transition from glassy to (viscoelastic) rubbery state or
from solid to (viscous) liquid state, where the former necessarily occurs at lower temperature than the
latter if both are possible; certain polymers can also thermally decompose into various chemical products from a solid state, rather than liquify, with increasing temperature \cite{beyler2002}.

The present research focuses on amorphous (i.e., non-crystalline) thermoplastics that are glassy at room temperature, and in particular the transparent polymer PMMA, though descriptive models and physical findings may translate to other polymers that share similar physical traits.
These materials are used in protection systems including windows (i.e., ``bullet-proof glass''), goggles, face shields, and layered transparent armors \cite{grujicic2012}. PMMA---also known as acrylic or acrylic glass with
trade names Plexiglas and Lucite, among others---is furthermore widely used as a window material in shock physics experiments \cite{barker1970}.  Over such a broad dynamic application space, strains, strain rates, pressures, and temperatures can be severe, inducing one or more coupled phenomena: nonlinear elastic and viscoelastic behaviors, plasticity and localization by shear yielding, brittle fracture by crazing and spall, melting, and shock or thermal decomposition. Crazes are cavitation zones arising normal to the direction of maximum principal stress in amorphous polymers \cite{argon1975}.

A detailed treatment of origins of physical behaviors at the molecular level is beyond the current scope, but a
 few remarks are given here for motivation of previous and current continuum theory. The glassy polymer macromolecular structure
 is comprised of a network of entangled molecular chains. The latter can be idealized as rigid links along the
 strongly bonded polymer chain \cite{arruda1995}. Temporal changes in network entanglement structures and secondary bonding (e.g., chain interactions) induced by strain and temperature histories affect thermo-mechanical
 properties controlling viscoelasticity, yielding, directional hardening, and fracture mechanisms (e.g., ductility).
 Viscoelastic and viscoplastic mechanisms are correlated with structural transitions (e.g., $\alpha$ and
 various $\beta$ processes \cite{fleck1990,mulliken2006,richeton2006a}).
 Entanglement density affects the tendency for ductile versus brittle failure, the former by shear yielding and the
 latter by crazing or network disentanglement with chain scission \cite{gearing2004,narayan2021}.
 Generally, the more entanglements, the greater potential for elongation (e.g., plastic flow) without fracture under tensile loading. The quantity ``free volume'', namely the unoccupied space in a solid polymer between molecular chains and their aggregates, has also been widely associated to changes in thermal, elastic, viscous, and plastic
 behaviors \cite{williams1955,anand2003,jatin2014,white2016,dal2022}. Though different physical and mathematical
 definitions for free volume exist \cite{white2016}, an increase in free volume correlates with strain-softening of plastic yield strength below the glass transition \cite{jatin2014} and decreased (e.g., elastic) stiffness across the glass transition itself \cite{white2016}.

A geometrically nonlinear theory is essential in the current applications: elastic and viscoelastic
deformations can be large, even in uniaxial shock compression \cite{menikoff2004}. 
Apparent shear and bulk moduli vary strongly with temperature and pressure
in PMMA, even far below the glass transition \cite{asay1969}. 
Bulk and shear moduli are time or frequency dependent \cite{kono1960,sane2001},
suggesting a viscoelastic treatment should be included in a complete theory that seeks to
span a wide range of loading rates and temperatures \cite{mulliken2006,richeton2007,clements2009,clements2012a}.
Innumerable theories have been constructed to capture viscoelasticity of thermoplastic polymers.
Some models simply assign the elastic coefficients as functions of strain rate in addition to
temperature \cite{mulliken2006,richeton2007}.
Others do not resolve viscoelasticity distinctly, but rather implicitly embed all dissipative deformation
mechanisms in an inelastic or plastic model component \cite{arruda1995,menikoff2004,anand2009,bouvard2010,bouvard2013}.
The stress-history functional approach based on fading memory concepts and finite linear viscoelasticity  \cite{coleman1961} has described one-dimensional (1-D) shock waves \cite{schuler1970,schuler1973,nunziato1973p}.
A sophisticated model spanning static to shock regimes incorporated non-equilibrium history functionals
for bulk and shear response \cite{clements2012a}.
A viscoelastic-plastic model for thermoplastics \cite{jinaga2025} used tensor-valued internal variables \cite{holzapfel1996a,holzapfel1996b} for configurational energy changes.

In the current work, viscoelastic response is distinguished from (visco)plastic response following Ref.~\cite{sadik2024}, wherein plastic deformation is one form of ``anelastic'' deformation in that reference. 
Total elastic deformation can be decomposed into that which is recovered instantaneously
(i.e., perfectly glassy response without dissipation) and that which is recovered more slowly (i.e., dissipative viscoelastic response). Finite-strain theories decomposed the deformation gradient into
a product of (instantaneous) elastic and transient viscoelastic parts \cite{sidoroff1974,reese1998,sadik2024}. 
In this context, the intermediate configuration achieved upon local instantaneous unloading is
generally not stress-free.  In contrast, in finite elastoplasticity theory (cf. \cite{claytonNCM2011}), the intermediate configuration is locally stress-free upon removal of local elastic deformation.
In the theory developed in the present work, consistent with Ref.~\cite{sadik2024}, all elastic deformation (instantaneous and viscoelastic) is eventually recovered when load is removed, while plastic deformation
is residual and can be permanent in the absence of load reversal.
However, the current theory uses tensor-valued internal variables to address viscoelasticity following  Refs.~\cite{holzapfel1996a,holzapfel1996b}, and it does not deconstruct the total elastic deformation
into instantaneous and dissipative parts.  The deformation gradient is decomposed into a total thermoelastic component (encompassing thermal expansion and viscoelasticity) and a residual plastic component incorporating irreversible chain motions, free volume changes, and dilatational crazing mechanisms.
The current approach simplifies the kinematic and thermodynamic analysis 
over that which would be incurred from explicitly resolving thermal, instantaneous elastic, viscous, plastic, and damage mechanisms
via distinct terms in the deformation gradient \cite{arruda1995,bouvard2013,francis2014}, though all mechanisms are properly encompassed in rate form through the velocity gradient.

Finite-deformation constitutive models for amorphous glassy polymers have been under
development since at least the 1980s \cite{boyce1988}, building on seminal earlier studies from materials science, for example, Ref.~\cite{argon1973}.
Inelastic models with a basis in polymer physics 
include those of the Boyce and Arruda research groups \cite{arruda1993b,arruda1995,mulliken2006,dupaix2007}. 
Other sophisticated thermodynamics or molecular physics-motivated models include those of the Anand research group~\cite{anand2003,anand2009,ames2009}, Richeton et al.~\cite{richeton2005,richeton2006b,richeton2006a,richeton2007} and Bouvard et al. \citep{bouvard2010,bouvard2013}.
These models generally use scalar and tensor-valued internal or kinematic variables 
for capturing hardening and softening associated with polymer chain interactions, rotations, and internal stretches. Large-strain elastic and inelastic behavior has been related to chain orientation distributions \cite{wu1993}; statistical mechanics
frequently motivates the former (e.g., \cite{arruda1993a,khand2023} and references therein).
While many focused only on the glassy response \cite{arruda1993a,arruda1993b,arruda1995,anand2003,anand2009,ames2009}, theories have addressed drastic softening witnessed as temperature approaches and exceeds the glass transition \cite{palm2006,mulliken2006,dupaix2007,srivastava2010,richeton2005,richeton2006b,richeton2006a,richeton2007}.
Aforementioned models did not, however, address the high pressures and extreme rates achieved in shock loading. Equations-of-state coupled to less sophisticated inelasticity or rheological models (e.g., no anisotropic hardening) have been developed for this regime and were used predominantly to study the 1-D shock response \citep{schuler1974,menikoff2004,clements2012a,merz2012,popova2015,popova2018}. 
Empirical plasticity models such as Drucker-Prager \citep{dorogoy2010} and Johnson-Cook \citep{holmquist2016,lawlor2024} have been calibrated to depict shock or ballistic impact of PMMA, but
these engineering approaches do not provide insight into relative importance of fundamental physical mechanisms originating at the macromolecular scale, nor do they formally consider continuum thermodynamic relations and restrictions \cite{coleman1963,coleman1967}. 

Brittle fracture of glassy polymers such as PMMA originates from crazing, even
at high rates pertinent to dynamics and spallation \cite{christman1972,fleck1990,washabaugh1993}.
Crazing is discussed from a materials physics perspective in Refs.~\cite{argon1975,argon1977,dal2022}
and can incur significant inelastic volume increase \cite{gsell2002}. 
However, the tendency for ductile versus brittle failure tends to increase with temperature \cite{fleck1990} and confining pressure \cite{satapathy2000,rittel2008,zhang2021}.
Shear banding in PMMA is highly rate sensitive \cite{archer2010}: diffuse bands form at lower rates and
sharp bands at higher rates leading to mode II fracture.
Temperature rise during ductile or brittle mode II fracture can be large enough to approach or surpass the glass transition, leading to localized melting \cite{bjerke2002,rittel2008}.

Continuum damage models for crazing have been coupled to finite polymer inelasticity, albeit at moderate rates and low pressures \cite{gearing2004,francis2014}.
These methods, devoid of an intrinsic length scale, can produce mesh-sensitive results,
as can phenomenological approaches such as the Johnson-Cook fracture model used for ballistic failure response of PMMA \cite{holmquist2016}.
Cohesive finite element (FE) models \cite{estevez2000,arias2007,schluter2016} have been used for ductile and brittle fractures.
Spallation has been modeled in other material systems (e.g., metallic and ceramic polycrystals) with
cohesive models \cite{claytonJMPS2005,claytonIJSS2005,foulk2010}, but apparently not in monolithic thermoplastics.
Regularized phase-field models have been implemented for shear yielding and craze fractures
in polymers such as PC and PMMA \cite{miehe2015b,narayan2021,dal2022,li2024}. These implementations, however, have been limited to relatively low rates, low pressures, and isothermal conditions. Therefore, they are unsuitable for ballistic loading.
Spall fracture experiments on amorphous thermoplastics suggest an increase in spall strength with increasing loading (i.e., pressure release) rate \cite{geraskin2009,diamond2025} and and decrease with increasing temperature, the latter notably severe across the glass transition \cite{zaretsky2019,cherepanov2024}.
A micromechanics-based model for spall fracture of PC was developed in Ref.~\cite{curran1973}.
In a computational study of spall fracture in high density polyethylene \cite{dattelbaum2024}, a volumetric damage component was appended to the dynamic polymer framework of
Clements \cite{clements2012a}, though details of that damage model and parameters were not reported.
Molecular dynamics simulations \cite{dewapriya2022} provide some insight into mechanisms of spallation in polymers, but length and time scales are exceedingly small for comparison to standard experiments. 
Herein, a phase-field approach extending that of Refs.~\cite{miehe2015b,claytonAMECH2025} is extended
to study rate- and temperature-dependent crazing pertinent to spall fracture of thermoplastics.

Under planar shock compression, many thermoplastic as well as thermosetting polymers display a variation along their principal Hugoniot in the form of a local volume collapse at pressures around or exceeding the order of 20 GPa
\cite{carter1995,dattelbaum2019}. Such behavior was originally attributed to a generic kind of phase transition
\cite{hauver1964,hauver1965,carter1995} but has been more recently confirmed as a chemical decomposition
from a condensed (solid, melt, or fluid) reactant phase to a combination of condensed elemental and gaseous products
\cite{dattelbaum2019,lentz2020,bordz2021,coe2022,huber2023}.
Transformation of PMMA to a more fluid phase at impact stresses between 20 and 30 GPa was apparently first discovered by shock-induced polarization measurements of Hauver \cite{hauver1964,hauver1965}.
This now-designated shock decomposition is accompanied by a volume decrease of 3.4\% \cite{carter1995,dattelbaum2019}.
A correlation between tendencies for shock-induced polarization and shock-induced conduction in specific polymers was noted in the context of a bond scission model \cite{graham1979,chan1981}.
Modeling of shock decomposition in polymers has been limited to date. Recent methods include distinct tabular equations-of-state for reactant and product phases,
thermochemical modeling of products with free energy minimization, and Arrhenius kinetics for
reaction rates \cite{maerzke2019,lentz2020,coe2022,huber2023}.
For PMMA, the former have been compared with ab initio molecular dynamics simulations \cite{coe2022}.
Although temperatures at the decomposition threshold can exceed 1000--2000 K, it
is unclear if the material is solid or fluid prior to decomposition due to the immense pressures that should suppress
glass-to-melt transition.  Shock decomposition at extreme pressures, temperatures, and strain rates should be distinguished from  more understood thermal decomposition
and combustion processes in polymers at lower pressures \cite{kashiwagi1985,beyler2002,stoliarov2003,korob2019}.
Experimental principal Hugoniot data for PMMA exceeding shock pressures of 100 GPa are available \cite{vanthiel1977}.

The objective of the present study is construction and demonstration of accuracy and utility of a constitutive model for amorphous polymers with all features listed below:
\begin{itemize}[nosep] %\setlength{\itemsep}{-5pt}
\item Nonlinear thermo-viscoelasticity applicable to high rates, temperatures, and pressures
\item Finite-strain plasticity with isotropic and anisotropic hardening/softening
\item Ductile and brittle fracture applicable to high rates, temperatures, and pressures
\item Behavior changes across the glass transition and shock decomposition thresholds
\end{itemize}
No theory known to the author synthesizes all of these features in a manner consistent with established continuum mechanical and thermodynamic laws.  The proposed framework achieves its objective
using a combined internal variable and phase-field formulation of finite-strain thermoviscoelasticity and viscoplasticity.
The current fully three-dimensional (3-D) theory adapts the established Holzapfel-Simo approach \cite{holzapfel1996a,holzapfel1996b}
with tensor configurational variables, as this appears more straightforward to integrate
with formal internal state variable theory \cite{coleman1967} than finite linear viscoelastic functionals \cite{coleman1961},
non-equilibrium energy functionals \cite{clements2012a}, and Prony series approximations \cite{scheidler2009}.
Invariants of a logarithmic thermoelastic strain measure \cite{claytonIJES2014,claytonNEIM2019} enter the free and internal energy functions, contributing to 
an equation of state valid at high pressure.
For viscoplasticity, a combination of scalar and tensor internal state variables (the latter linked to plastic stretch) is adopted to account for isotropic and anisotropic hardening and softening mechanisms
\cite{arruda1995,richeton2007,anand2009,bouvard2013}.
A phase-field model resolving ductile and brittle fracture contributions,
with dilatation linked to void growth, adapts ideas from Refs.~\cite{miehe2015b,narayan2021,dal2022}.
Features are newly augmented account for extreme 
pressures, rates, and temperatures pertinent to shock compression and spall.

Scalar order parameters, of local rather than gradient type, are used to interpolate the material response across the
glass transition and shock-decomposition thresholds. Depending on the kinetic laws and functions, the response can 
be smooth, abrupt or even discontinuous. Phase-field models have been used for
interfacial physics of melting and solidification \cite{levitas2011b} and polymer crystallization \cite{bahloul2020}. Here, as morphological details of the solid-melt and condensed-vaporized material interfaces are not of primary interest, gradient regularizations of solid-melt and condensed-decomposed order parameters are omitted, as elsewhere in modeling fracture of liquids \cite{levitas2011a} and soft solids \cite{claytonPRE2024} and shear localized melting in metals \cite{claytonARX2025}.
The glass-melt transition in amorphous polymers can be interpreted as a second-order phase transformation \cite{gibbs1958,wu1999}, whereas shock decomposition involves a mass density jump \cite{hauver1965,carter1995} so is of first order.

Given the immense scope of physics, it is unreasonable to expect detailed functions and parameters be obtained for all possible sub-features and loading regimes in a single publication, even for one material.
Unique features are demonstrated for 1-D problems in high-rate and shock regimes. 
These problems, here considering PMMA, can be treated analytically or with basic numerical integration.
Notable physics observed in shock experiments include a rapid slope change or inflection in the Hugoniot indicative of a Hugoniot Elastic Limit (HEL) for plastic yielding at a longitudinal stress around 0.7 GPa \cite{barker1970,millett2000},
lack of an elastic precursor \cite{barker1970,schuler1974}, rounding of the particle velocity profile due to
viscoelastic relaxation
\cite{barker1970,schuler1970,schuler1973}, and increase or maintenance of shear strength to at least moderate shock pressures on the order of 6--8 GPa \cite{gupta1980a,gupta1980b,millett2000,jordan2020}, followed by a decay above 10 GPa \cite{batkov1996}.
Shear strength is thought to persist at shock-elevated temperatures above the ambient static glass transition (e.g., $\theta_G \approx 375-395$ K for common kinds of PMMA) \cite{menikoff2004}. 
The aforementioned physics result from a combination of dependence of viscoelastic moduli, plastic flow strength, and melting on pressure and time scale (e.g., loading rate) \cite{rosenberg1994,menikoff2004,clements2012a}.
 Steady wave forms are observed within a certain impact velocity range and after a minimum run distance \cite{nunziato1973e}, at least below the Hugoniot inflection that otherwise can cause dispersion \cite{schuler1970}. Shock evolution and steady waves at stresses below the HEL were described in theoretical works by
 Schuler, Nunziato, and Walsh \cite{schuler1970,nunziato1973am,nunziato1973e,nunziato1973p,schuler1973}.
 Their solutions for shock evolution adopted the analysis of Chen and Gurtin \cite{chen1970a,chen1972} on shock waves in (thermo)viscoelastic materials with fading memory and finite linear viscoelasticity.
 Numerical studies of shock evolution encompassing higher shock pressures in PMMA followed thereafter using
 various theoretical and computational schemes \cite{schuler1974,menikoff2004,clements2012a,merz2012,popova2015,popova2018}.
 In the current study, methods of Ref.~\cite{claytonPRE2024} for viscoelastic soft solids
 described by internal variables \cite{holzapfel1996a,holzapfel1996b}, Ref.~\cite{chen1971} for fluids with
 internal variables, and Refs.~\cite{claytonJMPS2021,claytonPROCA2022} for steady waves in the context of phase-field fracture are adapted to derive equations for shock evolution in a different constitutive setting for amorphous viscoelastic polymers.

General continuum principles are presented in Section \ref{sec2}.
These include balance laws and governing equations deemed pertinent to any thermoviscoelastic-inelastic material
with internal state variables of scalar, tensor, and gradient (e.g., phase-field) type demonstrating isotropic elastic response.
The constitutive framework is directed toward amorphous polymers, with possible plastic anisotropy, in Section \ref{sec3}.
Energy functions and kinetic laws for thermoelasticity, viscoelasticity, plasticity, fracture,
melting across the glass transition, and structural decomposition are constructed.
Applications to PMMA follow in Section \ref{sec4}. Known physical properties and parameters are
reviewed. 
The principal ``equilibrium'' shock Hugoniot, including pressure, temperature, shear strength, and structural changes, is calculated via an iterative analytical method in Section \ref{sec4.1} to shock stresses exceeding the experimental decomposition threshold. Shock evolution and steady waves are analyzed in Section \ref{sec4.2} for the lower-pressure viscoelastic regime.  Spall fracture is analyzed via 1-D dynamic calculations in Section \ref{sec4.3} for different starting temperatures and loading rates. Comparisons with experiments are included throughout Section \ref{sec4} where possible, providing complementary insight into observed physics. Conclusions appear in Section \ref{sec5}.

\section{Continuum principles}
\label{sec2}

\subsection{Standard relations}
\label{sec2.1}
Denote the spatial volume occupied by a body by $\Omega(t)$, where $t$ is time.
In a reference configuration at $t= t_0$, the body occupies $\Omega_0 = \Omega(t_0)$.
External boundaries are $\partial \Omega (t)$ and $\partial \Omega_0$ with respective unit normal
vector fields $\bm{n}(\bm {x})$ and $\bm{n}_0 (\bm {X})$. Spatial position is $\bm{x}$ measured
by coordinate chart(s) $\{ x^k \}$ where generally $k = 1,2,3$.
Similarly, reference position is $\bm X$ with charts $\{X^K\}$ and $K = 1,2,3$.
Let $\square$ be a differentiable function of position and time.
Basis vectors are ${\bm g}_k = \partial_k {\bm x}$ and ${\bm G}_K = \partial_K {\bm X}$
where $\partial_k \square = \partial \square / \partial x^k$ and $\partial_K \square = \partial \square / \partial X^K$.
Usual metric tensors have components $g_{ij} = {\bm g}_i \cdot {\bm g}_j$
and $G_{IJ} = {\bm G}_I \cdot {\bm G}_J$ with determinants $g = \det (g_{ij})$ and $G = \det (G_{IJ})$.
Particle motion and particle velocity are, respectively,
\begin{align}
\label{eq:motion}
{\bm x} = {\bm \chi}({\bm X},t), \qquad {\bm \upsilon}({\bm X},t) = \partial {\bm \chi} ({\bm X},t) / \partial t.
\end{align}
The material time derivative (i.e., derivative at fixed $\bm X$) is denoted by $\dot{\square}$.
The partial space-time derivative  (i.e., at fixed $\bm x$) is denoted by $\partial_t \square$.
Material and spatial covariant derivatives are denoted, respectively, by $\nabla_0 \square$ and
$\nabla \square$.
The deformation gradient and Jacobian determinant are
\begin{align}
\label{eq:defgrad}
{\bm F} = \nabla_0 {\bm \chi} = {\partial x^i}/{\partial X^J} {\bm g}_i \otimes {\bm G}^J,
\qquad J = \sqrt{g/G} \det (F^i_J) = \d \Omega / \d \Omega_0 > 0.
\end{align}
The spatial velocity gradient, deformation rate, spin, and rate of volume change obey
\begin{align}
\label{eq:velgrad}
{\bm l} = \nabla {\bm \upsilon} = \dot{\bm F}{\bm F}^{-1} = {\bm d} + {\bm \omega},
\quad {\bm d} = {\textstyle{\frac{1}{2}}} ( {\bm l} + {\bm l}^{\mathsf T}),
\quad {\bm \omega} = {\textstyle{\frac{1}{2}}} ( {\bm l} - {\bm l}^{\mathsf T}),
\quad
\dot{J} = J \nabla \cdot {\bm \upsilon} = J {\rm tr} {\bm d}.
\end{align}

Mass densities are $\rho ({\bm X},t)$ and $\rho_0 = \rho({\bm X},t_0)$. Cauchy stress is ${\bm \sigma}$,
body force per unit mass is $\bm{b}$, internal energy per unit mass is $u$, spatial heat flux is ${\bm q}$, 
and point heat sources per unit mass are $r$. Classical local conservation laws for mass, momentum, and energy
are (cf. \cite{malvern1969,marsden1983,claytonNCM2011}) 
\begin{align}
\label{eq:conslaws}
\dot{\rho} + \rho  {\rm tr} {\bm d} = 0,
\quad
\nabla \cdot {\bm \sigma} + \rho {\bm b} = \rho \dot{\bm \upsilon},
\quad
{\bm \sigma} = {\bm \sigma}^{\mathsf T},
\quad
\rho \dot{u} = {\bm \sigma}: {\bm d} - \nabla \cdot {\bm q} + \rho r.
\end{align}
The first of \eqref{eq:conslaws} corresponds to $\rho_0 = \rho J$.
The last of \eqref{eq:conslaws} is derived from the usual, standard global balance of continuum mechanics.
No independent contributions from rates of order parameters or their gradients are ascribed to the internal energy rate, and no micro-force balances are imposed, differing from many other phase-field models \cite{gurtin1996,borden2016,narayan2021,claytonJMPS2021}. Rather, following Refs.~\cite{miehe2015a,miehe2015b,dal2022,claytonAMECH2025}, the order parameter for fracture obeys conservation and kinetic laws that are independent from the first and second laws of thermodynamics, as will be shown in Section \ref{sec2.3}.  The local form of the latter is, with $\theta$ absolute temperature and $s$ entropy per unit mass,
\begin{align}
\label{eq:seclaw}
\rho \theta \dot{s} + \nabla \cdot {\bm q} - \rho r - ({\bm q} / \theta) \cdot \nabla \theta \geq 0.
\end{align}

Let $\psi$ be Helmholtz free energy per unit mass and ($U,\Psi,\eta$) be (internal energy, free energy, entropy) per unit reference volume.  If heat conduction always produces non-negative dissipation, \eqref{eq:seclaw}
with the last of \eqref{eq:conslaws} provides the non-negative internal dissipation per unit volume $\mathfrak D$:
\begin{align}
\label{eq:energies}
& \psi = u - \theta s \quad \Leftrightarrow \quad \Psi = U - \theta \eta; \qquad
U = \rho_0 u, \quad \Psi = \rho_0 \psi, \quad \eta = \rho_0 s;
\\
\label{eq:Diss}
& {\mathfrak D} = \theta {\dot \eta} + J \nabla \cdot {\bm q} - \rho_0 r = J {\bm \sigma}:{\bm d} - \dot{\Psi} - \dot{\theta} \eta \geq 0.
\end{align}

\subsection{Kinematics, state variables, and thermodynamics}
\label{sec2.2}
The deformation gradient is decomposed into a thermoelastic part ${\bm F}^E$ and a plastic part ${\bm F}^P$.
Note ${\bm F}^E$ encompasses instantaneous hyperelasticity, thermal expansion, and dissipative viscoelastic deformation. Note ${\bm F}^P$ encompasses deviatoric plastic flow (e.g., from irreversible molecular chain motions),
changes in residual free volume, residual deformation from crack opening (e.g., porosity from crazing),
and potential volume changes from melting and decomposition. Specifically,
\begin{align}
\label{eq:FeFp0}
& {\bm F} = {\bm F}^E {\bm F}^P,  \qquad F^i_J = (F^E)^i_\alpha (F^P)^\alpha_J; 
\\ 
\label{eq:FeFp}
& {\bm l} = \dot{\bm F}^E ( {\bm F}^E)^{-1} + {\bm F}^E {\bm D}^P {\bm F^E}^{-1},
\qquad {\bm D}^P = ({\bm D}^P)^{\mathsf T} =  \dot{\bm F}^P ( {\bm F}^P)^{-1} = ( {\bm F}^E)^{-1} {\bm d}^P {\bm F}^E.
\end{align}
The spatial plastic deformation rate is ${\bm d}^P$. For isotropy, plastic velocity gradient ${\bm D}^P$ is symmetric; plastic spin vanishes \cite{anand2003,gurtin2005,anand2009,claytonAMECH2025}.
Implicit in \eqref{eq:FeFp0} is a locally stress-free intermediate configuration. Following \cite{claytonDGKC2014,claytonJMPS2021}, basis vectors on this anholonomic space are
chosen as ${\bm g}_\alpha = \delta^K_\alpha {\bm G}_K$ producing metric tensor components
$g_{\alpha \beta} = \delta^I_\alpha G_{IJ} \delta^J_\beta$ such that $\det (g_{\alpha \beta} ) = G$.
Volume ratios obey
\begin{align}
\label{eq:Jp}
J^E = \det ({F}^E)^i_\alpha \sqrt {g/G} > 0, \qquad J^P = \det ({F}^P)^\alpha_J > 0; \qquad \dot{J}^P = J^P {\rm tr} {\bm D}^P
= J^P {\rm tr} {\bm d}^P.
\end{align}
Thermoelastic deformation can be decomposed into rotation tensor ${\bm R}^E = ({\bm R}^E)^{-\mathsf T}$ and stretch tensors ${\bm U}^E, {\bm V}^E$ via polar decompositions. 
Logarithmic strain tensors and their deviatoric parts follow:
\begin{align}
\label{eq:kinvars1}
& {\bm F}^E  = {\bm R}^E {\bm U}^E, \qquad {\bm C}^E = ({\bm F}^E)^{\mathsf T} {\bm F}^E = ({\bm U}^E)^2,
\qquad {\bm \epsilon}^E = \ln {\bm U}^E = {\textstyle {\frac{1}{2}}} \ln {\bm C}^E, \quad \epsilon^E_V  = \ln J^E ; \\
\label{eq:kinvars1b}
& {\bm F}^E  = {\bm V}^E {\bm R}^E, \qquad {\bm B}^E = {\bm F}^E ({\bm F}^E)^{\mathsf T}  = ({\bm V}^E)^2,
\qquad {\bm e}^E = \ln {\bm V}^E = {\textstyle {\frac{1}{2}}} \ln {\bm B}^E, \quad e^E_V = \ln J^E ;\\
\label{eq:kinvars2}
& J^E 
=   \exp [( \epsilon^E)^\alpha_\alpha] = \exp [( e^E)^k_k] ,
\quad \bar{\bm \epsilon}^E = {\bm \epsilon}^E - {\textstyle {\frac{1}{3}}}\epsilon^E_V {\bm 1}, \quad
\bar{\bm e}^E = {\bm e}^E - {\textstyle {\frac{1}{3}}} e^E_V {\bm 1}.
\end{align}
Logarithmic thermoelastic strain in locally relaxed material coordinates \cite{claytonIJES2014} is ${\bm \epsilon}^E$,
with deviatoric (i.e., traceless) part $\bar{\bm \epsilon}^E$.  Its analogy in spatial coordinates \cite{anand1979,fitzgerald1980,criscione2000} is ${\bm e}^E$ with deviatoric part $\bar{\bm e}^E$.

Denote by $\{ \bm \alpha \}({\bm X},t)$ the set of conventional internal state variable fields, and denote by $\xi ({\bm X},t)$ an order parameter field of gradient type for phase-field fracture that does not belong to $\{ \bm \alpha \}({\bm X},t)$. Properties of $\xi$ are discussed in Section \ref{sec2.3}. Individual entries of $\{ {\bm \alpha} \}$
encompassing configurational viscoelasticity, isotropic and anisotropic plastic hardening,
porosity, free volume, and phase fractions of melt and decomposed material are introduced in Section \ref{sec3.1}.
For isotropic elasticity, strain energy can depend on up to three independent scalar invariants of ${\bm e}^E$ \cite{criscione2000}. A logarithmic theory for anisotropic elasticity would require dependence on $\bm{\epsilon}^E$ instead. While more broadly applicable to crystalline \cite{claytonIJES2014,claytonNEIM2019,claytonJMPS2021} solids, use
of $\bm{\epsilon}^E$ is more mathematically cumbersome than ${\bm e}^E$.
Energy potentials are assigned the following dependencies, omitting argument $\bm X$ for brevity:
\begin{align}
\label{eq:psigen}
\Psi = \Psi ({\bm e}^E ({\bm F}^E), \theta, \{ {\bm \alpha} \}, \xi), \qquad U = U ({\bm e}^E ({\bm F}^E), \eta, \{ {\bm \alpha} \}, \xi).
\end{align}
Total Cauchy stress is split into an elastic part $\bm{\sigma}^E$ and a viscous part ${\bm{\sigma}}^V$ \cite{coleman1963,claytonPRE2024}, both symmetric:
\begin{align}
\label{eq:totalstress}
{\bm \sigma}({\bm e}^E, \theta, \{ {\bm \alpha} \}, \xi, {\bm d}) = 
{\bm \sigma}^E ({\bm e}^E, \theta, \{ {\bm \alpha} \}, \xi)  +
{\bm \sigma}^V ({\bm e}^E, \theta, \{ {\bm \alpha} \}, \xi, {\bm d}).
\end{align}
Deviatoric parts are denoted by $\bar{\square}$, and spherical Cauchy pressures are $p$, $p^E$, and $p^V$:
\begin{align}
\label{eq:devstresses}
& \bar{\bm \sigma} = {\bm \sigma} + p {\bm 1}, \qquad \bar{\bm \sigma}^E = {\bm \sigma}^E + p^E {\bm 1}, \qquad \bar{\bm \sigma}^V = {\bm \sigma} + p^V {\bm 1}; \\
\label{eq: pressures}
& p = - {\textstyle{\frac{1}{3}}} {\rm tr} {\bm \sigma}, \qquad p^E = - {\textstyle{\frac{1}{3}}} {\rm tr} {\bm \sigma}^E,
\qquad p^V = - {\textstyle{\frac{1}{3}}} {\rm tr} {\bm \sigma}^V.
\end{align}
Expanding the first of \eqref{eq:psigen} with abbreviated notation $\partial_\square \Psi = \partial \Psi / \partial \square$ gives
\begin{align}
\label{eq:psidot}
\dot{\Psi} = (\partial_{ {\bm F}^E} \Psi ): \dot {\bm F}^E + (\partial_\theta \Psi)  \dot{\theta} + 
(\partial_{ \{ \bm \alpha \} } \Psi) \{ \dot{\bm \alpha} \} + (\partial_\xi \Psi)  \dot{\xi} .
\end{align}
From \eqref{eq:FeFp} and \eqref{eq:totalstress}, the stress power per unit reference volume is
\begin{align}
\label{eq:stresspow}
J {\bm \sigma}: {\bm d} = J [ {\bm \sigma}^E ({\bm F}^E)^{- {\mathsf T}}]: \dot{\bm F}^E + 
J [ ({\bm F}^E)^{ {\mathsf T}} {\bm \sigma}^E ({\bm F}^E)^{- {\mathsf T}}]: {\bm D}^P +
 {\bm \tau}^V : {\bm d}, \qquad {\bm \tau}^V = J {\bm \sigma}^V.
\end{align}
Inserting \eqref{eq:psidot} and \eqref{eq:stresspow}  into \eqref{eq:Diss} produces
\begin{align}
& {\mathfrak D}  =   [  J {\bm \sigma}^E ({\bm F}^E)^{- {\mathsf T}} -  \partial_{ {\bm F}^E} \Psi ] : \dot {\bm F}^E
- [ \eta + \partial_\theta \Psi]  \dot{\theta}  +  {\bm \tau}^V : {\bm d} + {\bm M}^E: {\bm D}^P
+ \{ \bm \pi  \} \cdot  \{ \dot{\bm \alpha} \} + \zeta  \dot{\xi} \geq 0;
\nonumber
\\
\label{eq:diss2}
& {\bm M}^E  = {\textstyle{\frac{1}{2}}} [ { \bm C}^E {\bm S}^E + {\bm S}^E { \bm C}^E], \quad {\bm S}^E = J ({\bm F}^E)^{-1} {\bm \sigma}^E ({\bm F}^E)^{- \mathsf T};
\quad  \{ \bm \pi  \} = -\partial_{\{ {\bm \alpha} \}} \Psi   , \quad \zeta = -\partial_\xi \Psi.
\end{align}
Denoted by ${\bm M}^E$ and ${\bm S}^E$ are a symmetrized Mandel stress \cite{claytonNCM2011,claytonJMPS2019} and an elastic second Piola-Kirchhoff stress, herein energy per unit reference rather than per unit intermediate-configuration volume.
Conjugate forces to internal variables and the fracture parameter are $ \{ {\bm \pi}  \} $ and $\zeta$.

Regarding the inequality in \eqref{eq:diss2}, quantities in square brackets are independent of rates of thermoelastic deformation and temperature and thus should be zero per classical arguments \cite{coleman1963,coleman1967,claytonNCM2011,claytonJMPS2021}.
Constitutive relations and a reduced dissipation inequality follow as a result:
\begin{align}
\label{eq:const1}
& {\bm \sigma}^E = J^{-1} ( \partial_{ {\bm F}^E} \Psi )({\bm F}^E)^{ {\mathsf T}}
= J^{-1}  ( \partial_{ {\bm V}^E} \Psi ){\bm V}^E, \qquad \eta = -  \partial_\theta \Psi ; \\
\label{eq:Dissr}
& \mathfrak{D} = {\bm \tau}^V:{\bm d} +  {\bm M}^E: {\bm D}^P  + \{ \bm \pi  \} \cdot  \{ \dot{\bm \alpha} \} + \zeta  \dot{\xi} \geq 0.
\end{align}
In isotropic logarithmic elasticity, the second identity for $\bm \sigma^E$ in  \eqref{eq:const1} can be derived from \eqref{eq:psigen} with ${\bm e}^E = {\bm e}^E ( {\bm V}^E)$ and ${\bm V}^E = ({\bm V}^E)^{\mathsf T} = {\bm F}^E ({\bm R}^E)^{\mathsf T}$. See Ref.~\cite{fitzgerald1980} where ${\bm \sigma}^E : \dot{\bm e}^E = {\bm \sigma}^E: [\dot{\bm V}^E ({\bm V}^E)^{-1}]$ is confirmed as the rate of elastic strain energy. Then the free energy rate at fixed $(\theta,\{ {\bm \alpha} \}, \xi)$ arises from \eqref{eq:const1}, leading to a third equality for stress in isotropic logarithmic elasticity \cite{fitzgerald1980,criscione2000,claytonAMECH2025}:
\begin{align}
\label{eq:log1}
(\partial_{ {\bm V}^E} \Psi ):\dot{\bm V}^E = [ J {\bm \sigma}^E ({\bm V}^E)^{-1}] :\dot{\bm V}^E
%= J {\bm \sigma}: [\dot{\bm V}^E ({\bm V}^E)^{-1}] 
= J {\bm \sigma}^E : \dot{\bm e}^E = (\partial_{ {\bm e}^E} \Psi ):\dot{\bm e}^E \quad \Rightarrow \quad
{\bm \sigma}^E = J^{-1} \partial_{ {\bm e}^E} \Psi.
\end{align}
Letting $\theta = \theta({\bm e}^E ({\bm F}^E), \eta, \{ {\bm \alpha} \}, \xi)$, a Legendre transform \cite{claytonNCM2011,marsden1983} produces analogs of \eqref{eq:const1} in terms of internal energy:
\begin{align}
\label{eq:const2}
& {\bm \sigma}^E = J^{-1} ( \partial_{ {\bm F}^E} U )({\bm F}^E)^{ {\mathsf T}}
= J^{-1}  ( \partial_{ {\bm V}^E} U ){\bm V}^E, \qquad \theta =   \partial_\eta U.
\end{align}

Heat conduction and viscous response are isotropic.
Thermal conductivity is $k_\theta \geq 0$. 
Bulk viscosity is $\kappa_V \geq 0$, and shear viscosity is $\mu_V \geq 0$.
Any or all of $k_\theta, \kappa_V, \mu_V$ can depend on isotropic invariants of thermodynamic state
variables $({\bm e}^E, \theta, \{ {\bm \alpha} \}, \xi, {\bm d})$ including deformation rate to enable non-Newtonian flow. 
More definitive forms are given in Section \ref{sec3.8}.
Spatial heat flux and spatial viscous stress are assigned traditionally as \cite{malvern1969}
\begin{align}
& {\bm q } = -k_\theta \nabla \theta; \qquad {\bm \tau}^V = J {\bm \sigma}^V = \kappa_V ({\rm tr} {\bm d}) {\bm 1} +2  \mu_V \bar{\bm d},
\qquad \bar{\bm d} = {\bm d} -{\textstyle{\frac{1}{3}}} ({\rm tr} {\bm d}) {\bm 1},
\nonumber
\\
& p^V = - J^{-1} \kappa_V ({\rm tr} {\bm d}), \qquad \bar{\bm \sigma}^V = 2 J^{-1} \mu_V \bar{\bm d}.
\label{eq:fluxvisc1}
\end{align}
 Entropy production from conduction and viscous stress is always non-negative:
\begin{align}
\label{eq:fluxvisc2}
-J {\bm q} \cdot \nabla \theta = J k_\theta |\nabla \theta|^2 \geq 0, \qquad
 {\bm \tau}^V : {\bm d} =  \kappa_V ({\rm tr} {\bm d})^2 + 2 \mu_V \bar{\bm d}: \bar{\bm d}  \geq 0.
\end{align}

Following classical thermodynamic procedures (cf.~\cite{claytonNCM2011}), 
the local energy balance is transformed into a temperature rate equation.
Specifically, using the last of \eqref{eq:conslaws} in conjunction with \eqref{eq:const1}, 
chain-rule expansion of $\dot{\eta}({\bm e}^E, \theta, \{ {\bm \alpha} \}, \xi)$, and $\dot{U} = \dot{\Psi} + \dot{\eta}\theta + \eta \dot{\theta}$ from \eqref{eq:energies} gives
\begin{align}
c \dot{\theta} = & {\bm M}^E: {\bm D}^P - c \theta {\bm \Gamma}: \dot{\bm e}^E   
+ \{ \bm \pi  -  \theta \partial_\theta  \bm \pi \} \cdot  \{ \dot{\bm \alpha} \} + (\zeta - \theta \partial_\theta \zeta) \dot{\xi}
\nonumber 
\\
\label{eq:temprate}
& \qquad \qquad \qquad \qquad
+  [ \kappa_V ({\rm tr} {\bm d})^2 + 2 \mu_V \bar{\bm d}: \bar{\bm d} ] + J \nabla \cdot (k_\theta \nabla \theta) + \rho_0 r;
\\
\label{eq:cv1}
c = & - \theta \partial^2_{\theta \theta} \Psi = \partial_\theta U, \qquad {\bm \Gamma} = - [ \partial^2_{\theta {\bm e}^E} \Psi]/c = - [ \partial^2_{\eta {\bm e}^E} U ]/ \theta.
\end{align}
Specific heat, as energy per unit reference volume, at constant strain is $c$. Gr\"uneisen's tensor is $\bm \Gamma$. 

\subsection{Phase-field fracture mechanics}
\label{sec2.3}
 The present phase-field representation of ductile and brittle fracture, including mechanisms of localized shear yielding, chain scission, and crazing for polymers, follows many notions of Refs.~\cite{schanzel2012,miehe2015b,dal2022,bas2022,claytonAMECH2025}.
 Herein, damage order parameter $\xi \in [0,1]$ takes a value of 0 for the pristine state and 1 for the fully degraded state.
Healing is forbidden in the current application, meaning $\dot{\xi} \geq 0$.

The forthcoming general treatment follows Refs.~\cite{miehe2015b,claytonAMECH2025} for isotropic damage. 
Denote by
$l \geq 0$ a constant regularization length approximating the half-width of the significantly damage-softened zone \cite{miehe2010}. An integral over the domain of isotropic crack surface density $\check{\gamma}(\xi({\bm X},t),\nabla_0 \xi({\bm X},t))$ then defines the regularized crack surface area functional $\Upsilon[\xi(t)]$:
\begin{align}
\label{eq:Ups}
\Upsilon[\xi]  = \int_{\Omega_0} \check{\gamma} \, \d \Omega_0 =  \int_{\Omega_0} \left[ \frac{1}{2 l} \xi^2 
+ \frac{l}{2} | \nabla_0 \xi |^2 \right] \d \Omega_0.
\end{align}
A kinetic law for $\Upsilon$ is prescribed as follows, along with admissible natural or essential boundary conditions, recalling that ${\bm n}_0({\bm X})$ is the unit normal vector to $\partial \Omega_0$:
\begin{align}
\label{eq:xiglob}
l \frac{\d}{\d t} \Upsilon = \int_{\Omega_0} [ (1-\xi) {\mathcal H} - {\mathcal R} ] \dot{\xi} \, \d \Omega_0 , \quad
{\mathcal R} = {\textstyle{\frac{1}{2}}} \beta_\xi \dot{\xi} ; \quad [\nabla_0 \xi \cdot {\bm n}_0 = 0 \, \, \, \text{or} \, \, \, \xi = \xi_0 \, \, \forall \, {\bm X} \in \partial \Omega_0].
\end{align}
Dimensionless forcing and rate functions are respectively written as $\mathcal H$ and $\mathcal R$.
A scalar viscosity for damage kinetics, with dimensions of time, is denoted by $\beta_\xi(\bullet) \geq 0 $;
this can be a function of thermodynamic state and rate variables \cite{gurtin1996,claytonMRC2023}. 
Differentiation of \eqref{eq:Ups}, integration by parts, and invocation of the divergence theorem
with either choice of boundary conditions in \eqref{eq:xiglob} gives
\begin{align}
\label{eq:Upsdot}
\frac{ \d }{\d t} \Upsilon = \int_{\Omega_0} \left[ \frac{\xi}{l} \dot{\xi} + l \nabla_0 \xi \cdot \nabla_0 \dot{\xi} \right] \d \Omega_0 = \int_{\Omega_0} \left[ \frac{\xi}{l} - l \nabla_0^2 \xi \right] \dot{\xi} \d \Omega_0.
\end{align}
Setting \eqref{eq:xiglob} equal to \eqref{eq:Upsdot} and localizing, the kinetic law
with standard initial conditions is
\begin{align}
\label{eq:xidot1}
\beta_\xi \dot{\xi} = 2(1 - \xi) {\mathcal H} - 2[\xi - l^2 \nabla_0^2 \xi], \qquad \xi({\bm X},t_0) = \xi_0({\bm X}) = 0.
\end{align}
Irreversibility is enforced following arguments established in Refs.~\cite{miehe2015a,miehe2015b}. Require the history function $\mathcal H ({\bm X},t) = \mathcal H ( \Phi(\{ {\bm \alpha} ({\bm X},t) \}), t)$ to be a continuous non-decreasing function of function $\Phi \geq 0$ . Function $\Phi$, in turn, can depend on any or all of the state variables $\{ \bm \Lambda \}$ that exclude $\xi$:
\begin{align}
\label{eq:Hzeta}
{\mathcal H} = \underset{s \in [0, \, t]} {\max} \Phi(\{ {\bm \Lambda} \} ({\bm X},s)) \geq 0,
\qquad 
\{ {\bm \Lambda} \} ({\bm X},t)= \{{\bm e}^E, \theta, \{ {\bm \alpha} \} \} ({\bm X},t), \qquad
\frac{\d}{\d t} \Upsilon \geq 0.
\end{align}
Henceforth, $s$ is a dummy time variable not to be confused with entropy per unit mass. Following from \eqref{eq:Hzeta}, rate dependent ($\beta_\xi > 0$) and independent ($\beta_\xi \rightarrow 0$)
versions of \eqref{eq:xidot1} are \cite{miehe2015a,miehe2015b}
\begin{align}
\label{eq:ratedep}
& \dot{\xi} = (1/\beta_\xi) \langle 2 (1-\xi) \Phi - 2\xi + 2 l^2 \nabla_0^2 \xi \rangle, \qquad [\beta_\xi > 0];
\\
\label{eq:rateind}
&  \dot{\xi} [(1-\xi) \Phi - \xi + l^2 \nabla_0^2 \xi] = 0, \qquad \dot{\xi} \geq 0, \qquad  (1-\xi) \Phi - \xi + l^2 \nabla_0^2 \xi \leq 0, \quad [\beta_\xi = 0].
\end{align}
Rate-dependent law \eqref{eq:ratedep} is used primarily in what follows.
Denote right-continuous Heaviside function by ${\mathsf H}(\square)$, whereby ${\mathsf H}(0) = 1$.
Set $\beta_\xi = \beta^\xi_0 / {\mathsf H} (\zeta)$ where $\beta_0^\xi = {\rm const} > 0$ and $\zeta = - \partial_\xi \Psi$.
It follows from \eqref{eq:ratedep} that any dissipation from fracture in \eqref{eq:Dissr} must be non-negative:
\begin{align}
\label{eq:Dxi}
\xi({\bm X},t) \in [0,1], \quad \dot{\xi} \geq 0;  
\quad {\mathfrak D}^\xi = \zeta \dot{\xi}   = (\zeta {\mathsf H} (\zeta) /\beta^\xi_0)  \langle 2 (1-\xi) \Phi - 2\xi + 2 l^2 \nabla_0^2 \xi \rangle \geq 0.
\end{align}
The needed function $\Phi$ is assigned for amorphous polymer physics in Section~\ref{sec3.5}.

\section{Constitutive theory: amorphous polymers}
\label{sec3}

\subsection{Internal state variables, inelastic deformation, and free energy}
\label{sec3.1}

The framework of Section~\ref{sec2} is specialized to amorphous thermoplastic polymers
in a glassy state at room temperature and atmospheric pressure.
Internal state variables consist of the fracture phase-field parameter field $\xi({\bm X},t)$ of Section~\ref{sec2.3} and
the following dimensionless fields:
\begin{align}
\label{eq:alpha}
\{ {\bm \alpha} \}({\bm X},t) \ni 
\begin{cases}
 \{ {\bm \Upsilon} \} ({\bm X},t) &: \text{tensor-valued viscoelastic configurational variables},
\\  \{ \varsigma \} ({\bm X},t) &:  \text{scalar-valued plastic isotropic hardening/softening variables},
\\  {\bm A} ({\bm X},t) &: \text{tensor-valued plastic anisotropic hardening variable},
\\  \varphi ({\bm X},t) &: \text{scalar-valued residual free volume change},
\\  \phi ({\bm X},t) &: \text{scalar-valued void volume fraction},
\\  \omega ({\bm X},t) &: \text{scalar-valued melt fraction of non-dissociated material},
\\  \Xi ({\bm X},t)&: \text{scalar-valued decomposition fraction}.
\end{cases}
\end{align}
Governing equations for $\{ {\bm \Upsilon} \}$
are discussed in Section~\ref{sec3.3}, ($\{ \varsigma \},{\bm A}, \varphi$ ) in Section~\ref{sec3.4},
$\phi$ in Section~\ref{sec3.5}, $\omega$ in Section~\ref{sec3.6}, and $\Xi$ in Section~\ref{sec3.7}.
The following constraints are noted:
\begin{align}
\label{eq:alphac}
{\bm \Upsilon} = {\bm \Upsilon}^{\mathsf T}; \quad
{\bm A} = {\bm A}^{\mathsf T}, \quad \det {\bm A} = 1; \quad
0 \leq \varphi \ll 1; \quad \phi \in [0,1); \quad \omega \in [0,1]; \quad \Xi \in [0,1]. 
\end{align}
Henceforth, the state variable $\omega$ refers to a rubbery melt phase rather than a liquid melt.
The latter is not formally distinguished with a state variable, but rather is achieved in practice by a temperature-softening shear modulus.
Melting is not applicable for the decomposed phase; $\omega (\bm X,t)  + \Xi (\bm X,t) > 1$ is possible.
When the material is partially melted or partially decomposed (e.g., $\omega (\bm X,t) \in (0,1)$ or $\Xi (\bm X,t) \in (0,1)$), all phases at $(\bm X,t)$
share the same deformation gradient $\bm F$, the same values of ${\bm F}^E$
and ${\bm F}^P$, and the same values of temperature $\theta (\bm X,t)$.
For example, if material undergoes plastic deformation prior to liquification, 
${\bm F}^P$ loses physical meaning but is not automatically reversed when the solid liquifies. 
Similarly, $\xi$ and the internal state variables in \eqref{eq:alpha} are defined only for
the entire volume element $\d \Omega$, even in a mixed-phase state,
and are not formally distinguished among phases.
Stress, entropy, and other response functions for the volume element, if of mixed phases,
are obtained from thermodynamic relations in Section~\ref{sec2} where free energy density
is a weighted sum of contributions from each phase as assigned in what follows in \eqref{eq:Helmdecomp}.

Total plastic deformation rate ${\bm D}^P$ defined in  \eqref{eq:FeFp} and rate of plastic volume change $J^P$ of \eqref{eq:Jp} contain contributions
from isochoric shear-plastic flow, generally anisotropic dilatational crazing or scission, 
residual free volume change, and isotropic structural or chemical decomposition:
\begin{align}
\label{eq:Dpexp}
& {\bm D}^P = \dot{\bar{\bm \epsilon}}^P + \frac{\dot \phi}{1-\phi} {\bm N}^\phi + \frac{\dot{\varphi} c_\varphi }{3(1-\varphi c_\varphi)} {\bm 1} % + \frac{\dot{\omega} \delta_\omega }{3(1-\omega \delta_\omega)} {\bm 1}
+  \frac{\dot{\Xi} \delta_\Xi }{3(1-\Xi \delta_\Xi)} {\bm 1},
\\
\label{eq:Jpexp}
& \dot{J}^P (J^P)^{-1} = \dot{\phi}/(1-\phi) + \dot{\varphi} c_\varphi /(1-\varphi c_\varphi) 
% + \dot{\omega} \delta_\omega /(1-\omega \delta_\omega) 
+ \dot{\Xi} \delta_\Xi /(1-\Xi \delta_\Xi) ,
\\
\label{eq:kinP}
& \dot{\bar{\bm \epsilon}}^P = (\dot{\bar{\bm \epsilon}}^P)^{\mathsf T}, 
\quad {\rm tr} \, \dot{\bar{\bm \epsilon}}^P = 0; \quad {\bm N}^\phi = {\bm n}_\phi \otimes {\bm n}_\phi,
\quad {\bm n}_\phi \cdot {\bm n}_\phi = 1, \quad {\bar{\bm N}}^\phi = {\bm N}^\phi - {\textstyle{\frac{1}{3}}} {\bm 1}; \\
\label{eq:deltas}
& \bar{\bm D}^P =  \dot{\bar{\bm \epsilon}}^P + [\dot{\phi}/(1-\phi)] \bar{\bm N}^\phi
; \quad c_\varphi = {\rm const} \in [0,1), \quad \delta_\Xi  = {\rm const} \in (-\infty , 1).
\end{align}
The non-negative parameter $c_\varphi$ quantifies the amount of free volume change that contributes
to net plastic dilatation. If all local free volume in a material element is offset by increased density from of
polymer chains, local mass density is unchanged and $c_\varphi \rightarrow 0$.
Relative volume change from complete shock decomposition is quantified by $\delta_\Xi$.
The deviatoric part of the total plastic deformation rate is $\bar{\bm D}^P$.
The unit vector in the direction of craze or scission crack opening, normal to a mode I crack plane, is ${\bm n}_\phi$. 
From \eqref{eq:Dpexp}--\eqref{eq:deltas} and  noting ${\rm tr} \, \bm{M}^E = - 3Jp^E$, total dissipation from inelasticity entering \eqref{eq:Dissr} is
\begin{align}
\label{eq:Dissp}
{\bm M}^E: {\bm D}^P = \bar{\bm M}^E : \dot{\bar{\bm \epsilon}}^P +   \left[ \frac{\bar{\bm M}^E: {\bar {\bm N}}^\phi
- J p^E}{1-\phi} \right] \dot{\phi} - \left[ \frac{ J p^E  c_\varphi }{1-\varphi c_\varphi} \right] \dot{\varphi}
% -  \left[ \frac{ J p^E  \delta_\omega }{1-\omega \delta_\omega} \right] \dot{\omega}
- \left[  \frac{J p^E  \delta_\Xi }{1-\Xi \delta_\Xi} \right] \dot{\Xi}.
\end{align}

If $\omega$ were to describe a transition from crystalline to amorphous state, or solid to liquid, then nonzero volume
change,  and possible shape change \cite{levitas2011b,claytonARX2025}, would be expected \cite{loo2001}. However, specific volume remains continuous across the glass transition
in amorphous polymers of interest (e.g., \cite{richeton2007,clements2012a,srivastava2010}),
so \eqref{eq:Dpexp} contains no explicit contribution from $\dot{\omega}$.
Inelastic shape changes from shock decomposition are physically possible but are omitted here for brevity and lack of quantifiable data on their magnitudes.
The change in free volume $\varphi$ that affects $J^P$ is very small \cite{dal2022} 
and therefore often omitted for brevity \cite{anand2003}. This quantity is somewhat analogous to
the volume or mass density change from dislocations or other lattice defects in crystalline solids \cite{claytonNCM2011,claytonAMECH2025} usually omitted in crystal plasticity theory. Assignment of ${\bm n}_\phi ({\bm X},t)$ is discussed in Section~\ref{sec3.5}.

A standard uniform reference state at $t=t_0$ is defined as follows. The material is undeformed, undamaged, fully
dense (i.e., null porosity), and fully glassy (i.e., no melting or decomposition). Corresponding initial conditions on 
deformation gradient fields, temperature, $\xi$, and $\{ {\bm \alpha } \} $ are
\begin{align}
\nonumber
& {\bm F}({\bm X},t_0) = {\bm F}^E({\bm X},t_0) = {\bm F}^P({\bm X},t_0) = {\bm 1}, \qquad {\theta}({\bm X},t_0) = \theta_0 = {\rm const} \in (0,\theta_G);
\\
\nonumber
& \xi({\bm X},t_0) = 0, \qquad  \{ {\bm \Upsilon} \} ({\bm X},t_0) = \{ {\bm 0} \}, 
\qquad  \{ \varsigma \} ({\bm X},t_0) = \{ 0 \},  \qquad {\bm A} ({\bm X},t_0) = {\bm 1}, 
\\
& \varphi ({\bm X},t_0) = \phi ({\bm X},t_0)  = \omega({\bm X},t_0)  = \Xi ({\bm X},t_0) = 0.
\label{eq:ICS}
\end{align}
The glass transition temperature at atmospheric pressure and null mechanical loading is $\theta_G = {\rm const}$.

Helmholtz free energy per unit reference volume $\Psi$ comprises functions for glassy phase $\Psi^G$,
melt phase $\Psi^M$, and decomposed products $\Psi^D$,
recalling from \eqref{eq:kinvars2} that $e^E_V$ and $\bar{\bm e}^E$ are finite logarithmic elastic volume and shape strains:
\begin{align}
\nonumber 
\Psi ({\bm e}^E ,\theta,\xi, \{ {\bm \Upsilon} \}, \{ \varsigma \}, {\bm A},  \varphi, \phi, \omega, \Xi)  = &
(1-\Xi) [ (1 - \omega) \Psi^G (\bar{\bm e}^E, e^E_V, \theta,\xi, \{ {\bm \Upsilon} \}, \{ \varsigma \}, {\bm A}, \varphi, \phi) 
\\
& +   \omega  \Psi^M (\bar{\bm e}^E, e^E_V, \theta,\xi, \{ {\bm \Upsilon} \}, \{ \varsigma \}, {\bm A}, \varphi, \phi) ]
 + \Xi \Psi^D (e^E_V, \theta).
\label{eq:Helmdecomp}
\end{align}
From the rightmost term in \eqref{eq:Helmdecomp}, shock-decomposed products are dissociated such that shear strain and internal state variables contribute no free energy. Energies can be transformed to per unit mass dimensions via $\psi =  \Psi / \rho_0$, where $\rho_0 = {\rm const}$ is
  mass density of glassy solid in state \eqref{eq:ICS}.
Internal energy can be obtained as a function of strain, internal state, and temperature via
$U = \Psi - \theta \partial_\theta \Psi$.

Free energies are deconstructed into volumetric and shear
elastic strain energies $\Psi_V^E$ and $\Psi_S^E$, 
bulk thermoelastic coupling energy $\Psi_\beta^E$,
volumetric and shear viscoelastic configurational
energies $\Psi_V^\Upsilon$ and $\Psi_S^\Upsilon$, 
energy of plastic microstructure evolution $\Psi^P$ (e.g., cold work \cite{rittel1999,shao2017}), and specific and latent heat energy $\Psi^\theta$. Denoting degradation functions from damage by $f^E_V$ and $f^E_S$,
\begin{align}
\Psi^G ({\bm e}^E, \theta,\xi, \{ {\bm \Upsilon} \}, \{ \varsigma \}, {\bm A}, \varphi, \phi) =
\quad & f^E_V(\xi,e^E_V) [ \Psi^{G,E}_V(e^E_V) +
\Psi^{G,E}_\beta (e^E_V,\theta) + \Psi^{G,\Upsilon}_V(e^E_V,\{ {\bm \Upsilon}_V \},\theta )] 
\nonumber
 \\
 + & f^E_S (\xi) [ \Psi^{G,E}_S(\bar{\bm e}^E,e^E_V,\theta) + \Psi^{G,\Upsilon}_S(\bar{\bm e}^E,e^E_V,\{ {\bm \Upsilon}_S \},\theta )] 
 \nonumber 
 \\
  + & f^E_S(\xi) [ \Psi^{G,P} (\{ \varsigma \}, {\bm A}, \varphi, \phi, \theta) ]
  + \Psi^{G,\theta}(\theta),
\label{eq:HelmG}
\end{align}
\begin{align}
\Psi^M ({\bm e}^E, \theta,\xi, \{ {\bm \Upsilon} \}, \{ \varsigma \}, {\bm A}, \varphi, \phi) =
\quad & f^E_V(\xi,e^E_V) [ \Psi^{M,E}_V(e^E_V) + 
 \Psi^{M,E}_\beta (e^E_V,\theta) + 
\Psi^{M,\Upsilon}_V(e^E_V,\{ {\bm \Upsilon}_V \},\theta )] 
\nonumber
\\
  + & f^E_S (\xi) [ \Psi^{M,E}_S(\bar{\bm e}^E,e^E_V,\theta) + \Psi^{M,\Upsilon}_S(\bar{\bm e}^E,e^E_V,\{ {\bm \Upsilon}_S \},\theta )] 
 \nonumber 
 \\
  + & f^E_S(\xi) [ \Psi^{M,P} (\{ \varsigma \}, {\bm A}, \varphi, \phi, \theta) ]
  + \Psi^{M,\theta}(\theta),
  \label{eq:HelmM}
\end{align}
\begin{align}
\Psi^D (e^E_V, \theta) =   \Psi^{D,E}_V(e^E_V) +  \Psi^{D,E}_\beta (e^E_V,\theta)  + \Psi^{D,\theta}(\theta);
  \label{eq:HelmD}
 \end{align}
 Degradation functions $f^E_S(\xi)$ and $f^E_V(\xi,e^E_V)$ depend on $\xi$ 
in the usual quadratic sense \cite{miehe2015b,claytonJMPS2021,narayan2021,dal2022} and $f^V_E$ on volume change $e^E_V$ ensuring a positive bulk modulus in compression \cite{claytonIJF2014}:
 \begin{align}
 \label{eq:fE}
 f^E_S(\xi) = (1-\xi)^2, \qquad f^E_V (\xi,e^E_V) = (1 - \xi {\mathsf H} (e^E_V) )^2.
\end{align}
Recall the right-continuous Heaviside unit step function is ${\mathsf H}(\square)$; $f^E_V(\xi,e^E_V \geq 0) = f^E_S(\xi)$. As in Refs.~\cite{narayan2021,dal2022,claytonAMECH2025}, functions \eqref{eq:fE} do not depend explicitly on $\phi$, but porosity can affect kinetics of $\xi$. Superscripts $\square^{G,*}$, $\square^{M,*}$, $\square^{D,*}$ label free energies of  glass, melt, and decomposed states. Notation $ {\bm \Upsilon}_V$ and ${\bm \Upsilon}_S$ labels
configurational variables for volumetric and shear deformation. Forms for other functions on right sides of 
\eqref{eq:HelmG}--\eqref{eq:HelmD} are posited in Sections ~\ref{sec3.2}, \ref{sec3.3}, and \ref{sec3.4}.

\subsection{Thermoelasticity}
\label{sec3.2}

Thermoelastic free energy describes an equilibrium response with no viscoelastic, plastic, or rate effects.
Dependence on strain is through two independent invariants $e^E_V = {\rm tr} \bm{e}^E = \ln J^E$ and 
$|\bar{\bm{e}}^E|^2 = \bar{\bm{e}}^E:\bar{\bm{e}}^E = {\rm tr} [(\bar{\bm{e}}^E)^2]$. A third invariant is admissible for nonlinear isotropic elasticity \cite{marsden1983} but  is not implemented. For the decomposed substance, all free energy  $\Psi^D$ in \eqref{eq:HelmD} is thermoelastic.  For the solid in glassy \eqref{eq:HelmG} and rubbery \eqref{eq:HelmM} states, thermoelastic free energy $\Psi^{*,E}$ comprises terms from
purely volumetric deformation, shear deformation, thermoelastic coupling, and specific heat.

Volumetric contributions are each akin to a ``cold curve'' \cite{davison2008} in the limit that $\theta_0 \rightarrow 0$.
These contribute an isotropic elastic pressure $p^{*,E}_V$ that depends only on volume change:
\begin{align}
\label{eq:coldG}
& \Psi^{G,E}_V =  \begin{cases}
& B_0^{G,E} (e^E_V)^2 \left[ {\textstyle{\frac{1}{2!}}} - {\textstyle{\frac{1}{3!}}}  B_1^{G,E} e^E_V + {\textstyle{\frac{1}{4!}}} B_2^{G,E} (e^E_V)^2 - {\textstyle{\frac{1}{5!}}} B_3^{G,E} (e^E_V)^3 + \cdots \right],  \qquad [e^E_V \leq 0],
\\
& {\textstyle{\frac{1}{2}}} B_0^{G,E} (e^E_V)^2,  \qquad \qquad \qquad \qquad \qquad  \qquad \qquad \qquad \qquad \qquad \qquad \, \, \,  [e^E_V > 0];
\end{cases} 
\\
\label{eq:pG}
& 
p^{G,E}_V = -  \frac{1}{J} \frac{\partial \Psi^{G,E}}{ \partial e^E_V }  = 
\begin{cases}
& - {\textstyle{\frac{1}{J}}} B_0^{G,E} e^E_V \left[ 1 - {\textstyle{\frac{1}{2!}}}  B_1^{G,E} e^E_V + {\textstyle{\frac{1}{3!}}} B_2^{G,E} (e^E_V)^2 - {\textstyle{\frac{1}{4!}}} B_3^{G,E} (e^E_V)^3 \cdots \right],  \quad [e^E_V \leq 0],
\\
&- {\textstyle{\frac{1}{J}}} B_0^{G,E} e^E_V, \qquad \qquad \qquad \qquad  \qquad \qquad \qquad \qquad \qquad \qquad \quad \! \!  \! \! [e^E_V > 0];
\end{cases} 
\end{align}
\begin{align}
%\\
\label{eq:coldM}
& \Psi^{M,E}_V =  \begin{cases}
& B_0^{M,E} (e^E_V)^2 \left[ {\textstyle{\frac{1}{2!}}} - {\textstyle{\frac{1}{3!}}}  B_1^{M,E} e^E_V + {\textstyle{\frac{1}{4!}}} B_2^{M,E} (e^E_V)^2 - {\textstyle{\frac{1}{5!}}} B_3^{M,E} (e^E_V)^3 + \cdots \right],  \qquad [e^E_V \leq 0],
\\
& {\textstyle{\frac{1}{2}}} B_0^{M,E} (e^E_V)^2,  \qquad \qquad \qquad \qquad \qquad  \qquad \qquad \qquad \qquad \qquad \qquad \quad \! [e^E_V > 0];
\end{cases} 
\\
\label{eq:pM}
& 
p^{M,E}_V = -  \frac{1}{J} \frac{\partial \Psi^{M,E}}{ \partial e^E_V }  = 
\begin{cases}
& - {\textstyle{\frac{1}{J}}} B_0^{M,E} e^E_V \left[ 1 - {\textstyle{\frac{1}{2!}}}  B_1^{M,E} e^E_V + {\textstyle{\frac{1}{3!}}} B_2^{M,E} (e^E_V)^2 - {\textstyle{\frac{1}{4!}}} B_3^{M,E} (e^E_V)^3 \cdots  \right],  \, \, [e^E_V \leq 0],
\\
&- {\textstyle{\frac{1}{J}}} B_0^{M,E} e^E_V, \qquad \qquad \qquad \qquad  \qquad \qquad \qquad \qquad \qquad \qquad \, \,  [e^E_V > 0].
\end{cases} 
%\end{align}
%\begin{align}
\\
\label{eq:coldD}
& \Psi^{D,E}_V =  \begin{cases}
& B_0^{D,E} (e^E_V)^2 \left[ {\textstyle{\frac{1}{2!}}} - {\textstyle{\frac{1}{3!}}}  B_1^{D,E} e^E_V + {\textstyle{\frac{1}{4!}}} B_2^{D,E} (e^E_V)^2 - {\textstyle{\frac{1}{5!}}} B_3^{D,E} (e^E_V)^3 + \cdots \right],  \qquad [e^E_V \leq 0],
\\
& 0,  \qquad \qquad \qquad \qquad \qquad  \qquad \qquad \qquad \qquad \qquad \qquad \qquad \qquad  \, \, \, \, [e^E_V > 0];
\end{cases} 
\\
\label{eq:pD}
& 
p^{D,E}_V = -  \frac{1}{J} \frac{\partial \Psi^{D,E}}{ \partial e^E_V }  = 
\begin{cases}
& - {\textstyle{\frac{1}{J}}} B_0^{D,E} e^E_V \left[ 1 - {\textstyle{\frac{1}{2!}}}  B_1^{D,E} e^E_V + {\textstyle{\frac{1}{3!}}} B_2^{D,E} (e^E_V)^2 - {\textstyle{\frac{1}{4!}}} B_3^{D,E} (e^E_V)^3 \cdots \right],  \, \, [e^E_V \leq 0],
\\
& 0, \qquad \qquad \qquad \qquad  \qquad \qquad \qquad \qquad \qquad \qquad \qquad \qquad \! \! \! [e^E_V > 0].
\end{cases} 
\end{align}
A linear-type response is invoked for elastic tensile states in glassy and melt phases since volumetric expansion is constrained by limits of fracture or cavitation.
Decomposed products comprise gaseous fluid and disassociated particulate matter \cite{dattelbaum2019,coe2022}; products  are approximated as having null (tensile) cold pressure for $e^E_V > 0$.
The isothermal equilibrium bulk modulus at $e^E_V = 0$ and $\theta = \theta_0$ is $B_0^{*,E}$.
Dimensionless constants for nonlinear pressure-volume response are $B^{*,E}_1, B^{*,E}_2, B^{*,E}_3, \ldots$.
The first is related to the pressure derivative of isothermal equilibrium tangent bulk modulus $B^{*,E} = -J 
\partial p / \partial J$ at the reference
state via $B^{*,E}_1 + 2 = (\partial B^{*,E} / \partial p)|_0 $  \cite{claytonIJES2014,claytonNEIM2019}.
Energies $\Psi^{*,E}_V$ are non-negative so long as $B^{*,E}_k \geq 0 \, \forall \, k = 0,1,2, \ldots$. Only $B^{*,E}_0 > 0$ is essential for generic applications. Note coupling $\Psi^{*,E}_\beta$ and viscoelastic $\Psi^{*,\Upsilon}_V$ and $\Psi^{*,\Upsilon}_S$ energies, and not cold energy $\Psi^{*,E}_V$, all specified later, can affect temperature dependence of the tangent bulk modulus.

Shear-dominant contributions in condensed phases arise from an equibrium shear modulus $G^{*,E}(e^E_V,\theta)$ that depends at most linearly on temperature (omitted in Ref.~\cite{menikoff2004}) and up to quadratically on volume change as in Ref.~\cite{menikoff2004}.
A deviatoric Cauchy stress $\bar{\bm \sigma}^{*,E}_S$ is the primary result, but a pressure contribution
$p^{*,E}_S$ arises from volume dependence of $G^{*,E}$:
\begin{align}
\label{eq:shearG}
& \Psi^{G,E}_S = G_0^{G,E} \left[ 1 + (G^{G,E}_\theta / G_0^{G,E})(\theta - \theta_0) 
-  (B_0^{G,E} \{ G^{G,E}_p -G^{G,E}_2 e^E_V \} / G_0^{G,E}) e^E_V \right] \bar{\bm e}^E:\bar{\bm e}^E, \\
& \bar{\bm \sigma}^{G,E}_S = \frac{1}{J} \frac{\partial \Psi^{G,E}_S } {\partial \bar{\bm e}^E} = 
\frac{2}{J}  G_0^{G,E} \left[ 1 + \frac{G^{G,E}_\theta}{ G_0^{G,E}}(\theta - \theta_0) 
- \frac{B_0^{G,E}}{G_0^{G,E}} ( G^{G,E}_p - G^{G,E}_2  e^E_V )e^E_V \right] \bar{\bm e}^E,
\label{eq:shearGdev}
 \\ 
& p^{G,E}_S = - \frac{1}{J} \frac { \partial \Psi^{G,E}_S} {\partial e^E_V} = 
 \frac{1}{J} B_0^{G,E} ( G^{G,E}_p - 2 G^{G,E}_2 e^E_V) \bar{\bm e}^E:\bar{\bm e}^E;
 \label{eq:shearGp}
  \\
 \label{eq:shearM}
& \Psi^{M,E}_S = G_0^{M,E} \left[ 1 + (G^{M,E}_\theta / G_0^{M,E})(\theta - \theta_G) 
-  (B_0^{M,E} \{ G^{M,E}_p -G^{M,E}_2 e^E_V \} / G_0^{M,E}) e^E_V \right] \bar{\bm e}^E:\bar{\bm e}^E,
 \\
& \bar{\bm \sigma}^{M,E}_S = \frac{1}{J} \frac{\partial \Psi^{M,E}_S } {\partial \bar{\bm e}^E} = 
\frac{2}{J}  G_0^{M,E}  \left[ 1 + \frac{G^{M,E}_\theta}{ G_0^{M,E}}(\theta - \theta_G) 
- \frac{B_0^{M,E}}{G_0^{M,E}} ( G^{M,E}_p - G^{M,E}_2  e^E_V )e^E_V \right] \bar{\bm e}^E,
\label{eq:shearMdev}
  \end{align}
 \begin{align}
% \\
& p^{M,E}_S = - \frac{1}{J} \frac { \partial \Psi^{M,E}_S} {\partial e^E_V} = 
 \frac{1}{J} B_0^{M,E} ( G^{M,E}_p - 2 G^{M,E}_2 e^E_V) \bar{\bm e}^E:\bar{\bm e}^E.
 \label{eq:shearMp}
\end{align}
The equilibrium shear modulus in an unstrained state is $G^{*,E}_0 > 0$, referred to $\theta_0$ for glass and $\theta_G$ for melt.
Temperature and pressure derivatives of $G^{*,E}$ are respectively, to first order, 
$G^{*,E}_\theta$ and $G^{*,E}_p$. Constant $G^{*,E}_2$ gives a second-order volume dependency. 
In typical solids, $G^{*,E}_\theta < 0 $ and $G^{*,E}_p > 0$.
This would enable the tangent shear modulus to become negative at large temperatures or tensile states.
To circumvent this, the tangent shear modulus and $\Psi^{*,E}_S$ are forbidden to be negative:
\begin{align}
\label{eq:negG}
& \frac{G^{G,E}_\theta}{ G_0^{G,E}}(\theta - \theta_0) 
- \frac{B_0^{G,E}}{G_0^{G,E}} ( G^{G,E}_p - G^{G,E}_2  e^E_V )e^E_V
< - 1 \, \Rightarrow \, \Psi^{G,E}_S = 0 = p^{G,E}_S, \quad \bar{\bm \sigma}^{G,E}_S = {\bm 0};
\\
& \frac{G^{M,E}_\theta}{ G_0^{M,E}}(\theta - \theta_G) 
- \frac{B_0^{M,E}}{G_0^{M,E}} ( G^{M,E}_p - G^{M,E}_2  e^E_V )e^E_V
< - 1 \, \Rightarrow \, \Psi^{M,E}_S = 0 = p^{M,E}_S, \quad \bar{\bm \sigma}^{M,E}_S = {\bm 0}.
\end{align}
Transition from a solid to liquid state can be linked to a vanishing tangent shear modulus \cite{born1939}.
At null strain, the transition from rubber to liquid melt would occur at a liquification temperature 
$\theta_M$:
\begin{equation}
\label{eq:thetaM}
\theta_M  = \theta_G -G^{M,E}_0 / G^{M,E}_\theta > \theta_G,\qquad  \theta_G \leq \theta_0 - G^{G,E}_0 / G^{G,E}_\theta, \qquad [G^{M,E}_\theta < 0, \, G^{G,E}_\theta <  0].
\end{equation}
Inequalities in \eqref{eq:thetaM} ensure that, at null elastic volume strain, the glassy phase has non-negative shear modulus at $\theta = \theta_G$ and glass transition temperature $\theta_G$ cannot exceed the
liquification temperature of the glass. These are valid only when shear moduli decrease with increasing $\theta$.

Equilibrium thermoelastic coupling energy $\Psi^{*,E}_\beta$ accounts for interaction energy from temperature and elastic strain. For isotropic response, this energy depends only on the first strain invariant $e^E_V$.
The corresponding contribution to the Gr\"uneisen tensor ${\bm \Gamma}$ in \eqref{eq:cv1} is spherical and allowed to depend up to quadratically on volumetric strain \cite{menikoff2004}. Let ($c^G_{V0},c^M_{V0},c^D_{V0}$) denote specific heat per unit reference volume of
(glass, melt, decomposed) phases at $\theta = \theta_0$ and  ${\bm e}^E = {\bm 0}$. 
Then prescribe for each phase the following free energy and resulting pressure contribution $p^{*,E}_\beta = -{\frac{1}{J}} \partial \Psi^{*,E}_\beta / \partial e^E_V$:
\begin{align}
\label{eq:psiBetaG}
& \Psi^{G,E}_\beta = -c^G_{V0} (\theta - \theta_0) e^E_V [ \Gamma_0^{G,E} + {\textstyle{\frac{1}{2!}}} \Gamma_1^{G,E} e^E_V
+ {\textstyle{\frac{1}{3!}}} \Gamma_2^{G,E} (e^E_V)^2],
\\
&  p^{G,E}_\beta = {\textstyle{\frac{c^G_{V0}}{J}}}(\theta - \theta_0) [ \Gamma_0^{G,E} +  \Gamma_1^{G,E} e^E_V
+ {\textstyle{\frac{1}{2!}}} \Gamma_2^{G,E} (e^E_V)^2];
\label{eq:pGbeta}
\\
\label{eq:psiBetaM}
& \Psi^{M,E}_\beta = -c^M_{V0} (\theta - \theta_0) e^E_V [ \Gamma_0^{M,E} + {\textstyle{\frac{1}{2!}}} \Gamma_1^{M,E} e^E_V
+ {\textstyle{\frac{1}{3!}}} \Gamma_2 (e^E_V)^2],
\\
& p^{M,E}_\beta = {\textstyle{\frac{c^M_{V0}}{J}}}(\theta - \theta_0) [ \Gamma_0^{M,E} +  \Gamma_1^{M,E} e^E_V
+ {\textstyle{\frac{1}{2!}}} \Gamma_2^{M,E} (e^E_V)^2];
\label{eq:pMbeta}
\\
\label{eq:psiBetaD}
& \Psi^{D,E}_\beta = -c^D_{V0} (\theta - \theta_0) e^E_V [ \Gamma_0^{D,E} + {\textstyle{\frac{1}{2!}}} \Gamma_1^{D,E} e^E_V
+ {\textstyle{\frac{1}{3!}}} \Gamma_2^{D,E} (e^E_V)^2],
\\
& p^{D,E}_\beta = {\textstyle{\frac{c^D_{V0}}{J}}}(\theta - \theta_0) [ \Gamma_0^{D,E} +  \Gamma_1^{D,E} e^E_V
+ {\textstyle{\frac{1}{2!}}} \Gamma_2^{D,E} (e^E_V)^2].
\label{eq:pDbeta}
\end{align}
The equilibrium Gr\"uneisen parameter for respective glassy, rubbery, or decomposed phases in the unstrained condition is $\Gamma^{G,E}_0$, $\Gamma^{M,E}_0$, or $\Gamma^{D,E}_0$.
Constants $\Gamma^{*,E}_1$ and $\Gamma^{*,E}_2$ enable linear and quadratic volume dependence of the
spherical part of the equilibrium Gr\"uneisen tensor entering \eqref{eq:cv1}:
\begin{align}
\label{eq:Gruntens}
{\bm \Gamma}^{*,E} & = - \frac{1}{c^*_V} \frac{\partial^2 ( f^E_V \Psi^{*,E}_\beta + f^E_S \Psi^{*,E}_S) }{\partial \theta \partial {\bm e}^E} \nonumber \\
& = \frac{c^*_{V0} }{c^*_V} \{ f^E_V [ \Gamma_0^{*,E} +  \Gamma_1^{*,E} e^E_V
+ {\textstyle{\frac{1}{2}}} \Gamma_2^{*,E} (e^E_V)^2] {\bm 1}
- 2 f^E_S \frac{G^{*,E}_\theta}{  c^*_{V0} } \bar{\bm e}^E \}.
\end{align}
% Constants $\Gamma^{*,E}_1$ and $\Gamma^{*,E}_2$ are related to the temperature
% derivative of the relaxed isothermal bulk modulus of each phase under hydrostatic elastic compression:
% \begin{align}
% (\partial B^{*,E} / \partial \theta)|_{J,\bullet} = - (c^*_{V0}/J) [\Gamma_1^{*,E} +\Gamma_2^{*,E} e^E_V].
%\end{align}

Lastly, the purely thermal energy $\Psi^{*,\theta}(\theta)$ comprises specific and possible latent heat. Denote by $\lambda^D_\theta$ a 
latent heat parameter \cite{levitas2011b,claytonCMT2022,claytonARX2025} that, when positive, increases the internal energy of decomposing material due to bond rearrangements, scissions, or dissociations \cite{graham1979,carter1995}. Denote by $c^*_\theta$
temperature derivatives of the specific heat at constant strain that herein varies linearly with temperature in each phase. 
Define ($c^G_{00}, c^M_{00}, c^D_{00}$) specific heat constants of (glass, melt, decomposed) phases linearly extrapolated downward to $\theta = 0$ from $\theta = \theta_0$.
Let $\Psi^{*,\theta}_0$ be datum constants accounting for free energy differences between phases coexisting at the same temperature and strain, vanishing by default for the glassy state. 
Contributions to specific heat from viscoelastic coupling are theoretically and physically possible \cite{nunziato1973am,nunziato1973g} but are assumed negligible here \cite{clements2012a}.
Denote by $\theta_D = {\rm const} > \theta_G$ a zero-strain decomposition temperature \cite{kashiwagi1985,beyler2002}. 
No latent heat is associated with the glass-melt transition that is often viewed as a second-order transformation
\cite{gibbs1958,wu1999};
null volume change is correspondingly assigned for this melting process in \eqref{eq:Dpexp} and Section~\ref{sec3.4}.
In contrast, decomposition is modeled as a first-order phase transition \cite{hauver1965,carter1995} with potentially, but not necessarily, nonzero latent heat and volume change
as discussed in Section~\ref{sec3.7}.
Then, given these definitions and assumptions, 
\begin{align}
& \Psi^{G,\theta} = -c^G_{00} [ \theta \ln \frac{\theta}{ \theta_0} - (\theta - \theta_0)] -  {\frac{c^G_\theta}{2}} 
(\theta - \theta_0)^2,
\quad c^*_V = -\theta \frac{ \d^2  \Psi^{*,\theta}}{ \d \theta^2} = c^*_{V0} + c^*_\theta (\theta - \theta_0),
% \\ & (\theta_* = \theta_0, \theta_G, \text{ or } \theta_D);
\label{eq:thermG}
\\
\label{eq:thermM}
& \Psi^{M,\theta} = -c^M_{00} [ \theta \ln \frac{\theta}{ \theta_0} - (\theta - \theta_0)] -  {\frac{c^M_\theta}{2}} 
(\theta - \theta_0)^2 + \Psi^{M,\theta}_0; \qquad c^*_{00} = c^*_{V0} - c^*_\theta \theta_0;
 \\
 \label{eq:thermD}
& \Psi^{D,\theta} = -c^D_{00} [ \theta \ln \frac{\theta}{ \theta_0} - (\theta - \theta_0)] -  {\frac{c^D_\theta}{2}} 
(\theta - \theta_0)^2 - \frac{\lambda^D_\theta}{\theta_D} (\theta - \theta_D) + \Psi^{D,\theta}_0.
% \quad c^D_V = c^D_{V0} + c^D_\theta (\theta - \theta_D).
\end{align}

\subsection{Viscoelasticity}
\label{sec3.3}
Strain-like configurational state variables for viscoelasticity
comprise volumetric and shear contributions $\{ {\bm \Upsilon} \} = (\{ {\bm \Upsilon}_V \},\{ {\bm \Upsilon}_S \})$.
These are further decomposed into $l = 1,\ldots,L$ bulk relaxation modes and
$m = 1,\ldots,M$ shear modes: $\{ {\bm \Upsilon}_V \} = ( {\bm \Upsilon}^V_1, \ldots, {\bm \Upsilon}^V_L)$ and $\{ {\bm \Upsilon}_S \} = ( {\bm \Upsilon}^S_1, \ldots, {\bm \Upsilon}^S_M)$.
Recall that for a local volume element of mixed phases, these internal state variables, like kinematic variables and temperature,  are not distinguished among glassy, melt, and decomposed phases.
Instead, their kinetic equations are influenced by local melt and decomposition fractions $\omega$ and $\Xi$.

The present treatment extends Refs.~\cite{holzapfel1996a,holzapfel1996b,holzapfel2001,holzapfel2002,claytonPRE2024}
to possible coexisting phases and inelastic deformations.
Assigned to each volumetric $l$ and shear $m$ relaxation process are 
% a constant scalar energy multiplier $\beta^V_l \geq 0$ and $\beta^S_m \geq 0$. 
respective relaxation times $\tau^V_l$ and $\tau^S_m$, possibly temperature dependent: 
\begin{align}
\label{eq:taurelax}
\tau^V_l (\theta) = a^V_T(\theta ; \theta_R^V) \tau^V_{Rl} (\theta_R^V) > 0, 
\qquad \tau^S_m (\theta) a^S_T(\theta ; \theta_R^S) \tau^S_{Rm} (\theta_R^S) > 0.
\end{align}
Volumetric and shear time-temperature shift functions are $a^V_T$ and $a^S_T$, for which
 inverse logarithmic and Arrhenius forms are typical for polymers \cite{williams1955,clements2012a,hu2016,federico2018}.
Reference temperatures for bulk and shear relaxation are $\theta_R^V$ and $\theta_R^S$, at 
which respective  reference times $\tau^V_{Rl}$ and $\tau^S_{Rm}$ are measured.

For shearing modes, stresses $\{ {\bm Q}^S_m  \}$
 conjugate to configurational variables and total free energy $\Psi^\Upsilon_S = \Psi^{G,\Upsilon}_S = \Psi^{M,\Upsilon}_S$ 
 indistinguishable from glass and melt phases are, summing over $m =1, \ldots, M$,
 \begin{align}
 \label{eq:Qsm}
 {\bm Q}^S_m  = 
 - \frac{ \partial \Psi^{\Upsilon}_S }{\partial {\bm \Upsilon}^S_m }
 = 2 \frac{\partial \Psi^S_m} {\partial {\bm C}^E},
\qquad
  \Psi^\Upsilon_S (\{ {\bm \Upsilon}_S \}, {\bm F}^E, \theta)  =  \sum_m \Psi^S_m ({\bm \Upsilon}^S_m,{\bm F}^E,\theta) =  \sum_m \int {\textstyle{\frac{1}{2}}}  {\bm Q}^S_m :{\d } {\bm C}^E.
 \end{align}
 Recall ${\bm C}^E = ( {\bm F}^E)^{\mathsf T} {\bm F}^E $.
Dependence of $\Psi^\Upsilon_S$ and $\Psi^S_m$ on phase fractions $(\omega,\Xi)$ is implicit through
$\{ {\bm \Upsilon}^S_m \} $ and not written out in \eqref{eq:Qsm}. 
Denote $D_t (\square) = \dot{(\square)}$. 
Introduce energy functions $\hat{\Psi}^S_m$, $\hat{\Psi}^{G,S}_m$, and $\hat{\Psi}^{M,S}_m$ that are independent of configurational variables. Rate equations for internal stresses are 
\begin{align}
\label{eq:dotQS}
  \dot{\bm Q}^S_m + \frac{{\bm Q}^S_m}{ \tau^S_m}
 = 2 D_t \left( \frac{ \partial \hat{\Psi}^S_m }{ \partial {\bm C}^E} \right), 
\quad
  \hat{\Psi}^S_m({\bm F}^E, \theta,\omega) = (1-\omega ) \hat{\Psi}^{G,S}_m ({\bm F}^E, \theta) + \omega \hat{\Psi}^{M,S}_m ({\bm F}^E, \theta).
\end{align}
If relaxation times $\tau^S_m$ are constants, then a solution to \eqref{eq:dotQS} is the convolution integral \cite{holzapfel1996a,holzapfel1996b}
\begin{align}
\label{eq:convint}
 {\bm Q}^S_m (t) = 
 {\bm Q}^S_{m0} \exp \biggr{[} \frac{t_0-t}{ \tau^S_m} \biggr{]}
 + 
 \int_{t_0^+}^t \exp \biggr{[} \frac{s-t+t_0}{ \tau^S_m} \biggr{]}
 D_s \biggr{(} 2 \frac{\partial \hat{\Psi}^S_m}{ \partial {\bm C}^E} \biggr{)}  {\d } s, \qquad
{\bm Q}^S_{m0} = 2 \frac{ \partial \hat{\Psi}^S_m }{ \partial {\bm C}^E} \bigr{\rvert}_{t = t_0}.
\end{align}
Cauchy stress contributions from shear viscoelasticity are, for isotropic logarithmic response and recalling ${\bm e}^E = \frac{1}{2} \ln  [{\bm F}^E ({\bm F}^E)^{\mathsf T}]$, sums over $m$:
\begin{align}
\label{eq:QC1}
& {\bm \sigma}^\Upsilon_S 
= \frac{1}{J}  \frac{\partial \Psi^\Upsilon_S }{\partial {\bm e}^E }
=  \frac{2}{J} {\bm F}^E   \frac{\partial \Psi^\Upsilon_S }{\partial {\bm C}^E } ({\bm F}^E)^{\mathsf{T}}
  = \frac{1}{J} \sum_m  {\bm F}^E  {\bm Q}^S_m ({\bm F}^E)^{\mathsf{T}}
 =  \bar{\bm \sigma}^\Upsilon_S - p^\Upsilon_S {\bm 1}.
 \end{align}
Let $t_R$ be the time duration for a load history. For infinitely slow loading, $t_R / \tau^S_m \rightarrow \infty$, 
 $ {\bm Q}^S_m \rightarrow {\bm 0}$, and ${\bm \sigma}^\Upsilon_S \rightarrow {\bm 0}$
 is the equilibrium shear response if this holds for all $m$.
For extremely rapid loading, $t_R / \tau^S_m \rightarrow 0$; thereby,
$ {\bm \sigma}^\Upsilon_S$ reduces to an instantaneous (i.e., glassy) Cauchy stress contribution:
\begin{align}
&  {t_R }/{\tau^S_m} \rightarrow 0 \quad [\, \forall \, m = 1, \ldots, M] \, \Rightarrow \,
% \nonumber \\ &  
 {\bm \sigma}^\Upsilon_S   =   
 % \frac{2}{ J } \sum_m {\bm F}^E \frac{ \partial \hat{\Psi}^S_m }{\partial {\bm C}^E}  ({\bm F}^E)^{\mathsf{T}} =
  \frac{1}{ J } \sum_m  \frac{ \partial \hat{\Psi}^S_m }{\partial {\bm e}^E}  
   = \frac{1-\omega}{J} \sum_m  \frac{\partial \hat{\Psi}^{G,S}_m } {\partial {\bm e}^E} 
   + \frac{\omega}{J}  \sum_m  \frac{\partial \hat{\Psi}^{M,S}_m } {\partial {\bm e}^E}.
    \label{eq:QC2}
    \end {align}
Kinetics from shear viscoelasticity is prescribed in what follows, where
 conjugate forces $\{ {\bm \pi}^S_m \}$ are a subset of $\{ {\bm \pi} \} = - \partial \Psi / \partial \{ {\bm \alpha} \}$.
 Forces, evolution equations,
 and dissipation are \cite{holzapfel1996a,holzapfel1996b,claytonPRE2024}
 \begin{align}
 \label{eq:pivisc}
& {\bm \pi}^S_m = - \frac{\partial \Psi}{\partial {\bm \Upsilon}^S_m} =  f^E_S (1- \Xi) {\bm Q}^S_m
, \qquad
 \dot{{\bm \Upsilon}}^S_m  = \frac{ {\bm Q}^S_m} {  \tau^S_m \mu^S_\Upsilon }, 
 \quad
 {\mathfrak D}^\Upsilon_S % =  f_S^E \sum_m {\bm Q}^S_m :  \dot{\bar{\bm \epsilon}}^S_m 
= f_S^E (1-\Xi) \sum_m \frac{  {\bm Q}^S_m : {\bm Q}^S_m}
 { \tau^S_m \mu^S_\Upsilon } \geq 0;
\\
\label{eq:viscint}
& 
  \Psi^{\Upsilon}_S =    \Psi^{G,\Upsilon}_S =   \Psi^{M,\Upsilon}_S
=  \sum_m  \left[ \hat{\Psi}^S_m  - \int_{t_0}^t  {\bm Q}^S_m : D_s  {\bm \Upsilon}^S_m \, {\d } s \right];
\qquad \mu^S_\Upsilon = {\rm const} > 0;
\end{align}
  Integration over time obtains shear configurational energy in \eqref{eq:viscint}. Initial conditions at $t =t_0$ are
 ${\bm \Upsilon}^S_m = {\bm 0}$,
$\Psi^\Upsilon_S (\{ {\bm 0} \} ,{\bm F}^E,\theta)= \sum_m \hat{\Psi}^S_m ({\bm F}^E,\theta; \omega = 0)$. 

Volumetric (i.e., bulk) viscoelasticity is addressed similarly to shear viscoelasticity. 
Thermo-viscoelastic coupling is enabled, namely, relaxation of Gr\"uneisen's parameters
 \cite{nunziato1973g,nunziato1973am,clements2012a}.
For volumetric modes, conjugate stresses $\{ {\bm Q}^V_l  \}$ and 
  free energy $\Psi^\Upsilon_V = \Psi^{G,\Upsilon}_V = \Psi^{M,\Upsilon}_V$ 
  are as follows: %, summing over $l =1, \ldots, L$,
 \begin{align}
 \label{eq:QVl}
 {\bm Q}^V_l  = 
 - \frac{ \partial \Psi^{\Upsilon}_V }{\partial {\bm \Upsilon}^V_l }
 = 2 \frac{\partial \Psi^V_l} {\partial {\bm C}^E},
\qquad
  \Psi^\Upsilon_V (\{ {\bm \Upsilon}_V \}, {\bm F}^E, \theta)  =  \sum_l \Psi^V_l ({\bm \Upsilon}^V_l,{\bm F}^E,\theta) =  \sum_l \int {\textstyle{\frac{1}{2}}}  {\bm Q}^V_l :{\d } {\bm C}^E.
 \end{align}
Dependence of $\Psi^\Upsilon_V$ and $\Psi^V_l$ on $(\omega,\Xi)$ enters implicitly through kinetics for
$\{ {\bm \Upsilon}^V_l \} $. 
Rate equations are as follows, including the bulk and thermoelastic coupling energies in $\hat{\Psi}^V_l$, $\hat{\Psi}^{G,V}_l$,
and $\hat{\Psi}^{M,V}_l$:
\begin{align}
\label{eq:dotQV}
  \dot{\bm Q}^V_l + \frac{{\bm Q}^V_l}{ \tau^V_l} 
 = 2 D_t \left( \frac{ \partial \hat{\Psi}^V_l }{  \partial {\bm C}^E} \right), 
\quad
  \hat{\Psi}^V_l({\bm F}^E, \theta,\omega) = (1-\omega ) \hat{\Psi}^{G,V}_l ({\bm F}^E, \theta) + \omega \hat{\Psi}^{M,V}_l ({\bm F}^E, \theta).
  \end{align}
  In the special case that relaxation times $\tau^V_l$ are constants, then a solution to \eqref{eq:dotQV} is the integral 
  \begin{align}
\label{eq:convintV}
& {\bm Q}^V_l (t) = 
 {\bm Q}^V_{l0} \exp \biggr{[} \frac{t_0-t}{ \tau^V_l} \biggr{]}
 + 
 \int_{t_0^+}^t \exp \biggr{[} \frac{s-t+t_0}{ \tau^V_l} \biggr{]}
 D_s \biggr{(} 2 \frac{\partial \hat{\Psi}^V_l}{ \partial {\bm C}^E} \biggr{)}  {\d } s, \qquad
{\bm Q}^V_{l0} = 2 \frac{ \partial \hat{\Psi}^V_l }{ \partial {\bm C}^E} \bigr{\rvert}_{t = t_0}.
\end{align}
Cauchy stress contributions from bulk viscoelasticity are the following sums over $l = 1,\ldots,L$:
\begin{align}
\label{eq:QC1V}
& {\bm \sigma}^\Upsilon_V 
= \frac{1}{J}  \frac{\partial \Psi^\Upsilon_V }{\partial {\bm e}^E }
=  \frac{2}{J} {\bm F}^E   \frac{\partial \Psi^\Upsilon_V }{\partial {\bm C}^E } ({\bm F}^E)^{\mathsf{T}}
  = \frac{1}{J} \sum_l  {\bm F}^E  {\bm Q}^V_l ({\bm F}^E)^{\mathsf{T}}
  = -p^\Upsilon_V {\bm 1}.
 \end{align}
As previously explained for shear, under infinitely slow loading $t_R / \tau^V_l \rightarrow \infty$, 
 $ {\bm Q}^V_l \rightarrow {\bm 0}$, so ${\bm \sigma}^\Upsilon_S \rightarrow {\bm 0}$ if all modes are relaxed.
 For instantaneous loading, $t_R / \tau^V_l \rightarrow 0$ so that 
$ {\bm \sigma}^{\Upsilon}_V$ becomes a glassy stress:
\begin{align}
&  {t_R }/{\tau^V_l} \rightarrow 0 \, \, \, [\, \forall \, l = 1, \ldots, L] \, \Rightarrow \, 
   {\bm \sigma}^\Upsilon_V   =   
  \frac{1}{ J } \sum_l  \frac{ \partial \hat{\Psi}^V_l }{\partial {\bm e}^E}  
   = \left[  \frac{1 - \omega}{J} \sum_l  \frac{\partial \hat{ \Psi}^{G,V}_l } {\partial e^E_V} + 
   \frac{\omega}{J} \sum_l  \frac{\partial \hat{ \Psi}^{M,V}_l } {\partial e^E_V} \right] {\bm 1}.
  \label{eq:QC2V}
    \end {align}
  Initial conditions at $t =t_0$ are
 ${\bm \Upsilon}^V_l = {\bm 0}$,
$\Psi^\Upsilon_V ( \{ {\bm 0} \}, {\bm F}^E,\theta)= \sum_l \hat{\Psi}^V_l ({\bm F}^E,\theta; \omega  = 0)$.
  Thermodynamic forces, evolution equations,
 dissipation, and bulk configurational energies are obtained from
 \begin{align}
 \label{eq:piviscV}
& {\bm \pi}^V_l = - \frac{\partial \Psi}{\partial {\bm \Upsilon}^V_l} =  f^E_V (1- \Xi) {\bm Q}^V_l
, \qquad
 \dot{ \bm \Upsilon}^V_l  = \frac{ {\bm Q}^V_l} { \tau^V_l \mu^V_\Upsilon }, 
 \quad
 {\mathfrak D}^\Upsilon_V % =  f_S^E \sum_m {\bm Q}^S_m :  \dot{\bar{\bm \epsilon}}^S_m 
= f_V^E (1-\Xi) \sum_l \frac{  {\bm Q}^V_l : {\bm Q}^V_l}
 { \tau^V_l \mu^V_\Upsilon } \geq 0;
\\
\label{eq:viscintV}
& 
  \Psi^{\Upsilon}_V =    \Psi^{G,\Upsilon}_V =   \Psi^{M,\Upsilon}_V
=  \sum_l  \left[ \hat{\Psi}^V_l  - \int_{t_0}^t  {\bm Q}^V_l : D_s  {\bm \Upsilon}^V_l \, {\d } s \right];
\qquad \mu^V_\Upsilon = {\rm const} > 0.
\end{align}
Extending Ref.~\cite{holzapfel1996a}, contributions to temperature rate arise from shear and bulk viscoelasticity:
\begin{align}
\label{eq:shear2nd}
 \sum_m \theta ({\partial^2 \Psi^\Upsilon_S }/{\partial {\bm \Upsilon}^S_m  \partial \theta}):\dot{\bm \Upsilon}^S_m 
& = - 2 \theta \sum_m ({\partial^2 {\Psi}^S_m }/{\partial {\bm C}^E  \partial \theta}):\dot{\bm \Upsilon}^S_m 
\\ & = -  \theta \sum_m  [({\bm F}^E)^{-1}({\partial^2 {\Psi}^S_m }/{\partial {\bm e}^E  \partial \theta}) ({\bm F}^E)^{- \mathsf{T} }]:\dot{\bm \Upsilon}^S_m ,
\nonumber
\\
\label{eq:bulk2nd}
 \sum_l \theta ({\partial^2 \Psi^\Upsilon_V }/{\partial {\bm \Upsilon}^V_l  \partial \theta}):\dot{\bm \Upsilon}^V_l 
& = - 2 \theta \sum_l ({\partial^2 {\Psi}^V_l }/{\partial {\bm C}^E  \partial \theta}):\dot{\bm \Upsilon}^V_l 
\\ & = -  \theta \sum_l [({\bm F}^E)^{-1}({\partial^2 {\Psi}^V_l }/{\partial {\bm e}^E  \partial \theta}) ({\bm F}^E)^{- \mathsf{T} }]:\dot{\bm \Upsilon}^V_l .
\nonumber 
\end{align}
   
\subsection{Plasticity}
\label{sec3.4}
The plasticity model comprises three main components.
Described first are free energy, anisotropic hardening, and evolution of tensor internal variable $\bm A$ associated
with alignment and stretching of molecular chains in both glassy and rubbery regimes. 
Described second are energy and flow stress contributions, and coupled kinetic laws, for free volume $\varphi$ and internal variable $ \varsigma $ associated with deformation induced disordering in the glassy phase.
Described third are the flow rule for traceless symmetric strain rate $\dot {\bar {\bm \epsilon}}^P$ and commensurate dissipation from shear yielding. 

Multiple models exist that, depending on numbers of governing equations and calibrated parameters, can match complex finite stress-strain behavior of amorphous polymers across ranges of strain, strain rate, and temperature,
most often for uniaxial compressive stress states \cite{arruda1993a,arruda1993b,arruda1995,richeton2005,richeton2006a,richeton2006b,richeton2007,palm2006,mulliken2006,anand2003,anand2009,ames2009,miehe2009,srivastava2010,bouvard2010,clements2012a,bouvard2013}.
Many recent works draw heavily from their immediate predecessors, and differences among theories can be subtle.
Here, representative model features follow from this literature, balancing descriptive capability with model complexity and numbers of parameters. A single scalar hardening parameter $\varsigma$ is used \cite{anand2003,richeton2007,narayan2021}, noting
multiple entries of $\{ \varsigma \}$ have been proposed to address crystallization or enhanced hardening at large strain
\cite{ames2009,bouvard2010,bouvard2013} or distinct mechanisms across the glass transition \cite{srivastava2010}.

As in cited works, many ``constants'' in what follows can be made temperature-  or strain-rate dependent to better fit complex data.  
However, nonlinear temperature dependence of inelastic free energy is discouraged to avoid affecting specific heat.
Dependence of free energy on strain rate is discouraged to avoid thermodynamic inconsistencies.
Previous works invoked a glass transition temperature $\theta_G$ that depends on strain rate and also enters the
definitions of elastic constants and other contributors to free energy \cite{palm2006,richeton2007,srivastava2010}.
This approach seems acceptable when the strain rate is constant, but thermodynamic inconsistencies and unusual viscoelastic effects can arise if the strain rate varies during the load history. Such issues are avoided in the present approach by using
an order parameter $\omega$ to account for rate- and pressure-dependent melting, rather than a transient $\theta_G$.

Plastic free energy densities in \eqref{eq:HelmG} and \eqref{eq:HelmM} are partitioned, without $\phi$ dependence, as
\begin{align}
\label{eq:psiPpart}
\Psi^{*,P} ( \varsigma, {\bm A}, \varphi, \phi, \theta) = \Psi^{*,P}_A ({\bm A},\theta) + \Psi^{*,P}_\varsigma (\varsigma,\varphi).
\end{align}
The evolution law for the tensor $\bm A$, initial condition, and network plastic stretch $\lambda^P$ are \cite{anand2003,narayan2021}
\begin{align}
\label{eq:Adot}
\dot{\bm A} = {\bm A} \dot{\bar{\bm \epsilon}}^P + \dot{\bar{\bm \epsilon}}^P {\bm A}, \qquad {\bm A}({\bm X},t_0) = {\bm 1},
\qquad \lambda^P = ({\textstyle{\frac{1}{3}}} {\rm tr} {\bm A})^{1/2};
\qquad {\bar{\bm A}} = {\bm A} - (\lambda^P)^2 {\bm 1}.
\end{align}
When ${\bm D}^P({\bm X},t) = \dot{\bar{\bm \epsilon}}^P({\bm X},t) \, \forall t$, then ${\bm A} = {\bm F}^P ({\bm F}^P)^{\mathsf T}$.
Denote $N^{*,P}(\theta) > 0 $ the temperature dependent number of statistical links per polymer chain \cite{arruda1995} with
reference value $N^P_0 = N^{G,P}(\theta_0)$. Isothermal network locking stretch is defined as $\lambda^P_L = (N^P_0)^{1/2}$. 
Denote Langevin function ${\mathcal L} (\square) = {\rm coth} (\square) - 1/ \square$ and approximate inverse
by ${\mathcal L}^{-1} (\square) \approx \square (3 - \square^2)/(1- \square^2)$ \cite{cohen1991}. 
Normalized stretch is $\lambda^P_N = \lambda^P/\lambda^P_L$. Free energy contributions
from chain alignment and finite inelastic stretch processes in glassy and melt phases extend Refs.~\cite{anand2003,gearing2004,richeton2006b,richeton2007,dal2022}:
\begin{align}
\nonumber
& \Psi^{G,P}_A (\lambda^P_N ({\bm A}), \theta) = \mu_A^G (\theta) N^{G,P}(\theta)
\biggr{[} \lambda^P_N {\mathcal L}^{-1} ( \lambda^P_N )  + \ln  \frac{ {\mathcal L}^{-1} (\lambda^P_N) }{{\rm sinh}  
\{ {\mathcal L}^{-1}( \lambda^P_N) \}  }
\\
& \qquad \qquad \qquad  \qquad \qquad \qquad \qquad \, -  ( {1}/{\lambda^P_L}) {\mathcal L}^{-1} ( 1/\lambda^P_L )  - \ln  \frac{ {\mathcal L}^{-1} (1/\lambda^P_L) }{{\rm sinh}  
\{ {\mathcal L}^{-1}( 1/\lambda^P_L) \}  }\biggr{]},
\label{eq:psiPG}
\\
\nonumber
& \Psi^{M,P}_A (\lambda^P_N ({\bm A}), \theta) = \mu_A^M (\theta) N^{M,P}(\theta)
\biggr{[} \lambda^P_N {\mathcal L}^{-1} ( \lambda^P_N )  + \ln  \frac{ {\mathcal L}^{-1} (\lambda^P_N) }{{\rm sinh}  
\{ {\mathcal L}^{-1}( \lambda^P_N) \}  }
\\
& \qquad \qquad \qquad  \qquad \qquad \qquad \qquad \, -  ( {1}/{\lambda^P_L}) {\mathcal L}^{-1} ( 1/\lambda^P_L )  - \ln  \frac{ {\mathcal L}^{-1} (1/\lambda^P_L) }{{\rm sinh}  
\{ {\mathcal L}^{-1}( 1/\lambda^P_L) \}  }\biggr{]};
\label{eq:psiPM}
\\ 
& \label{eq:muP}
\mu^G_A = \langle \mu_{A0} - \mu_{A \theta} (\theta - \theta_0) \rangle, \qquad
\mu^M_A = \langle \mu_{A0} - \mu_{A \theta} (\theta_G - \theta_0) \rangle \theta / \theta_G;
\\
\label{eq:NP}
& N^{G,P} = \langle N^P_0 + N^P_\theta (\theta- \theta_0) \rangle, \qquad
N^{M,P} = \langle N^P_0 + N^P_\theta (\theta_G - \theta_0) \rangle.
\end{align}
Angled brackets preclude negative values:
$\langle \square \rangle = \frac{1}{2} (\square + | \square |)$.
Material constants are $\mu_{A0} \geq 0$, $\mu_{A \theta}$, $N^P_0 > 0$, and $N^P_\theta$. 

Denote by $\tau^P = \tau^E - \tau^A \geq 0$ the net shear stress in the flow direction, with applied elastic part $\tau^E$ and back-stress part $\tau^A$.
Define $\dot{\gamma}^P \geq 0$ as the plastic shearing rate and $\bar{\bm N}^P$ the flow direction tensor that is symmetric, traceless, and of unit magnitude.
Net dissipation from shear flow $\dot{\bar{\bm \epsilon}}^P$ and $\dot{\bm A}$, using \eqref{eq:Adot}
and isotropy \cite{anand2009} of $\Psi^{*,P}_A ({\rm tr} {\bm A},\theta)$, is the following:
\begin{align}
\nonumber
  {\mathfrak D}^P_A  & =
\bar{\bm M}^E: \dot{\bar{\bm \epsilon}}^P - (\partial_{\bm A} \Psi) : \dot{\bm A} 
\\   & = 
 f^E_S (1-\Xi) [ (1-\omega) \bar{\bm M}^{G,E}_S + \omega \bar{\bm M}^{M,E}_S
+ \bar{\bm M}^\Upsilon_S ]: \dot{\bar{\bm \epsilon}}^P \nonumber
\\ & \qquad \qquad \qquad 
 - \{ 2 f^E_S (1-\Xi) [ (1-\omega) \partial_{\bm A} \Psi^{G,P} + \omega \partial_{\bm A} \Psi^{M,P}] {\bm A} \}: \dot{\bar{\bm \epsilon}}^P 
\nonumber
\\ 
& =   (\bar{\bm M}^E - \bar{\bm M}^A) : \dot{\bar{\bm \epsilon}}^P = {\bar{\bm M}}^P : \dot{\bar{\bm \epsilon}}^P
= \tau^P \dot{\gamma}^P = | \tau^E - \tau^A | \dot{\gamma}^P \geq 0;
\label{eq:plasdiss}
\end{align}
\begin{align}
\nonumber
&  \bar{\bm M}^{*,E}_S   = {\textstyle{\frac{1}{2}}} J [({\bm F}^E)^{\mathsf T} \bar{\bm \sigma}^{*,E}_S ({\bm F}^E)^{- \mathsf T} 
+ ({\bm F}^E)^{-1} \bar{\bm \sigma}^{*,E}_S {\bm F}^E ],
\\
&  \bar{\bm M}^{\Upsilon}_S    = {\textstyle{\frac{1}{2}}} J [({\bm F}^E)^{\mathsf T} \bar{\bm \sigma}^{\Upsilon}_S ({\bm F}^E)^{- \mathsf T} 
+ ({\bm F}^E)^{-1} \bar{\bm \sigma}^{\Upsilon}_S {\bm F}^E ];
\nonumber 
\\ 
& {\bar{\bm M}}^P  = {\textstyle{\sqrt 2}} \tau^P \bar{\bm N}^P, \qquad
\bar{\bm N}^P =  \bar{\bm M}^P  / (\bar{\bm M}^P: \bar{\bm M}^P) ^{1/2},  \qquad
\dot{\bar{\bm \epsilon}}^P = {\textstyle{\frac{1}{\sqrt 2}}}\dot{\gamma}^P \bar{\bm N}^P;
\nonumber
\\ 
&   \tau^P  = {\textstyle{\frac{1}{\sqrt 2}}} \bar{\bm M}^P: \bar{\bm N}^P =  {\textstyle{\frac{1}{\sqrt 2}}}  (\bar{\bm M}^P: \bar{\bm M}^P) ^{1/2}, \qquad
 \dot{\gamma}^P  = {\textstyle{\sqrt 2}} (\dot{\bar{\bm \epsilon}}^P:\dot{\bar{\bm \epsilon}}^P)^{1/2};
\label{eq:taus}
\end{align}
\begin{align}
\label{eq:MA}
 \bar{\bm M}^A & = 2 \left[ \frac{\partial \Psi}{\partial {\bm A}} {\bm A} - \frac{1}{3} {\rm tr} \left( \frac{  \partial \Psi}{ \partial {\bm A}} {\bm A}
 \right) {\bm 1} \right] = 
 f_V^E (1-\Xi) \left[ \frac{ (1- \omega) }{3 \lambda^P} \frac{\partial \Psi^{G,P}_A}{\partial \lambda^P}
+ \frac{ \omega }{3 \lambda^P} \frac{\partial \Psi^{M,P}_A}{\partial \lambda^P} \right] \bar{\bm A} \nonumber
\\
& = \frac{f^E_S (1-\Xi) }{3 N^P_0 \lambda^P_N} [ (1- \omega) \mu^G_A N^{G,P} + \omega \mu^M_A N^{M,P} ] {\mathcal L}^{-1}(\lambda^P_N) \,  \bar{\bm A}
\nonumber
\\ &
\approx  \frac{f^E_S (1-\Xi)}{3 N^P_0} [ (1- \omega) \mu^G_A N^{G,P} + \omega \mu^M_A N^{M,P} ] 
\left[ \frac{3 - (\lambda^P_N)^2}{1-(\lambda^P_N)^2} \right]  \bar{\bm A}.
\end{align}

Following Refs.~\cite{anand2003,anand2009,ames2009,bouvard2010,bouvard2013,narayan2021,dal2022},
 $\varsigma$ and $\varphi$ account for initial hardening, a yield peak, and subsequent strain
softening associated with network disorder.  Free volume $\varphi$ in this context only measures plastic
structural disorder and its effect on plastic softening \cite{jatin2014}. It does not account for
changes in some thermo-physical properties such as thermal expansion and compressibility. The latter are widely correlated to  definitions of ``free volume'' that increases across the glass transition \cite{williams1955,white2016}. Rather, such correlations are captured here by different thermoelastic properties for glass and melt.
Because residual free volume change does not appreciably affect mass density \cite{anand2003,anand2009},
\begin{align}
\label{eq:deltaomega}
c_\varphi \rightarrow 0.
\end{align}

The cooperative-model flow rule of Ref.~\cite{richeton2007} is used, whereby the isotropic shear resistance associated
with molecular defect rearrangements vanishes identically in the melt phase. Thus, from \eqref{eq:deltaomega},
values of $\varsigma$ and $\varphi$ are inconsequential for the melt, and likewise for decomposed products, and their time evolution is controlled by properties of
the glassy solid. 
Material parameters $\tau^\varsigma_0,\mu^G_\varsigma, a_\varphi, b_\varphi,\tau^\varsigma_\infty$, and $\varphi_\infty$ are
now introduced.
Under steady flow,
$\tau^\varsigma \rightarrow \tau^\varsigma_\infty$ and $\varphi \rightarrow \varphi_\infty$ as $t \rightarrow \infty$.
The following system  \cite{anand2003,gearing2004,dal2022} of ordinary differential equations (ODEs) then suffices
for kinetics of $\varsigma$ and $\varphi$,
where $\tau^\varsigma$ is flow resistance associated with $\varsigma$:
\begin{align}
\label{eq:tausig}
\tau^\varsigma & = \tau^\varsigma_0 + \mu^G_\varsigma \varsigma, \qquad
\varsigma ({\bm X},t_0) = 0, \qquad \varphi ({\bm X},t_0) = 0; \qquad
\tau^\varsigma_\varphi  = \tau^\varsigma_\infty [ 1 + b_\varphi (\varphi_\infty - \varphi) ];
\\
\label{eq:tausigdot}
\dot{\tau}^\varsigma &=  \mu^G_\varsigma (1 - \tau^\varsigma / \tau^\varsigma_\varphi) \dot{\gamma}^P;
\qquad
\dot{\varphi}  = a_\varphi ( \tau^\varsigma / \tau^\varsigma_\infty - 1) \dot{\gamma}^P; 
\qquad
\dot{\varsigma} = \dot{\tau}^\varsigma  / {\mu}^G_\varsigma = (1 - \tau^\varsigma / \tau^\varsigma_\varphi) \dot{\gamma}^P. %- \dot{\mu}^G_\varsigma \varsigma) ;
%\\
%\tau^\varsigma_\varphi & = \tau^\varsigma_\infty [ 1 + b^\varphi_0 (\varphi_\infty - \varphi) ]; \qquad
%\mu^G_\varsigma = (1 - \omega)(1 - \Xi) \mu^\varsigma_0, \qquad a_\varphi =   (1 - \omega)(1- \Xi) a^\varphi_0.
%\label{eq:muvarsigma}
\end{align}

The scalar plastic shearing rate is furnished by the cooperative model of Refs.~\cite{richeton2006b,richeton2007},
modified to account for mixed phase regions $\omega \in (0,1)$ and possible decomposition $\Xi \in (0,1]$.
When $\omega = \Xi = 0$, the present flow rule reduces to that of the glassy material in the original reference \cite{richeton2005,richeton2006a}, and
when $\omega = 1$, it reduces to that of the rubbery material \cite{richeton2007}. As $\Xi \rightarrow 1$, the polymer decomposes into chemical products and plastic flow ceases to have physical relevance. Herein, 
\begin{align}
\label{eq:richeton}
\dot{\gamma}^P  = (1-\Xi) \dot{\gamma}_0 \exp \left( \frac{ - H_\beta}{k_B \theta} \right)
\left[(1 - \omega) + \omega \exp \biggr{ \{ }  \frac{ (\ln 10) c_1^G (\theta - \theta_G)}{c_2^G + (\theta - \theta_G)} \biggr{ \} }  \right] \left[ {\rm sinh} \left( \frac{\tau^F V_F }{2 k_B \theta} \right) \right]^{1/m},
\end{align}
where $\dot{\gamma}_0 > 0$ and $m \geq 0$ are reference strain rate
and rate sensitivity, $k_B$ is Boltzmann's constant, and activation energy for secondary relaxation is $H_\beta$. Activation volume is $V_F$, and $c^G_1$ and $c^G_2$ are parameters. For rate independence, $m \rightarrow 0$.
The effective flow stress is defined akin to Ref.~\cite{bouvard2013}:
\begin{align}
\label{eq:tauF}
\tau^F = \langle {\tau}^E - \{ \tau^A  + f^E_S (1-\omega)(1 - \Xi) (1-\phi) \langle \tau^\varsigma - \alpha^F_\theta (\theta - \theta_0) + \alpha^F_p  B_0^{G,E} \langle - e^E_V \rangle \rangle \} f^P_R \rangle. \end{align}
Reduction in flow resistance by $f^E_S$ and $1-\phi$ accounts for softening from fractures \cite{narayan2021} and voids \cite{dal2022,claytonZAMM2024}, respectively.
Thermal softening is accelerated by constant $\alpha^F_\theta \approx \tau^\varsigma_0 / (\theta_G - \theta_0) > 0$ \cite{richeton2007}. Pressure hardening in compression \cite{rosenberg1994,mulliken2006} is enabled by constant $\alpha^F_p > 0 $.
Scalar function $f^P_R = f^P_R (e^E_V,\theta) \geq 0$ scales the total flow resistance to account for complex strength response under shock loading \cite{batkov1996,jordan2020} as explained in Section~\ref{sec4.1} for PMMA.
A more elaborate thermodynamic theory spanning the glass transition has been implemented \cite{srivastava2010} that captures some physics missing in \eqref{eq:richeton} for complex load histories, albeit with more parameters. See Ref.~\cite{srivastava2010}. 

The plasticity model is complete upon choosing the rightmost terms in \eqref{eq:psiPpart}.
For simplicity, assumed here is $\Psi^{G,P}_\varsigma = \Psi^{M,P}_\varsigma = \Psi^P_\varsigma$. These functions are otherwise left generic. In principle, they could be calibrated to temperature data (e.g., Taylor-Quinney factors \cite{shao2017}) if of sufficient fidelity. Weighted contributions to internal dissipation from rates of $\varsigma$ and
$\varphi$,
thus affecting $\dot{\theta}$, are
\begin{align}
\label{eq:Dissigma}
{\mathfrak D}^\varsigma & = - \partial_\varsigma \Psi \dot{\varsigma} = - (1 - \Xi) (\partial_{\varsigma} \Psi^P_\varsigma) 
f^E_S
(1 - \tau^\varsigma / \tau^\varsigma_\varphi) \dot{\gamma}^P,
% - (\dot{\mu}^G_\varsigma /  {\mu}^G_\varsigma) \varsigma ],
 \\
 \label{eq:Dissvarphi}
 {\mathfrak D}^\varphi & = - \partial_\varphi \Psi \dot{\varphi} = - (1 - \Xi) (\partial_{\varphi} \Psi^P_\varsigma) 
f^E_S a_\varphi ( \tau^\varsigma / \tau^\varsigma_\infty - 1) \dot{\gamma}^P.
\end{align}

\subsection{Fracture}
\label{sec3.5}

The fracture framework consists of the phase-field formulation of Section~\ref{sec2.3} for order parameter $\xi$, where
the net driving force $\Phi$ remains to be prescribed, and the kinetic law for porosity $\phi$ whose rate enters 
kinematic relation \eqref{eq:Dpexp} with direction $\bm{N}_\phi$ also to be prescribed.
Herein, the phase-field model, where $\xi$ simultaneously addresses degradation from brittle or ductile fractures, draws from prior phase-field descriptions of amorphous polymers undergoing combined inelastic shear-yielding and crazing 
\cite{schanzel2012,miehe2015b,narayan2021,bas2022,dal2022}. 
The same sources, along with continuum damage mechanics models~\cite{gearing2004,francis2014}, are
consulted for the porosity representation. A damage instantiation criterion based on maximum principal stress
 is used to activate brittle fracture from crazing, and an instantiation criterion based on local plastic stretch is used to
instantiate ductile tearing or chain scission \cite{gearing2004,narayan2021}.
For a polymer with low entanglement density such as PMMA, crazing applies for the nominally brittle glassy
regime. The scission mechanism is enabled for the rubbery melt since crazing is typically irrelevant above the glass transition. 
The decomposed phase of dissociated products is essentially already ``fractured" and ``porous''; 
$\xi$ and $\phi$ are unaffected by energy density of the decomposed phase that, according to Section~\ref{sec3.2}, cannot
support tensile pressure regardless.

First consider the phase-field formulation. Define a surface energy density per unit area $\Gamma_\xi$, akin to twice the fracture toughness, a length constant $l_\xi \geq 0$ that may or may not differ from gradient regularizer $l$ of Section~\ref{sec2.3},
and a threshold resistance energy per unit volume inhibiting damage initiation $R^*_\xi \geq 0$. Letting $\Gamma_\xi^G, R_\xi^G, \Gamma_\xi^M$, and  $R_\xi^M$ denote non-negative temperature-dependent functions for glass and melt,
\begin{align}
\label{eq:Gammaxi}
& \Gamma_\xi(\theta,\omega,\Xi) = (1-\Xi) [ (1-\omega) \Gamma_\xi^G (\theta) + \omega \Gamma_\xi^M(\theta)],
\\
& \label{eq:Rxi}
R_\xi(\theta,{\bm e}^E,\omega,\Xi) = (1-\Xi)[ (1-\omega) R_\xi^G (\theta,{\bm e}^E) + \omega R_\xi^M(\theta,{\bm e}^E)].
\end{align}
Thresholds $ R_\xi^G$ and $R_\xi^M$ can depend on strain ${\bm e}^E$ as well as $\theta$ to enable fracture resistance at high confining pressure \cite{satapathy2000,rittel2008}. The elastic driving force contribution to $\Phi$ allows brittle fracture even in the absence of porosity or shear yielding.
It comprises usual dilatational and deviatoric elastic strain energy densities \cite{claytonIJF2014,miehe2014,claytonJMPS2021,narayan2021} as well as viscoelastic configurational energies
\cite{claytonPRE2024}:
\begin{align}
\label{eq:Felxi}
F_\xi^E = (1-\Xi) \{ (1 -\omega) [\Psi^{G,E}_V  {\mathsf H}(e^E_V) + \Psi^{G,E}_S ]
+ \omega [\Psi^{M,E}_V  {\mathsf H}(e^E_V) + \Psi^{M,E}_S ]
+ [\Psi^\Upsilon_V {\mathsf H}(e^E_V) + \Psi^\Upsilon_S ] \}.
\end{align}
Denote $c_\xi^\phi \geq 0$ and  $c_\xi^P \geq 0$ respective constants for quadratic dependence of order-parameter kinetics on porosity\cite{schanzel2012,miehe2015b,dal2022} and for ductile fracture induced
quadratically by isochoric plastic stretch \cite{dal2022}. The net driving force function is finally posited as
\begin{align}
\label{eq:Phifunc}
 \Phi & ({\bm e}^E,\theta, \{ {\bm \Upsilon} \}, \lambda^P ({\bm A}), \phi, \omega, \Xi) =  \nonumber \\
& \qquad \frac { l_\xi \langle
F^E_\xi ({\bm e}^E,\theta, \{ {\bm \Upsilon} \},\omega, \Xi) + (1-\Xi)[c_\xi^\phi \phi^2 + c_\xi^P (\lambda^P - 1)^2] - R_\xi({\bm e}^E,\theta,\omega,\Xi) \rangle}{\Gamma_\xi(\theta,\omega,\Xi)} .
\end{align}

Now consider inelastic flow from dilatational deformations. Similar to Ref.~\cite{francis2014}, the porosity
rate is decomposed additively into contributions from several mechanisms, here crazing in the
brittle glass phase and chain scission in the ductile melt. Following Refs.~\cite{gearing2004,schanzel2012,miehe2015b,narayan2021,dal2022}, a non-negative scalar strain rate $\dot{\epsilon}^\phi$
is used to prescribe, in power-law form, the porosity rate $\dot{\phi}$:
\begin{align}
\label{eq:epsphidot}
\dot {\epsilon} ^\phi = \frac{\dot{\phi}}{1-\phi } = (1-\Xi) (1 - \omega) \varkappa_G \dot{\epsilon}_0 \left( \frac{ \langle M^E_\phi \rangle }{f^E_V S^G_\phi} \right)^{1/n}
+ (1-\Xi) \omega   \varkappa_M \dot{\epsilon}_0 \left( \frac{ \langle M^E_\phi \rangle }{ f^E_V S^M_\phi} \right)^{1/n} \geq 0.
\end{align}
Reference strain rate $\dot{\epsilon}_0 > 0$ and rate sensitivity $n > 0$ are assigned as the same constants for
each mechanism; distinct values could be used, if needed, to match data \cite{narayan2021}.
Indicator functions $\varkappa_G$ and $\varkappa_M$ acquire values of 0 or 1 depending on local stress or plastic stretch state. A spectral decomposition of Mandel stress ${\bm M}^E$ is, with principal components $M_I^E$ and
directions ${\bm n}_I^E$,
\begin{align}
\label{eq:Mspectral}
{\bm M}^E = \sum_{I = 1}^3 M^E_I {\bm n}^E_I \otimes {\bm n}^E_I; 
\qquad M^E_1 \geq M^E_2 \geq M^E_3; \qquad {\rm tr}{\bm M}^E =
\sum_{I = 1}^3 M^E_I = - 3 J p^E.
\end{align}
In \eqref{eq:kinP} and \eqref{eq:epsphidot}, the opening direction and driving force are prescribed as the direction 
and magnitude of maximum principal (Mandel) stress, similar to Refs.~\cite{gearing2004,schanzel2012,miehe2015b,narayan2021,dal2022}:
\begin{align}
\label{eq:phidir}
{\bm n}_\phi ({\bm X},t) = {\bm n}^E_1 ({\bm X},t) ; \qquad M^E_\phi = {\bm M}^E:{\bm n}_\phi  \otimes {\bm n}_\phi  =
{\bm M}^E:{\bm N}_\phi = M^E_1.
\end{align}
Unlike Refs.~\cite{narayan2021,dal2022}, but like other cited works, ${\bm n}_\phi ({\bm X},t)$ can evolve with time.
If ${\bm M}^E$ has repeated maximum eigenvalues, one of these must be chosen arbitrarily for \eqref{eq:phidir}. 
Indicator functions for crazing \cite{schanzel2012,miehe2015b,dal2022}
and ductile fracture \cite{gearing2004,narayan2021} are, with scaling $f^E_V \tilde{M}^E_\phi = M^E_\phi$ 
and $f^E_V \tilde{p}^E = p^E$, 
\begin{align}
\label{eq:indicatorG}
\varkappa_G & = {\mathsf H}(M^E_\phi) {\mathsf H}(-p^E) {\mathsf H}(f^G_\phi),
\quad f^G_\phi = \tilde{M}^E_\phi - [ c_1^\phi - c^\phi_2 / (J \tilde{p}^E) - 3 \nu_0^{G,E} J \tilde{p}^E / (1 + \nu_0^{G,E})];
\\
\label{indicatorM}
\varkappa_M & =  {\mathsf H}(M^E_\phi)  {\mathsf H}(\lambda^P - \lambda^P_\phi); \qquad
\lambda^P_\phi = {\rm const} > 1.
\end{align}
Material constants for craze initiation are $c^\phi_1$ and $c^\phi_2$, and $\nu_0^{G,E} = (3B_0^{G,E} - 2G_0^{G,E} )/(6B_0^{G,E} + 2G_0^{G,E})$ is Poisson's ratio. Resisting stresses in \eqref{eq:epsphidot} are driving stresses at instantiation times $t^G_\phi$ and $t^M_\phi$ of each mechanism \cite{narayan2021}:
\begin{align}
\label{eq:phiresG}
S^G_\phi ({\bm X},t \geq t_\phi^G ) & = \tilde{M}_\phi^E({\bm X}, t_\phi^G({\bm X})), 
\qquad t^G_\phi ( {\bm X}) = \underset{s \in (t_0,t]}{\min} \{ {\rm arg} 0 [ \varkappa_G ({\bm X},s) - 1] \}; 
\\
\label{eq:phiresG2}
S^M_\phi ({\bm X},t \geq t_\phi^M ) & = \tilde{M}_\phi^E({\bm X}, t_\phi^M({\bm X})),
\qquad t^M_\phi ( {\bm X}) = \underset{s \in (t_0,t]}{\min} \{ {\rm arg} 0 [ \varkappa_M ({\bm X},s) - 1] \}.
\end{align}
Dissipation from porosity is non-negative, noting $\Psi^{G,P}_\varsigma = \Psi^{M,P}_\varsigma$ in \eqref{eq:psiPpart} does not depend on $\phi$:
\begin{align}
% \nonumber 
{\mathfrak D}^\phi   % = [  {\bm M}^E: {\bm N}^\phi -  (1-\phi) \partial_\phi \Psi ] [{\dot{\phi}}/(1-\phi)] 
= (1-\Xi) M^E_\phi\dot{\epsilon}_0 \left[ (1 - \omega) \varkappa_G  \left( \frac{ \langle M^E_\phi \rangle }{f^E_V S^G_\phi} \right)^{1/n}
+ \omega   \varkappa_M \left( \frac{ \langle M^E_\phi \rangle }{ f^E_V S^M_\phi} \right)^{1/n} \right] \geq 0.
%\nonumber 
%\\ & \geq 0; \qquad [\partial_\phi \Psi^{G,P}_\varsigma = \partial_\phi \Psi^{M,P}_\varsigma \leq 0] .
\label{eq:Dissphi}
\end{align} 
To avoid computational complexity, as in prior treatments \cite{narayan2021,dal2022},
\eqref{eq:richeton} can be appended with multiplier $(1 - \varkappa_G)(1-\varkappa_M)$ to prevent shear yielding and
crazing or scission simultaneously at $\bm X$.

\subsection{Melting}
\label{sec3.6}
Recall ``melting'' refers to transition from a glassy to rubbery state.  At ambient pressure and null loading rate, this
second-order transition occurs as temperature is raised to a value in the vicinity of a constant $\theta_G$ characteristic of the polymer. The present framework also captures the reverse transition, from rubbery to glassy state, beginning when the polymer melt is cooled from high temperature to the vicinity of $\theta_G$. 
Typically the rubbery phase has a higher specific heat, a higher thermal expansion coefficient, and lower mechanical stiffness than the glass. Representative curves are available in Refs.~\cite{srivastava2010,clements2012a,clements2012b}. Calorimetry and mechanical experiments show that rapid thermo-physical property variations in amorphous polymers occur over a temperature window $\theta_G \pm \Delta \theta$, where $\Delta \theta$ is on the order of 5 to 10 K \cite{palm2006,dupaix2007}.
Evidence that the reduction in stiffness decreases with increasing loading rate \cite{palm2006,dupaix2007,richeton2007} implies a time scale is involved in the melt process. However, this time scale $\tau_\omega$ is expected to be small (e.g., relative to 1 second) since specimens preheated above $\theta_G$ respond to very high-rate loading in at least a partially glassy manner  \cite{richeton2006b,richeton2007,zaretsky2019}, meaning the melt acquires glassy stiffness very early in the dynamic deformation process.
Experiments also show that the melt process is inhibited by compressive pressure \cite{olabisi1975,clements2012a}, presumably because the vibrational free volume increase reflected by the increase in thermal expansivity is restricted.

Previous plasticity models for amorphous polymers traversing the glass transition 
assigned a variable glass transition temperature that depends on local effective strain rate \cite{palm2006,dupaix2007,srivastava2010}.
This transition temperature then enters thermodynamic (e.g., hyperelastic) energy potentials via elastic coefficients that depend explicitly on the variable glass transition temperature. Such an approach appears most theoretically sound if strain rate is fixed during a loading history. However, if strain rate varies with time, then thermodynamic potentials acquire implicit dependence on rate that could become incompatible with original modeling assumptions such as hyperelastic response. 

Circumventing this issue, in the present formulation the transition from glassy to rubbery response is described by an order parameter $\omega(t) \in [0,1]$ that evolves with thermodynamic state (including temperature, pressure, and strain rate) and enters free energy $\Psi$ of \eqref{eq:Helmdecomp}, along with various kinetic laws, using a constant $\theta_G$.
The kinetic law used for $\omega$ has similarities to prior models for first-order phase or structural transformations in crystalline materials \cite{forbes1977,boettger1997,claytonCMT2022,claytonZAMM2024,claytonARX2025,claytonAMECH2025}.
However, the glass-melt transition is viewed as a second-order solid state transformation \cite{gibbs1958,wu1999} in
that density $\rho$ and energy $\Psi$ stay constant if transformation occurs in the ideal limit of fixed $\theta = \theta_G$ under null elastic and plastic straining. Entropy and pressure need not remain constant, however, if specific heat and Gr\"uneisen parameters vary independently across the transition.

Net dissipation ${\mathfrak D}_\omega$ and thermodynamic driving force for melting $F_\omega$ (i.e., contribution from $\omega$ to $\{ {\bm \pi} \}$) are defined 
consistently with \eqref{eq:Dissr}, $\Psi^{G,\Upsilon}_V = \Psi^{M,\Upsilon}_V$,
$\Psi^{G,\Upsilon}_S = \Psi^{M,\Upsilon}_S$, and $\Psi^{G,P}_\varsigma = \Psi^{M,P}_\varsigma$ as
\begin{align}
\nonumber
{\mathfrak D}^\omega = F_\omega \dot{\omega}; \qquad
F_\omega = - \partial_\omega \Psi = & \, (1-\Xi) \{  f_V^E [\Psi^{G,E}_V + \Psi^{G,E}_\beta ] + f_S^E [\Psi^{G,E}_S
+ \Psi^{G,P}_A ] + \Psi^{G,\theta}  
\\
& - ( f_V^E [\Psi^{M,E}_V + \Psi^{M,E}_\beta ] + f_S^E [\Psi^{M,E}_S
+ \Psi^{M,P}_A ] + \Psi^{M,\theta} ) \} .
\label{eq:Domega}
\end{align}
As analyzed in Refs.~\cite{claytonCMT2022,claytonZAMM2024}, $F_\omega = F_\omega ({\bm e}^E,\theta,{\bm A}, \xi,\Xi)$ is related to the Gibbs free energy
difference between coexisting phases.
Introduce by $\tau_\omega = \tau_\omega (\theta) > 0$ a characteristic relaxation time for transition kinetics that can
depend on temperature if warranted. Denote by $R_\omega \geq 0$ a transition resistance function of thermodynamic state and deformation rate, with dimensions of energy per unit volume. Denote a metastable melt fraction by $\tilde{\omega}({\bm X},t) \in [0,1]$.  The following kinetic law 
enforces ${\mathfrak D}^\omega \geq 0$ for forward (melting, $\dot{ \omega} > 0$) and reverse (freezing, $\dot{\omega} < 0$) transitions at $({\bm X},t)$:
\begin{align}
\label{eq:omegadot}
\tau_\omega \dot{\omega} = 
\begin{cases}
& ({\tilde \omega} - \omega) {\mathsf H}( F_\omega), \qquad \quad[ \, {\rm if} \, {\omega} < \tilde{\omega} \quad {\rm and} \quad F_\omega > R_\omega \geq 0 \, ];
\\
& ({\tilde \omega} - \omega) {\mathsf H}( - F_\omega), \qquad \, \, [ \, {\rm if} \,\,  {\omega} > \tilde{\omega} \quad {\rm and} \quad - F_\omega > R_\omega \geq 0 \, ];
\\ & 0, \qquad \qquad \qquad \qquad \, \, \, \, [\,  {\rm otherwise} \, ].
\end{cases} 
\end{align}
The first condition in \eqref{eq:omegadot} pertains to glass$\, \rightarrow \, $melt, the second to melt$\, \rightarrow \, $glass, and
the third to a stationary state.

Consider forward transitions wherein $F_\omega > R_\omega$. The local metastable value of $\omega$
is motivated from differential relationship \cite{forbes1977,boettger1997,claytonCMT2022,claytonZAMM2024} $\d \omega = (1-\omega) \d {\mathbb{F}}$, where dimensionless forward transition driving force ${\mathbb{F}} = F_\omega / \beta_\omega$.
Proportionality factor $\beta_\omega > 0 $ for forward transformation 
is constant for simplicity. Transition begins when driving force exceeds normalized resistance $R_\omega/ \beta_\omega$. Integrating over melt process $\omega: 0 \rightarrow \tilde{\omega}$ to the current value of driving force $F_\omega$ produces 
\begin{equation}
\label{eq:metfwd}
F_\omega > R_\omega \geq 0 \quad \Rightarrow \quad
\int_0^{\tilde{\omega}} \frac {\d \omega}{1 - \omega} = \int_{R_\omega/\beta_\omega }^{F_\omega/\beta_\omega} {\d {\mathbb{F}}} \quad \Rightarrow \quad
\tilde{\omega} = 1 - \exp \left[  -  \frac{F_\omega - R_\omega }{\beta_\omega} \right].
\end{equation}
Now consider reverse transformations: $-F_\omega > R_\omega$. The differential relation is
 $\d \omega = - \omega \d {\mathbb{F}}$ with ${\mathbb{F}} = - F_\omega /  \beta_\omega$. 
 Since $\theta_G$ should be unchanged when the polymer is heated to melt or cooled to freeze, 
$\beta_\omega$ and $R_\omega$ should be the same proportionality constant and energy barrier for
forward and reverse transformation. Integrating over the freezing process $\omega: 1 \rightarrow \tilde{\omega}$ gives\footnote{Upper integration limits should be 1 in (4.48) and (2.113) of Refs.~\cite{claytonCMT2022,claytonZAMM2024}; these misprints are corrected here.}
\begin{equation}
\label{eq:metbak}
- F_\omega > R_\omega \geq 0 \quad \Rightarrow \quad
\int_{{\tilde \omega} }^1 \frac {\d \omega}{ \omega} = - \int^{R_\omega/\beta_\omega }_{- F_\omega/\beta_\omega} {\d {\mathbb{F}}}  \quad \Rightarrow \quad
\tilde{\omega} =  \exp \left[   \frac{F_\omega + R_\omega }{\beta_\omega} \right].
\end{equation}

The following are obtained for specific polymers by using \eqref{eq:metfwd} or \eqref{eq:metbak} in conjunction with experimental
glass transition data: free energy difference $\Psi^{M,\theta}_0 = {\rm const}$ between coexisting glass and melt at the same temperature introduced in \eqref{eq:thermM}, proportionality factor $\beta_\omega = {\rm const} > 0 $, and resistance function $R_\omega \geq 0$.  The latter measures the increase in effective transition start temperature
with increasing pressure and strain rate \cite{olabisi1975,palm2006,dupaix2007,srivastava2010,clements2012a}
and is of the following assumed form:
\begin{align}
\label{eq:Romega}
R_\omega = R_\omega(\langle - e^E_V \rangle, \{ {\bm \Upsilon}_V \}, \theta,{\dot \epsilon});
\quad
\dot{\epsilon} = \sqrt{2}( {\bm d}:{\bm d})^{1/2};
\quad R_\omega(0, \{ {\bm 0} \}, \theta \in (\theta_1, \theta_2), {\dot \epsilon} \leq {\dot \epsilon}_\omega) = 0.
\end{align}
As in Refs.~\cite{palm2006,dupaix2007,srivastava2010}, rate dependence is encapsulated by the scalar total strain rate $\dot{\epsilon}$, where $\dot{\epsilon}_\omega = {\rm const} > 0$ is a cutoff below which viscous effects do not affect the transition. As typical data only probe effects of compressive pressure \cite{olabisi1975}, dependence on volume is restricted to negative elastic volumetric strain $e^E_V$. Volumetric configurational variables $ \{ {\bm \Upsilon}_V \}$ are included in arguments of $R_\omega$ to account for dynamic stiffening of the bulk modulus. Resistance vanishes at null deformation within range $\theta_1 < \theta < \theta_2$, where $\theta_1 \ll \theta_G$ and $\theta_2 \gg \theta_G$. 
Recalling $F_\omega = F_\omega ({\bm e}^E,\theta,{\bm A}, \xi,\Xi)$ from \eqref{eq:Domega}, 
and noting $R_\omega = 0$ at steady state and null elastic volume change, energy difference $\Psi^{M,\theta}_0$
is obtained from metastability condition $F_\omega ({\bm 0},\theta_G,{\bm 1}, 0, 0) = R_\omega =  0$. From \eqref{eq:thermG} and
\eqref{eq:thermM},
\begin{align}
\label{eq:psitheta0}
\Psi^{M,\theta}_0 = (c_{00}^M - c_{00}^G) [ \theta_G \ln (\theta_G / \theta_0) - (\theta_G - \theta_0)]
+  {\textstyle{\frac{1}{2}}} (c_\theta^M - c_\theta^G)(\theta_G - \theta_0)^2.
\end{align}

Lastly, $\beta_\omega$ is determined from the ``half-width''  $\delta \theta_G$ of the transition region 
in $\theta$ versus $\tilde{\omega}$ space, on the order of 5-10 K for polymers of interest \cite{palm2006,dupaix2007}. 
Again enforce steady and null strain conditions and $\theta \in  (\theta_1,\theta_2)$ so $R_\omega = 0$.
Let $1 - \delta \omega \approx \frac{19}{20}$ be the completion fraction
of the forward transition as $\theta: \theta_G \rightarrow \theta_G + \delta \theta_G$.
Substituting $\tilde{\omega} = 1 - \delta \omega$ at $\theta = \theta_G + \delta \theta_G$
into \eqref{eq:metfwd} and solving for $\beta_\omega$ gives
$ \beta_\omega =  -(1/ \ln \delta \omega) [\Psi^{G,\theta}(\theta_G + \delta \theta_G) - \Psi^{M,\theta}(\theta_G + \delta \theta_G)]$. 
From symmetry, $\beta_\omega =  -(1/\ln \delta_\omega) [\Psi^{M,\theta}(\theta_G - \delta \theta_G) - \Psi^{G,\theta}(\theta_G - \delta \theta_G)]$ is derived for reverse transition to residual fraction $\delta \omega$ via $\theta: \theta_G \rightarrow \theta_G-\delta \theta_G$ in \eqref{eq:metbak}. Averaging the two defines $\beta_\omega$ used henceforth:
\begin{align}
\nonumber
& \beta_\omega = \frac{-1}{ 2 \ln \delta \omega} \{ [\Psi^{G,\theta}(\theta_G + \delta \theta_G) - \Psi^{G,\theta}(\theta_G - \delta \theta_G)] - [\Psi^{M,\theta}(\theta_G + \delta \theta_G) - \Psi^{M,\theta}(\theta_G - \delta \theta_G)] \},
\\
& \delta \omega \approx {\textstyle{\frac{1}{20}}}, \qquad \delta \theta_G \approx 5 \, {\rm K}.
\label{eq:betaomega}
\end{align}

\subsection{Shock decomposition}
\label{sec3.7}
The present framework focuses on shock decomposition rather than thermal decomposition.
Depending on heating rate and ambient environment (e.g., air or some other gas), thermal decomposition
at ambient atmospheric pressure typically commences at temperatures on the order of several hundred K above $\theta_G$
\cite{kashiwagi1985,beyler2002,korob2019}. The
process is often modeled via Arrhenius kinetics, where time scales span seconds to minutes. Combustion can take place in the presence of oxygen, but thermal decomposition is possible without combustion. The temperature rise induces chemical changes as the polymer backbones and side groups somewhat gradually dissociate into constituent monomers at chain ends or internal scissions \cite{kashiwagi1985,beyler2002,stoliarov2003}. 

Shock decomposition, conversely, occurs over time scales of microseconds or less, and at much higher pressures
and temperatures, for example tens of GPa and over 1000 K.  It was originally thought to involve a solid-fluid 
\cite{hauver1965} or solid-solid \cite{carter1995} phase transformation. An early model \cite{graham1979} linked observed changes in dielectric polarization and electric conductivity to polymer chain scission. More recent interpretations
\cite{dattelbaum2019,maerzke2019,lentz2020,bordz2021,coe2022,huber2023}, supported by contemporary experimental evidence (e.g., recovered samples) and chemistry models, regard shock decomposition as an extremely rapid disintegration of the condensed
glass or melt phase (i.e., the reactants) into various gases and remnant solid particles such as carbon (i.e., the products).
See also Ref.~\cite{kormer1968}. 
Hugoniot data show a reduction in specific volume upon shock decomposition that may be small to large depending on the particular polymer \cite{carter1995,dattelbaum2019}. Therefore, compressive pressure necessarily contributes toward the driving force for shock decomposition. However, since decomposition does not occur during isothermal hydrostatic compression to extreme pressures \cite{chan1981,jeong2015}, thermal energy or some other localized energetic source in the shock front \cite{graham1979}, in addition to mechanical pressure-volume work, must promote decomposition. Lower Hugoniot decomposition pressures for porous polymers relative to their dense counterparts \cite{dattelbaum2019}, in addition to higher temperatures reached in porous materials at the same pressure, support this assertion. Experimental measurements \cite{kormer1968,bordz2021} and various model predictions \cite{coe2022} are conflicting with regard to temperature rise and dissipated energy during shock decomposition. Some details of the process, for example the extent of densification, correlate qualitatively with chain structure and degree of crystallinity \cite{dattelbaum2019}.

Previous continuum models for shock decomposition appear limited to piecewise equations-of-state relating pressure,  volume, free or internal energy, and temperature \cite{dattelbaum2019,maerzke2019,bordz2021,coe2022} below and above the decomposition threshold. Such equations-of-state are typically constructed via a combination of theoretical assumptions (e.g., cold and thermal contributions) and calibrated to various high-pressure datasets.   
Thermochemical modeling \cite{dattelbaum2019,maerzke2019,coe2022,huber2023} has been used to provide equations-of-state for decomposed products, whereby fractions of fluid and solid products are determined by minimizing Gibbs free energy of the product mixture. Potential chemical products are assigned based on physical assumptions, but no calibration to experiments is needed. Ab-initio molecular dynamics has been used to model the shock Hugoniot spanning decomposition \cite{coe2022}.  Equations-of-state alone do not address kinetics. An Arrhenius model \cite{huber2023}
has instilled a time scale for decomposition reaction progress in simulations of shock propagation.

Here, shock decomposition is treated as a first-order phase transition from solid and melt reactants to dissociated products.
The kinetic model parallels that for melting given in Section~\ref{sec3.6}. Decomposition is one-way (i.e., products do not recombine into a condensed material when pressure and temperature are reduced), so only forward transformations are modeled. Unlike the second-order glass transition, the first-order decomposition process incurs a volume change
$\delta_\Xi \leq 0$, negative for irreversible volume collapse \cite{carter1995}, affecting \eqref{eq:Dpexp}.
Net dissipation ${\mathfrak D}_\Xi$ and thermodynamic driving force for decomposition $F_\Xi$  are defined 
as entering \eqref{eq:Dissr}. From \eqref{eq:Dissp} with \eqref{eq:Helmdecomp},
\begin{align}
& {\mathfrak D}^\Xi =  F_\Xi \dot{\Xi}; \qquad
F_\Xi  = - [  {J p^E  \delta_\Xi }/(1-\Xi \delta_\Xi) ] - \partial_\Xi \Psi
% = - [  {J p^E  \delta_\Xi }/(1-\Xi \delta_\Xi) ] + (1-\omega) \Psi^G + \omega \Psi^M - \Psi^D
= p^E \Delta_\Xi + \Psi^{GM} - \Psi^D; \nonumber
\label{eq:DXi}
\\ & \Delta_\Xi = - J \delta_\Xi /(1-\Xi \delta_\Xi) \approx - \delta_\Xi,
\qquad \Psi^{GM} = (1-\omega) \Psi^G + \omega \Psi^M.
\end{align}
The approximation in \eqref{eq:DXi} is most valid for $|J-1| \ll 1$ and $|\delta_\Xi| \ll 1$, and $\Psi^{GM}$ is
the combined free energy density of the intact solid-liquid mixture of glass and melt phases.
Here, $F_\Xi$ can be related to a Gibbs free energy
difference between reactant and product phases, though it includes a dependence on $\Xi$ through the transformation work term.
Let $\tau_\Xi = \tau_\Xi (\theta) > 0$ be a characteristic relaxation time for decomposition that likely decreases with increasing $\theta$ \cite{dattelbaum2019,huber2023}. For the process to substantially manifest in a shock front \cite{graham1979}, $\tau_\Xi$ should not exceed the order of tens of nanoseconds. Denote by $R_\Xi$ a decomposition resistance function (when positive) having dimensions of energy density. The metastable decomposed fraction is $\tilde{\Xi}({\bm X},t) \in [0,1]$.  A kinetic law 
enforcing ${\mathfrak D}_\Xi \geq 0$ for forward  ($\dot{ \Xi} > 0$) transitions at $({\bm X},t)$ is
\begin{align}
\label{eq:Xidot}
\tau_\Xi \dot{\Xi} = 
\begin{cases}
& ({\tilde \Xi} - \Xi) {\mathsf H}( F_\Xi), \qquad \quad[ \, {\rm if} \, {\Xi} < \tilde{\Xi} \quad {\rm and} \quad F_\Xi - R_\Xi \geq 0 \, ];
\\ & 0, \qquad \qquad \qquad \qquad \, \, \, \, [\,  {\rm otherwise} \, ].
\end{cases} 
\end{align}
The first condition in \eqref{eq:Xidot} describes products$\, \rightarrow \, $reactants, the second to static equilibrium. %for thermodynamic admissibility.

For positive decomposition rates, $F_\Xi > \langle R_\Xi \rangle $. The local metastable value of $\Xi$
stems from differential relationship $\d \Xi = (1-\Xi) \d {\mathbb{F}}$ \cite{forbes1977,boettger1997,claytonCMT2022,claytonZAMM2024,claytonAMECH2025}, where dimensionless forward transition driving force ${\mathbb{F}} = F_\Xi / \beta_\Xi$.
The proportionality factor is $\beta_\Xi = {\rm const} > 0 $. Decomposition can start when driving force exceeds normalized resistance $R_\Xi/ \beta_\Xi$. Integrating over the dissociation process $\Xi: 0 \rightarrow \tilde{\Xi}$ to the current value of driving force $F_\Xi$ gives the metastable value $\tilde{\Xi}$:
\begin{equation}
\label{eq:Xifwd}
F_\Xi - R_\Xi \geq 0 \quad \Rightarrow \quad
\int_0^{\tilde{\Xi}} \frac {\d \Xi}{1 - \Xi }= \int_{R_\Xi /\beta_\Xi }^{F_\Xi/\beta_\Xi} {\d {\mathbb{F}}} \quad \Rightarrow \quad
\tilde{\Xi} = 1 - \exp \left[  -  \frac{F_\Xi - R_\Xi }{\beta_\Xi} \right].
\end{equation}
Free energy difference $\Psi^{D,\theta}_0$ between reactants and decomposed products coexisting at the same temperature, latent heat energy for decomposition $\lambda^D_\theta$, and transformation temperature $\theta_D$, all constants in \eqref{eq:thermD}, must be assigned for specific polymers. To complete the kinetic model, relaxation time $\tau_\Xi > 0$, proportionality factor $\beta_\Xi  > 0 $, and resistance function $R_\Xi (\square)$ must also be specified.

Properties and functions are determined from shock compression data, most abundant for plane-wave loading
providing the principal Hugoniot (e.g., shock stress versus specific volume and shock velocity versus particle velocity).
Such data, which include shock pressure, volume, and possibly temperature, can be experimental \cite{kormer1968,carter1995} or, if necessary, obtained from published results of alternative continuum or molecular theories \cite{coe2022}. The time scale $\tau_\Xi$ is assumed small enough that Hugoniot data depict metastable states, as in prior work on metallic phase transitions \cite{boettger1997,claytonCMT2022,claytonZAMM2024} and 
implied in equation-of-state modeling of polymers \cite{dattelbaum2019,maerzke2019,bordz2021,coe2022}.
A finite magnitude of volume reduction $|\delta_\Xi|$ enables identification of the transition shock pressure from shock velocity, and
possibly shock pressure, data \cite{carter1995,dattelbaum2019}. 

Denote by $P^H = p^H + \frac{2}{3} \bar{\sigma}^H$ the longitudinal shock stress or ``shock pressure'', $p^H$ the mean compressive pressure, and $\bar{\sigma}^H = [\frac{3}{2} (\bar{\bm \sigma}:\bar{\bm \sigma})]^{1/2}$ the Von Mises equivalent deviatoric stress, all measured along the principal Hugoniot in steady conditions behind the shock where viscosity of Section~\ref{sec3.8} is negligible.
Denote by $J^H = V^H/V_0$ and $\theta^H$ the corresponding volume ratio and Hugoniot temperature.
Prior to decomposition onset, the principal Hugoniot can be uniquely parameterized by any of $J^H$, $P^H$, shock velocity, or particle velocity via the Rankine-Hugoniot equations with the constitutive model for the non-decomposed continuum \cite{davison2008,claytonNEIM2019}.
Let $P^H_D$, $J^H_D$, and $\theta_D = \theta^H_D$ be defined as the initiation shock pressure and temperature at the onset of decomposition as identified by the Hugoniot data. 
Define resistance function $R_\Xi (\square)$ so that $R_\Xi = 0$ at this particular Hugoniot state.
At decomposition onset, $F_\Xi = 0$, and from \eqref{eq:thermD}, $\theta = \theta_D$ so latent heat does not affect $\Psi^D(J^H_D)$. Denoting $p^E = p^H_D$ in \eqref{eq:DXi}, the constant energy term then follows from $F_\Xi = 0$,
using $J^H$ to parameterize free energies:
\begin{align}
\nonumber
\Psi^{D,\theta}_0 =  - p^H_D J^H_D \delta_\Xi & + \Psi^{GM}(J^H_D) -[ \Psi^{D,E}_V (J^H_D) + \Psi^{D,E}_\beta(J^H_D) ]
\\ & 
 + c_{00}^D [ \theta_D \ln (\theta_D / \theta_0) - (\theta_D - \theta_0)]
+  {\textstyle{\frac{1}{2}}} c_\theta^D (\theta_D - \theta_0)^2.
\label{eq:psiD0}
\end{align}

Now consider thermodynamic states at temperatures different from $\theta_D$. If $\lambda^D_\theta > 0$, then the latent heat contribution to $F_\Xi$
increases linearly with temperature above $\theta_D$ and decreases linearly with $\theta$ below $\theta_D$. This
trend agrees with experiments \cite{chan1981,jeong2015} that show no evidence of high-pressure decomposition in conductivity or pressure-volume response under isothermal (room temperature) compression. If function $R_\Xi$ is
prescribed as independent of $\theta$, then $\lambda^D_\theta$ can be chosen as having a minimum positive value that
prevents ``cold'' decomposition at $\theta = \theta_0$ as $e^E_V$ is decreased and $p^E$ increased above a large isothermal pressure, for example $ p^E \gtrsim 100 \,$GPa in PMMA \cite{chan1981}.
Alternatively, $\lambda^D_\theta$ can be set to zero and $R_\Xi = R_\Xi(\theta)$ defined
to increase with decreasing $\theta < \theta_D$ to similarly preclude low-temperature decomposition.
The latter approach is taken in Section~\ref{sec4.1}.

Constant $\beta_\Xi$ can be calibrated to Hugoniot $P^H$ versus $J^H$ data, if of sufficient fidelity, in the transition region from $P^H_D$ to a (slightly) higher pressure state $P^H_\delta > P^H_D$ where transition appears complete.
Upon full transformation volume collapse, a change in bulk compressibility is often perceptible in shock stress and velocity data \cite{carter1995}.
Let $1 - \delta \Xi \approx \frac{19}{20}$ be the
 decomposition fraction as $P^H: P^H_D \rightarrow P^H_\delta$.
Setting $\tilde{\Xi} = 1 - \delta \Xi $ at $P^H = P^H_\delta$
in \eqref{eq:Xifwd}, $\beta_\Xi$ is obtained:
\begin{align}
\label{eq:betaXi}
 \beta_\Xi =({-1}/{\ln \delta \Xi}) [ p^H(P^H_\delta) \Delta_\Xi (P^H_\delta) + \Psi^{GM}(P^H_\delta) - \Psi^D(P^H_\delta) - R_\Xi (P^H_\delta) ]. 
\end{align}
In \eqref{eq:betaXi}, shock stress $P^H$ is more convenient than volume $J^H$ for parameterizing Hugoniot energy states. Quantities on the right that depend implicitly on $\Xi$ are evaluated at $\Xi(P^H_\delta)= 1- \delta \Xi$.
Note that $\beta_\Xi \rightarrow 0$ as $P^H_\delta \rightarrow P^H_D$, corresponding to coincident metastable and
fully equilibrated states and a flat plateau on the $P^H$ versus $J^H$ Hugoniot. Volume collapse on this plateau
is decrement $(J^H_\delta - 1)/ J^H_\delta = \delta_\Xi$, independent of $J^H_D$ and $\beta_\Xi$.
%Nominally, $R_\Xi = 0$ for simplicity. 
If data warrant, and depending on $\lambda^D_\theta$ prescribed, $R_\Xi$ can be any function of state
such as (volume, temperature) with constraints $R_\Xi(J^H_D,\theta_D) = 0$.

\subsection{Heat conduction and viscosity}
\label{sec3.8}

Isotropic heat flux in the spatial configuration, $\bm q = - k_\theta \nabla \theta$,  is prescribed akin to Fourier's law in the first of \eqref{eq:fluxvisc1}.  The effective thermal conductivity $k_\theta$ depends on temperature, phase fractions, and damage.
Temperature and phase dependencies follow functional forms in Refs.~\cite{richeton2007,bouvard2013}, noting
different trends persist in the glassy and melt regimes; see also Ref.~\cite{srivastava2010}.
Degradation from fracture follows Refs.~\cite{miehe2015a,miehe2015b,claytonAMECH2025}, where a scalar parameter
$\alpha^\xi_k \in [0,1]$ denotes the ratio of conductivity in the fully degraded state to that of the undamaged state.
The decomposed phase is simply assumed to have a constant conductivity $k^D_\theta$. Therefore, 
with $a_k^G,a_k^M, b_k^M$ material parameters and $k_\theta^G \geq 0$ and $k^M_\theta \geq 0$ conductivities of respective glass and melt phases at $\theta = \theta_G$,
\begin{align}
\label{eq:kcond}
k_\theta (\theta,\xi,\omega,\Xi) = (1-\Xi) [1 + (\alpha^\xi_k-1) \xi^2] \{ (1-\omega)   k_{\theta}^G \left( \frac{\theta}{\theta_G} \right)^{a_k^G}  \! \! \! \!+ \omega k^M_{\theta} \langle a_k^M - b_k^M \frac{\theta}{\theta_G} \rangle \} + \Xi k^D_\theta .
\end{align}

Viscous stress ${\bm \tau}^V = J {\bm \sigma}^V$ of \eqref{eq:fluxvisc1} for isotropic polymers depends on total deformation rate
${\bm d}$, shear viscosity $\mu_V$, and bulk viscosity $\kappa_V$.
These viscosities are distinguished from nonlinear effective viscosities from shear and bulk viscoelastic response in Section~\ref{sec3.3},
respectively, $ \mu^S_\Upsilon \tau^S_m (\theta)$ and $ \mu^V_\Upsilon \tau^V_l (\theta)$ for each shear relaxation mode $m$ and
bulk relaxation mode $l$.  The latter viscoelastic viscosities tend to zero
at highly elevated temperatures since relaxation times $\tau^S_m$ and $\tau^V_l$ become negligible as shift functions
$a^S_T$ and $a^V_T$ become vanishingly small in \eqref{eq:taurelax}.

Here, $\mu_V$ accounts for high-temperature Newtonian viscosity in the melt, liquid, and decomposed states,
as viscoelastic stiffening effects become negligible:  
\begin{align}
\label{eq:muv}
\mu_V(\theta,\omega,\Xi, \dot{\epsilon} ) = \mu_{V}^D (\dot{\epsilon})  \Xi  +  \mu_V^M (\dot{\epsilon}) \omega [ \min \{ 1, \langle \theta - \theta_G \rangle /(\theta_M - \theta_G) \} ].
\end{align}
Rate dependent viscosities of decomposed (mostly gaseous) and liquid melt states are
$\mu_V^D \geq 0$ and $\mu_V^M \geq 0$. Liquification temperature $\theta_M$, where elastic shear modulus vanishes, is defined in \eqref{eq:thetaM}. 

Bulk viscosity $\kappa_V$ is included here only for numerical reasons, in the setting of discretized modeling of
shock waves. A standard technique \cite{wilkins1980,benson2007} is assignment of bulk viscous pressure $p^V$ with
terms linear and quadratic in compressive deformation rate to minimize spurious oscillations and
spread shock fronts over multiple grid spacings $l_G$. Here, $\kappa_V$ depends on ${\bm d}$,
is zero for volumetric expansion, and is otherwise independent of material state for simplicity:
\begin{align}
\label{eq:kappav}
\kappa_V ({\bm d}) ={\rho_0} l_G  \{c_1^\kappa C_L^\kappa -  c_2^\kappa l_G ({\rm tr} \, {\bm d})  \}
{\mathsf H} (-{\rm tr} \, {\bm d}) \geq 0.
\end{align}
Viscous pressure $p^V$ is proportional to $\rho$ and always non-negative. Standard constants for linear and quadratic bulk viscosity are $c_1^\kappa \geq 0$ and $c_2^\kappa \geq 0$. The effective longitudinal sound speed is a conservative upper bound taken as that of the glassy solid with instantaneous viscoelastic response:
\begin{align}
\label{eq:CLkappa}
C_L^\kappa = \{ [\hat{B}_{\eta 0}^{G,E}  + {\textstyle{\frac{4}{3}}} \hat{G}_0^{G,E}]/ \rho_0 \}^{1/2}, \qquad \hat{B}_{\eta 0}^{G,E} = \hat{B}_{0}^{G,E} + (\hat{\Gamma}^{G,E}_0)^2  c^G_{V0} \theta.
\end{align}
The instantaneous isentropic bulk modulus $\hat{B}_{0 \eta}^{G,E}$ is related to the isothermal bulk modulus $\hat{B}_{0}^{G,E}$
via the usual identity \cite{claytonNCM2011}. Both $l_G$ and $C_L^\kappa$ are taken as fixed Lagrangian quantities for consistency.
Notation $\hat{\square}$ denotes thermoelastic properties in the limits $t/\tau^V_l \rightarrow 0$ and  $t/\tau^S_m \rightarrow 0$ for all $l,m$.

Fourier conduction, Newtonian shear viscosity, and bulk shock viscosity are necessarily excluded when
modeling shocks as singular surfaces \cite{morro1980b,claytonPRE2024}. Otherwise, dissipation across the shock front would become infinite due to jump discontinuities in strain and temperature (i.e., effectively infinite strain rate and temperature gradient in the zero-width shock front).

\subsection{Total stress and temperature rate}
\label{sec3.9}

Total Cauchy stress ${\bm \sigma} = {\bm \sigma}^E + {\bm \sigma}^V$ in \eqref{eq:totalstress} comprises
summed contributions from elasticity and viscoelasticity in ${\bm \sigma}^E$ and bulk and shear viscosity in ${\bm \sigma}^V$. The former are weighted by phase volume fractions ($\omega,\Xi$) and damage degradation functions 
($f^E_V,f^E_S$) via \eqref{eq:Helmdecomp} with \eqref{eq:HelmG}, \eqref{eq:HelmM}, \eqref{eq:HelmD}, and \eqref{eq:fE}. Recall notation $\bar{\bm \sigma}^*$ and $-p^* {\bm 1}$ for deviatoric and pressure contributions to stress quantity ${\bm \sigma}^*$. Total Cauchy stress is the weighted sum of cold pressures $p^{G,E}_V$, $ p^{M,E}_V$, $p^{D,E}_V$
in \eqref{eq:pG}, \eqref{eq:pM}, \eqref{eq:pD};  elastic shear and pressure-shear coupling stresses
$\bar{\bm \sigma}^{G,E}_S$, $p^{G,E}_S$, $\bar{ \bm \sigma}^{M,E}_S$, $p^{M,E}_S$
in \eqref{eq:shearGdev}, \eqref{eq:shearGp}, \eqref{eq:shearMdev}, \eqref{eq:shearMp};
thermoelastic coupling pressures $p^{G,E}_\beta$, $p^{M,E}_\beta$, $p^{D,E}_\beta$ in
\eqref{eq:pGbeta}, \eqref{eq:pMbeta}, \eqref{eq:pDbeta};
viscoelastic shear and pressure-shear coupling stresses $\bar{\bm \sigma}^\Upsilon_S$, $p^\Upsilon_S$
in \eqref{eq:QC1}; viscoelastic bulk pressure $p^\Upsilon_V$ in \eqref{eq:QC1V}; viscous shear
stress $\bar{\bm \sigma}^V$ with coefficient $\mu_V$ in \eqref{eq:fluxvisc1}, \eqref{eq:muv}; and
viscous bulk pressure $p^V$ with coefficient $\kappa_V$ in \eqref{eq:fluxvisc1}, \eqref{eq:kappav}:
\begin{align}
\nonumber 
{\bm \sigma}   = & \quad
(1-\Xi) (1-\omega)[f^E_S ( \bar{\bm \sigma}^{G,E}_S - p^{G,E}_S {\bm 1})  -f^E_V ( p^{G,E}_V + p^{G,E}_\beta) {\bm 1} ]
\\ 
\nonumber
& + (1-\Xi) \omega [f^E_S ( \bar{\bm \sigma}^{M,E}_S - p^{M,E}_S {\bm 1})  -f^E_V ( p^{M,E}_V + p^{M,E}_\beta) {\bm 1} ]
\\
& + (1-\Xi) [ f^E_S (\bar{\bm \sigma}^\Upsilon_S -p^\Upsilon_S {\bm 1}) -f^E_V p^\Upsilon_V {\bm 1}]
%\\ & 
 - \Xi (p^{D,E}_V + p^{D,E}_\beta) {\bm 1}
  + J^{-1} [  ( \kappa_V   {\rm tr} {\bm d}) {\bm 1} + 2 \mu_V \bar{\bm d}].
\label{eq:sumstress}
\end{align}

To evaluate the temperature rate, total internal dissipation $\mathfrak D$ of \eqref{eq:Dissr} and other terms in \eqref{eq:temprate} are needed.
Internal dissipation comprises phase-field fracture kinetics ${\mathfrak D}^\xi \geq 0$ in \eqref{eq:Dxi}; shear and bulk viscoelasticity ${\mathfrak D}^\Upsilon_S \geq 0$,
${\mathfrak D}^\Upsilon_V \geq 0$ in \eqref{eq:pivisc}, \eqref{eq:piviscV}; shear yielding
${\mathfrak D}^P_A \geq 0$ in \eqref{eq:plasdiss}; internal state and free volume ${\mathfrak D}^\varsigma$, ${\mathfrak D}^\varphi$ in \eqref{eq:Dissigma}, \eqref{eq:Dissvarphi};
porosity ${\mathfrak D}^\phi \geq 0$ in \eqref{eq:Dissphi};
melting ${\mathfrak D}^\omega \geq 0$ in \eqref{eq:Domega}; and
shock decomposition ${\mathfrak D}^\Xi \geq 0$ in \eqref{eq:DXi}.
With $\Psi^P_\varsigma$ yet unspecified, assume ${\mathfrak D}^\varsigma + {\mathfrak D}^\varphi \rightarrow 0$,
implying counteracting contributions to stored energy from $\dot{\varsigma}$ (e.g., isotropic hardening) and $\dot{\varphi}$ (e.g., softening from excess free volume). Then internal dissipation is always non-negative:
\begin{align}
\label{eq:Disum}
{\mathfrak D} = {\mathfrak D}^\xi + {\mathfrak D}^\Upsilon_S + {\mathfrak D}^\Upsilon_V + {\mathfrak D}^P_A + {\mathfrak D}^\varsigma
+ {\mathfrak D}^\varphi + {\mathfrak D}^\phi + {\mathfrak D}^\omega + {\mathfrak D}^\Xi \geq 0.
\end{align}
As entering \eqref{eq:Dxi}, the conjugate force to fracture is evaluated using \eqref{eq:Helmdecomp}--\eqref{eq:fE} as
\begin{align}
\zeta = -\partial_\xi \Psi  = &  \quad 2 (1-\Xi)(1-\omega)
[(1-\xi {\mathsf H}(e^E_V)) (\Psi^{G,E}_V + \Psi^{G,E}_\beta + \Psi^{G,\Upsilon}_V)
\nonumber
\\
& \qquad \qquad \qquad \quad + (1-\xi) (\Psi^{G,E}_S + \Psi^{\Upsilon}_S + \Psi^{G,P}_A + \Psi^P_\varsigma)]
\nonumber
\\ 
& 
+ 2 (1-\Xi) \omega 
[(1-\xi {\mathsf H}(e^E_V)) (\Psi^{M,E}_V + \Psi^{M,E}_\beta + \Psi^{M,\Upsilon}_V)
\nonumber
\\
& \qquad \qquad \qquad \quad + (1-\xi) (\Psi^{M,E}_S + \Psi^{\Upsilon}_S + \Psi^{M,P}_A + \Psi^M_\varsigma)].
\label{eq:zetasum}
\end{align}

The total specific heat per unit reference volume at constant strain $c = c_V(\theta,\omega,\Xi)$ combines contributions to $\Psi$ that are nonlinear in $\theta$ and, in the approximation that transient nonlinear contributions from viscoelastic configurational free energies are negligible \cite{clements2012a}, are isolated to \eqref{eq:thermG}, \eqref{eq:thermM}, and \eqref{eq:thermD}:
\begin{align}
\nonumber
c  & =  - \theta \partial^2_{\theta \theta} \Psi   = - \theta (1-\Xi) [  (1-\omega) \partial^2_{\theta \theta} ( \Psi^{G,\theta} 
+  \Psi^{G,\Upsilon}_V +  \Psi^{G,\Upsilon}_S)
\\
& \qquad \qquad \qquad \qquad \qquad \qquad
+ \omega  \partial^2_{\theta \theta} ( \Psi^{M,\theta} 
+  \Psi^{M,\Upsilon}_V +  \Psi^{M,\Upsilon}_S) ]
- \theta \Xi \partial^2_{\theta \theta} \Psi^{D,\theta}
\nonumber
\\
& \approx \, \, (1-\Xi)(1-\omega) [c^G_{V0} + c^G_\theta (\theta - \theta_0) ]
%\nonumber
%\\ \quad & \, \, \, 
+ (1-\Xi) \omega [c^M_{V0} + c^M_\theta (\theta - \theta_0) ]
+ \Xi [c^D_{V0} + c^D_\theta (\theta - \theta_0) ].
\label{eq:totalcv}
\end{align}
The total Gr\"uneisen tensor ${\bm \Gamma}$ is a weighted sum of thermoelastic coupling terms ${\bm \Gamma}^{G,E}$, ${\bm \Gamma}^{M,E}$
in \eqref{eq:Gruntens} and viscoelastic contributions from temperature dependence of $\Psi^\Upsilon_S$, $\Psi^\Upsilon_V$ in \eqref{eq:Qsm}, \eqref{eq:QVl}:
\begin{align}
\nonumber
&{\bm \Gamma} = - \frac{1}{c} \frac{\partial^2 \Psi}{\partial \theta \partial {\bm e}^E}
 = (1-\Xi) \left[ \frac{c^G_{V} }{c} (1-\omega) {\bm \Gamma}^{G,E}
+ \frac{c^M_{V} }{c} \omega {\bm \Gamma}^{M,E}
- \frac{1}{c} \frac{\partial^2 ( \Psi^\Upsilon_S + \partial \Psi^\Upsilon_V) }{\partial \theta \partial {\bm e}^E} \right]
+ \frac{c^D_{V} }{c} \Xi {\bm \Gamma}^{D,E}
\\
& \qquad \qquad \qquad \quad \approx
(1-\Xi) \left[ \frac{c^G_{V} }{c} (1-\omega) {\bm \Gamma}^{G}
+ \frac{c^M_{V} }{c} \omega {\bm \Gamma}^{M} \right]
+ \frac{c^D_{V} }{c} \Xi {\bm \Gamma}^{D};
 \label{eq:Grunsum}
 \\
 \nonumber 
 & {\bm \Gamma}^G = \frac{c^G_{V0} }{c^G_V} \biggr{\{}
 %(1 + \sum_{l = 1}^{\hat{l}} \beta^V_l)
  f^E_V 
 [ \Gamma_0^{G,E} +  \Gamma_1^{G,E} e^E_V
+ {\textstyle{\frac{1}{2}}} \Gamma_2^{G,E} (e^E_V)^2] {\bm 1}
- 2  f^E_S \frac{G^{G,E}_\theta}{c^G_{V0}} \bar{\bm e}^E
\\ & \qquad \qquad \qquad \qquad \qquad \qquad \qquad \qquad - \frac{f^E_V}{c_{V0}^G} \sum_{l=1}^{\hat{l}} \frac{\partial^2 \hat{\Psi}^{G,V}_l}{\partial \theta \partial {\bm e}^E} 
- \frac{f^E_S}{c_{V0}^G} \sum_{m=1}^{\hat{m}} \frac{\partial^2 \hat{\Psi}^{G,S}_m}{\partial \theta \partial {\bm e}^E} 
\biggr{ \} },
 \label{eq:grunG}
 \\
 \nonumber 
 & {\bm \Gamma}^M = \frac{c^M_{V0} }{c^M_V} \{  %(1 + \sum_{l = 1}^{\hat{l}} \beta^V_l) 
  f^E_V [ \Gamma_0^{M,E} +  \Gamma_1^{M,E} e^E_V
+ {\textstyle{\frac{1}{2}}} \Gamma_2^{M,E} (e^E_V)^2] {\bm 1}
- 2  f^E_S \frac{M^{G,E}_\theta}{c^G_{V0}} \bar{\bm e}^E
\\ & \qquad \qquad \qquad \qquad \qquad \qquad \qquad \qquad - \frac{f^E_V}{c_{V0}^M} \sum_{l=1}^{\hat{l}} \frac{\partial^2 \hat{\Psi}^{M,V}_l}{\partial \theta \partial {\bm e}^E} 
- \frac{f^E_S}{c_{V0}^M} \sum_{m=1}^{\hat{m}} \frac{\partial^2 \hat{\Psi}^{M,S}_m}{\partial \theta \partial {\bm e}^E} 
\biggr{ \} },
 \label{eq:grunM}
% \\ 
 \end{align}
 \begin{align}
 & {\bm \Gamma}^D = \frac{c^D_{V0} }{c^D_V}   %(1 + \sum_{l = 1}^{\hat{l}} \beta^V_l) 
  [ \Gamma_0^{D,E} +  \Gamma_1^{D,E} e^E_V
+ {\textstyle{\frac{1}{2}}} \Gamma_2^{D,E} (e^E_V)^2] {\bm 1}.
 \label{eq:grunD}
\end{align}
In the approximation in \eqref{eq:Grunsum}, ${\bm \Gamma}^G$ and ${\bm \Gamma}^M$ comprise
sums of relaxed elastic and instantaneous volume ($l \leq \hat{l}$) and shear ($m \leq \hat{m}$) modes
active for the dynamic regime \cite{clements2012a,claytonPRE2024}. Choices of $\hat{m}$ and $\hat{l}$
pertain to ratios that obey $ \tau^V_l / t_R \gg 1$ and $ \tau^S_m / t_R \gg 1$, where $t_R$ is the time scale of the problem. 

More derivatives with respect to temperature are needed to evaluate the third and fourth terms on the
right side of \eqref{eq:temprate}. First derivatives, with viscoelastic approximations akin to \eqref{eq:Grunsum}, are
\begin{align}
\label{eq:pthG}
& \partial_\theta \Psi^G = f^E_V (\partial_\theta \Psi^{G,E}_\beta +  \partial_\theta \Psi^{G,\Upsilon}_V)
 + f^E_S (\partial_\theta \Psi^{G,E}_S + \partial_\theta \Psi^{G,\Upsilon}_S +  \partial_\theta \Psi^{G,P}_A)
 +  \partial_\theta \Psi^{G,\theta};
 \\
%\nonumber
& \partial_\theta \Psi^{G,E}_\beta = -c^G_{V0} e^E_V [ \Gamma_0^{G,E} + {\textstyle{\frac{1}{2!}}} \Gamma_1^{G,E} e^E_V
+ {\textstyle{\frac{1}{3!}}} \Gamma_2^{G,E} (e^E_V)^2],
\qquad
\partial_\theta \Psi^{G,E}_S = G^{G,E}_\theta \bar{\bm e}^E: \bar{\bm e}^E,
\\
& \partial_\theta \Psi^{G,\theta} = -c^G_{00} \ln \frac{\theta}{\theta_0} - c^G_\theta (\theta - \theta_0), 
 \label{eq:appr1} 
\\
& \partial_\theta \Psi^{G,P}_A =\left[ \frac{ \d \ln \mu^G_A }{ \d \theta} +  \frac{ \d \ln N^{G,P} }{ \d \theta} \right]{\Psi^{G,P}_A}  , \quad \frac{ \d \mu^G_A }{ \d \theta} = -\mu_{A \theta} {\mathsf H}(\mu^G_A),
\quad \frac{\d N^{G,P} }{ \d \theta} = N^P_\theta  {\mathsf H}(N^{G,P});
\label{eq:partialTG}
%\end{align}
%\begin{align}
\\
\label{eq:pthM}
& \partial_\theta \Psi^M = f^E_V (\partial_\theta \Psi^{M,E}_\beta +  \partial_\theta \Psi^{M,\Upsilon}_V)
 + f^E_S (\partial_\theta \Psi^{M,E}_S + \partial_\theta \Psi^{M,\Upsilon}_S +  \partial_\theta \Psi^{M,P}_A)
 +  \partial_\theta \Psi^{M,\theta};
\\
%\nonumber
& \partial_\theta \Psi^{M,E}_\beta = -c^M_{V0} e^E_V [ \Gamma_0^{M,E} + {\textstyle{\frac{1}{2!}}} \Gamma_1^{M,E} e^E_V
+ {\textstyle{\frac{1}{3!}}} \Gamma_2^{G,M} (e^E_V)^2],
\qquad
\partial_\theta \Psi^{G,M}_S = G^{G,M}_\theta \bar{\bm e}^E: \bar{\bm e}^E,
\\
& \partial_\theta \Psi^{M,\theta} = -c^M_{00} \ln \frac{\theta}{\theta_0} - c^M_\theta (\theta - \theta_0), 
 %\sum_{l = 1}^{\hat{l}} \beta^V_l \partial_\theta \Psi^{M,E}_\beta,
 \label{eq:appr2} 
\\
& \partial_\theta \Psi^{M,P}_A = \frac{ \d \ln \mu^M_A }{ \d \theta}{\Psi^{M,P}_A}  , \quad \frac{ \d \mu^M_A }{ \d \theta} = \frac{1}{\theta_G} \langle \mu_{A0} - \mu_{A \theta}(\theta_G - \theta_0) \rangle;
\label{eq:partialTM}
%\end{align}
%\begin{align}
\\
\label{eq:pthGammaV}
& \partial_\theta \Psi^{\Upsilon}_V = \partial_\theta \Psi^{G,\Upsilon}_V = \partial_\theta \Psi^{M,\Upsilon}_V \approx (1-\omega) \sum_{l=1}^{\hat{l}} \partial_\theta \hat{\Psi}^{G,V}_l
+ \omega  \sum_{l=1}^{\hat{l}} \partial_\theta \hat{\Psi}^{M,V}_l,
\\
\label{eq:pthGammaS}
& \partial_\theta \Psi^{\Upsilon}_S = \partial_\theta \Psi^{G,\Upsilon}_S = \partial_\theta \Psi^{M,\Upsilon}_S \approx (1-\omega) \sum_{m=1}^{\hat{m}} \partial_\theta \hat{\Psi}^{G,S}_m
+ \omega  \sum_{m=1}^{\hat{m}} \partial_\theta \hat{\Psi}^{M,S}_m;
%\end{align}
\\
\label{eq:pthD}
& \partial_\theta \Psi^D = \partial_\theta \Psi^{D,E}_\beta +   \partial_\theta \Psi^{D,\theta};
 \\
%\nonumber
&  \partial_\theta \Psi^{D,E}_\beta = -c^D_{V0} e^E_V [ \Gamma_0^{D,E} + {\textstyle{\frac{1}{2!}}} \Gamma_1^{D,E} e^E_V
+ {\textstyle{\frac{1}{3!}}} \Gamma_2^{D,M} (e^E_V)^2],
\\
& \partial_\theta \Psi^{D,\theta} = -c^D_{00} \ln \frac{\theta}{\theta_0} - c^D_\theta (\theta - \theta_0) - \frac{\lambda^D_\theta}{\theta_D}.
\label{eq;partialTD}
\end{align}
Internal energy per unit reference volume $U$ and requisite entropy $\eta$ can thus be written as a function of elastic strain, internal state, fracture order parameter, and temperature:
\begin{align}
\label{eq:Usum}
& U({\bm e}^E,\theta, \{ {\bm \alpha} \},\xi) = \Psi ({\bm e}^E,\theta, \{ {\bm \alpha} \},\xi) - \theta \partial_\theta \Psi ({\bm e}^E,\theta, \{ {\bm \alpha} \},\xi); 
\\ &
\eta = -\partial_\theta \Psi = -(1-\Xi)[(1-\omega) \partial_\theta \Psi^G + \omega \partial_\theta \Psi^M] + \Xi \partial_\theta \Psi^D.
\label{eq:etasum}
\end{align}

The third term on the right of \eqref{eq:temprate} contains contributions from rates of ($\omega, \Xi, {\bm A}, \{ { \bm \Upsilon}_V \}, \{ {\bm \Upsilon}_S \}$):
\begin{align}
- ( \partial_\theta \pi_\omega) \theta \dot{\omega} = (\partial^2_{\omega \theta} \Psi) \theta \dot{\omega} = 
(1-\Xi) (\partial_\theta \Psi^M - \partial_\theta \Psi^G) \theta \dot{\omega},
\label{eq:Domegatheta}
\end{align}
\begin{align}
-(\partial_{\theta} \pi_\Xi) \theta \dot{\Xi} = (\partial^2_{\Xi \theta} \Psi) \theta \dot{\Xi} =  
\{ \partial_\theta \Psi^D - [(1-\omega) \partial_\theta \Psi^G + \omega \partial_\theta \Psi^M ] \} \theta \dot{\Xi},
\label{eq:DXitheta}
\end{align}
\begin{align}
& - ( \theta \partial_\theta \pi_{\bm A} ): \dot{\bm A} = (\theta \partial^2_{{\bm A} \theta} \Psi) : \dot{\bm A}  = 
\theta \partial _\theta \bar{\bm M}^A : \dot{\bar{\bm \epsilon}}^P 
\nonumber
\\ & 
\approx 
 \frac{(1-\Xi) f^E_S  \theta}{3 N^P_0} \left[ (1- \omega) 
 \left( \frac{ \d \ln \mu^G_A }{ \d \theta} +  \frac{ \d \ln N^{G,P} }{ \d \theta} \right)
 \mu^G_A N^{G,P} + \omega  \frac{ \d \ln \mu^M_A }{ \d \theta} \mu^M_A N^{M,P} \right] 
\left[ \frac{3 - (\lambda^P_N)^2}{1-(\lambda^P_N)^2} \right]  \bar{\bm A}: \dot{\bar{\bm \epsilon}}^P,
\label{eq:DAtheta}
\end{align}
\begin{align}
- \theta \sum_l \partial_\theta {\bm \pi}^V_l : \dot{\bm \Upsilon}^{V}_l =  \theta \sum_l \partial^2_{ {\bm \Upsilon}^V_l \theta} \Psi : \dot{\bm \Upsilon}^{V}_l = - (1-\Xi) f^E_V \theta \sum_l \frac{ {\bm Q}^V_l:  \partial_\theta {\bm Q}^V_l}{ \tau^V_l \mu^V_\Upsilon},
\label{eq:DpiVtheta}
\end{align}
\begin{align}
- \theta \sum_m \partial_\theta {\bm \pi}^S_m : \dot{\bm \Upsilon}^{S}_m =  \theta \sum_m \partial^2_{ {\bm \Upsilon}^S_m \theta} \Psi : \dot{\bm \Upsilon}^{S}_m =  - (1-\Xi) f^E_S \theta \sum_m \frac{ {\bm Q}^S_m:  \partial_\theta {\bm Q}^S_m}{\tau^S_m \mu^S_\Upsilon}.
\label{eq:DpiStheta}
\end{align}
If approximations \eqref{eq:Grunsum}, \eqref{eq:pthGammaV}, and \eqref{eq:pthGammaS} are used, contributions from \eqref{eq:DpiVtheta} and \eqref{eq:DpiStheta} to \eqref{eq:temprate} are negligible. 
%Viscoelastic heating is implicitly embedded in the total Gr\"uneisen tensor ${\bm \Gamma}$. 
In that case, viscoelastic dissipation vanishes: $ \{ \dot{\bm \Upsilon}^V_l \} \rightarrow \{ {\bm 0} \} \Rightarrow {\mathfrak D}^\Upsilon_V \rightarrow 0$ and $\{ \dot{\bm \Upsilon}^S_m \} \rightarrow \{ {\bm 0} \} \Rightarrow{\mathfrak D}^\Upsilon_S \rightarrow 0$.
The fourth term on the right of \eqref{eq:temprate} is, with $\dot{\xi}$ from \eqref{eq:ratedep} and \eqref{eq:Dxi},
\begin{align}
\label{eq:Dxitheta}
- (\partial_\theta \zeta) \theta \dot{\xi}   =  ( \partial^2_{\xi \theta} \Psi)  \theta \dot{\xi}
= - \{ & 2 (1-\Xi)(1-\omega)
[(1-\xi {\mathsf H}(e^E_V)) ( \partial_\theta \Psi^{G,E}_\beta + \partial_\theta \Psi^{G,\Upsilon}_V)
\nonumber
\\
& \qquad  + (1-\xi) ( \partial_\theta \Psi^{G,E}_S +  \partial_\theta \Psi^{G,\Upsilon}_S + \partial_\theta \Psi^{G,P}_A )]
\nonumber
\\ 
+ & 2 (1-\Xi) \omega 
[(1-\xi {\mathsf H}(e^E_V)) ( \partial_\theta \Psi^{M,E}_\beta + \partial_\theta \Psi^{M,\Upsilon}_V)
\nonumber
\\
& \qquad + (1-\xi) (\partial_\theta \Psi^{M,E}_S + \partial_\theta \Psi^{M,\Upsilon}_S + 
\partial_\theta \Psi^{M,P}_A ]
\}  \theta \dot{\xi}.
\end{align}
The spatial gradient of thermal conductivity $k_\theta = k_\theta (\theta,\xi,\omega,\Xi)$ in \eqref{eq:kcond} is also needed: 
\begin{align}
\nabla k_\theta   = & \quad \left[ (1-\Xi) [1 + (\alpha^\xi_k-1) \xi^2] \{ (1-\omega)   k_{\theta}^G
\frac{a_k^G}{\theta_G}
 \left( \frac{\theta}{\theta_G} \right)^{a_k^G - 1}  \! \! \! \! - \omega k^M_{\theta} \frac{b_k^M}{\theta_G}
 {\mathsf H }\left( a_k^M - b_k^M \frac{\theta}{\theta_G} \right)  \} \right] \nabla \theta
 \nonumber
\\ & +  \left[ (1-\Xi) [ 2 (\alpha^\xi_k-1) \xi] \{ (1-\omega)   k_{\theta}^G \left( \frac{\theta}{\theta_G} \right)^{a_k^G}  \! \! \! \!+ \omega k^M_{\theta} \langle a_k^M - b_k^M \frac{\theta}{\theta_G} \rangle \} \right] \nabla \xi
\nonumber
\\ & +\left[ (1-\Xi) [1 + (\alpha^\xi_k-1) \xi^2] \{ 
k^M_{\theta} \langle a_k^M - b_k^M \frac{\theta}{\theta_G} \rangle -   k_{\theta}^G \left( \frac{\theta}{\theta_G} \right)^{a_k^G}  \! \! \} \right] \nabla \omega
\nonumber
\\ & + \left[ k^D_\theta -  [1 + (\alpha^\xi_k-1) \xi^2] \{ (1-\omega)   k_{\theta}^G \left( \frac{\theta}{\theta_G} \right)^{a_k^G}  \! \! \! \!+ \omega k^M_{\theta} \langle a_k^M - b_k^M \frac{\theta}{\theta_G} \rangle \} \right] \nabla \Xi .
\label{eq:gradkcond}
\end{align}
Finally, temperature rate in \eqref{eq:temprate} is, with all terms defined above in Section~\ref{sec3.9}, the following:
\begin{align}
c \dot{\theta} = &  {\mathfrak D}  - c \theta {\bm \Gamma}: ({\bm d} - {\bm d}^P)   
+  [ \kappa_V ({\rm tr} {\bm d})^2 + 2 \mu_V \bar{\bm d}: \bar{\bm d} ] + J [k_\theta \nabla^2 \theta + 
\nabla k_\theta \cdot  \nabla \theta]
 + \rho_0 r
\nonumber 
\\
& 
- \theta
\{ (\partial_\theta \zeta)   \dot{\xi}
+ (\partial_\theta \pi_\omega) \dot{\omega} 
+ ( \partial_\theta \pi_\Xi) \dot{\Xi} 
+ (  \partial_\theta \pi_{\bm A} ): \dot{\bm A}
+  \sum_l \partial_\theta {\bm \pi}^V_l : \dot{\bm \Upsilon}^{V}_l
+ \sum_m \partial_\theta {\bm \pi}^S_m : \dot{\bm \Upsilon}^{S}_m \}.
\label{eq:tempratefin}
\end{align}

\section{Application: PMMA}
\label{sec4}
The constitutive framework of Section~\ref{sec3} is applied toward the uniaxial-strain shock response of the transparent amorphous polymer
PMMA. Properties and parameters are given in Tables~\ref{tableA1} and~\ref{tableA2} of Appendix~\ref{secA} with supporting references.
Refined or additional parameters for the structured shock propagation analysis of Section~\ref{sec4.2} are given in Table~\ref{tableA3}.
Certain values are calculated or calibrated to data in what follows.
Not all kinds of PMMA are identical, but properties of current relevance appear to vary only 
modestly among samples from different suppliers \cite{jordan2016,jordan2017,lacina2018}.

Relaxation times for viscoelasticity span many orders of magnitude (e.g., time constants
 from $10^{-7}$s to $10^4$s in Ref.~\cite{clements2012a}), whereas usual shock experiments 
conclude by  $10^{-5}-10^{-4}$s. Long-time relaxation modes are inconsequential.
Therefore, thermoelastic properties in Table~\ref{table1} applicable to the Hugoniot analysis in
Section~\ref{sec4.1} correspond to equilibrium states (far) behind the shock front attained at these time scales; axial strain rates in the shock front span the order of $10^4$/s to $10^9$/s, depending on shock strength.
Equilibrium shear and bulk moduli at Hugoniot states are significantly larger in magnitude than their quasi-static versions \cite{richeton2007,srivastava2010} but may be modestly smaller than instantaneous values
from very high-frequency ultrasonic measurements \cite{asay1969,nunziato1972,nunziato1973ac} or states within the initial elastic shock front \cite{schuler1970,nunziato1973g,schuler1974}. 
Relaxation to \textit{quasi-static} values is irrelevant and not addressed in Section~\ref{sec4}. Viscoelastic stiffening above
quasi-static properties is implicitly included in $\Psi^{G,E}_V$, $\Psi^{G,E}_\beta$, and $\Psi^{G,E}_S$
rather than $\Psi^{G,\Upsilon}_V$ and $\Psi^{G,\Upsilon}_S$. This simplifies the analysis of Section~\ref{sec4.1} while yielding the same sought results. Differently, modeled in Section~\ref{sec4.2} is viscoelastic relaxation from instantaneous response at the head of structured wave form to equilibrium Hugoniot response at its tail.
Similar viscoelasticity assumptions are used in Refs.~\cite{nunziato1973am,nunziato1973p,schuler1973,schuler1974,menikoff2004}.

Consideration of the glass $\rightarrow$ melt transition at null deformation produces offset energy $\Psi^{M,\theta}_0$ and kinetic parameter $\beta_\omega$ via \eqref{eq:psitheta0} and \eqref{eq:betaomega} of Section~\ref{sec3.6}.  Forward and reverse second-order transitions occur over an approximate 10 K window centered at $\theta_G$ in Fig.~\ref{figA1}(a). The decrease in free energy and increase in specific heat from glass to melt are shown in Fig.~\ref{figA1}(b) and Fig.~\ref{figA1}(c). Prescription of $R_\omega$ for high-pressure states applicable to shock loading is addressed in Section~\ref{sec4.1}. Ranges of static, ambient pressure transition temperatures spanning
$373\,{\rm K} \leq \theta_G \leq 393\,{\rm K}$ have been reported for PMMA \cite{palm2006,richeton2007,srivastava2010,clements2012a,zaretsky2019}; here the value of 378 K quoted in Ref.~\cite{menikoff2004} is used.

\subsection{Hugoniot response to high pressure}
\label{sec4.1}

Shocks are analyzed in the infinitesimal width (i.e., singular surface) limit in Section~\ref{sec4.1},
though the forthcoming Hugoniot jump equations also apply between any two states in a structured steady wave \cite{claytonNEIM2019,davison2008}.
Cartesian Lagrangian and Eulerian coordinates are $(X_1,X_2,X_3)$ and $(x_1,x_2,x_3)$. A planar shock
propagates at Lagrangian speed $\mathcal U > 0 $ in the $X = X_1$ direction. Let $\square^+$ and $\square^-$ label respective quantities at states immediately ahead of and trailing the shock front. 
Particle velocity is $\upsilon = \upsilon_1$.
Shear and bulk shock viscosities are excluded following
discussion in Section~\ref{sec3.8}; the Rankine-Hugoniot equations directly incorporate shock dissipation without need for the latter.  Therefore, ${\bm \sigma} = {\bm \sigma}^E$,   $\bar{\bm \sigma} = \bar{\bm \sigma}^E$, and $p = p^E$. Recall ${\bar \sigma} =[\frac{3}{2}  \bar{\bm \sigma}:\bar{\bm \sigma}]^{1/2}$
is the Von Mises equivalent shear stress. The shock stress $P$ is the longitudinal true stress, positive in compression and obeying
\begin{align}
\label{eq:Pshock}
P = -\sigma_{11} = p - \bar{\sigma}_{11} =  p + {\textstyle{\frac{2}{3}}} \bar{\sigma}; \qquad
\bar{\sigma} = \sigma_{22} - \sigma_{11} =  \sigma_{33} - \sigma_{11} = -{\textstyle{\frac{3}{2}}} \bar{\sigma}_{11}.
\end{align}
Lateral stresses are $\sigma_{22} = \sigma_{33}$. 
Adiabatic conditions are assumed.
Particle velocity, specific volume, mass density, stress, temperature, energy, and entropy are discontinuous across the front.
The jump in any such quantity is $\llbracket \square \rrbracket = \square^- - \square^+$.
Rankine-Hugoniot jump relations are discrete analogs of
the continuum balances of mass, linear momentum and energy in \eqref{eq:conslaws} and the
entropy imbalance in \eqref{eq:seclaw}. In Lagrangian form \cite{claytonNEIM2019,davison2008}
and recalling $u = U/\rho_0$,
\begin{align}
\label{eq:RHeqs}
\rho_0 {\mathcal U} \llbracket 1/\rho \rrbracket = - \llbracket \upsilon \rrbracket,
\quad \rho_0 {\mathcal U} \llbracket \upsilon \rrbracket = \llbracket P \rrbracket,
\quad \rho_0 {\mathcal U} \llbracket u + {\textstyle{\frac{1}{2}}} \upsilon^2 \rrbracket = \llbracket P \upsilon \rrbracket, \quad {\mathcal U} \llbracket  \eta \rrbracket \geq 0.
\end{align}
No split-wave structure (e.g., precursor wave) emerges in shock wave studies on PMMA 
\cite{barker1970,menikoff2004}. Therefore, for analysis of the principal Hugoniot in Section~\ref{sec4.1}, conditions ahead of the wave correspond to a resting state in the reference configuration, whereby
 according to the constitutive functions in Section~\ref{sec3} with the null (gauge) pressure state corresponding to atmospheric pressure,
 \begin{align}
 \label{eq:plusstate}
 \upsilon^+ = 0, \quad J^+ = 1, \quad \rho^+ = \rho_0, \quad \theta^+ = \theta_0,
 \quad P^+ = p^+ = 0, \quad U^+ = (\Psi - \theta \partial_\theta \Psi)^+ = U_0.
 \end{align}
 The $\square^-$ superscripts for the shocked state are safely omitted in Section~\ref{sec4.1} without chance of confusion.
 The constitutive model furnishes equations for stress and free energy of the following form, where $J = J^- = J^H = \rho_0 / \rho^H$
 is the Hugoniot compression ratio:
 \begin{align}
 \label{eq:Hugresp}
 P = P^H(J^H,\theta(J^H),\gamma^P(J^H),\{ {\bm \alpha} \}(J^H), \ldots), \quad
 \Psi = \Psi^H(J^H,\theta(J^H),\gamma^P(J^H),\{ {\bm \alpha} \}(J^H), \ldots).
 \end{align}
 Rankine-Hugoniot equations \eqref{eq:RHeqs} reduce to the energy equation and
 expressions for particle and shock velocities for $J = J^H < 1$
 ($\mathcal U \rightarrow C_L$, the isentropic sound speed akin to \eqref{eq:CLkappa} as $J \rightarrow 1$):
 \begin{align}
 \label{eq:RHeqs2}
 U - U_0 =  {\textstyle{\frac{1}{2}}} P ( 1-J), \quad
 \upsilon = [P (1-J)/ \rho_0]^{1/2}, \quad {\mathcal U} = \upsilon / (1 - J).
 \end{align}
 In the present analysis, \eqref{eq:Hugresp} and the first of \eqref{eq:RHeqs2} are solved 
 for $P$, $\Psi$, $\theta$, and other material response functions using an iterative method as $J$ is decremented from unity on the principal Hugoniot. In each decrement, $\upsilon$ and $\mathcal U$ are
 calculated a posteriori using the second and third of \eqref{eq:RHeqs2}.
 
 Principal stresses are all negative (i.e., compressive), so no crazing or porosity change occurs, meaning $\phi = 0$.
 Furthermore, following Refs.~\cite{menikoff2004,clements2012a}, neither melting nor fracture occur
 under planar uniaxial compression from a start temperature of $\theta_0 = 295$ K.
 Although melting is promoted by temperature rise, and Hugoniot temperatures far exceed
 $\theta_G$, the transition from glass to melt is strongly impeded by compressive pressure \cite{olabisi1975,saeki1992,clements2012a} and requires a finite time $t_\omega$ that may be longer than that witnessed in shock experiments. For example, the effective glass transition temperature of PMMA increases by
 236 K per GPa of static pressure in Ref.~\cite{olabisi1975}. Extrapolating, this is on the order of $5 \times$ faster than the rate of temperature increase on the principal Hugoniot \cite{kormer1968,rosenberg1984,bordz2016}.
 Confining pressures modest relative to those in shock compression have been shown to suppress fracture
 of PMMA in quasi-static and dynamic loading, promoting shear yielding or shear banding instead \cite{satapathy2000,rittel2008}. Kinetic resistance functions $R_\omega$ and $R_\xi$ needed in the model to suppress
 melting and shear-induced fracture for Hugoniot states are calculated later.
 
 Let $\bar{\bm{F}}^P$ denote plastic deformation from shear yielding and $J^\Xi$ the volumetric deformation
 from shock decomposition. In matrix form, the uniaxial deformation gradient can be written
 \begin{align}
 \label{eq:Funi}
 {\bm F} = 
  \begin{bmatrix}
J & 0 & 0 \\
 0 & 1 & 0 \\
 0 & 0 & 1
 \end{bmatrix} = 
 {\bm F}^E \bar{\bm F}^P (J^\Xi)^{1/3} = 
 \begin{bmatrix}
 F^E_{11} & 0 & 0 \\
 0 & F^E_{22} & 0 \\
 0 & 0 & F^E_{22} 
 \end{bmatrix}
 \begin{bmatrix}
 \bar{F}^P & 0 & 0 \\
 0 & \bar{F}^{P-1/2} & 0 \\
 0 & 0 & \bar{F}^{P-1/2}
 \end{bmatrix}
 \begin{bmatrix}
F^\Xi & 0 & 0 \\
 0 & F^\Xi & 0 \\
 0 & 0 & F^\Xi
 \end{bmatrix}.
 \end{align}
 The following kinematic and stress relations then apply, where $F^\Xi = (J^\Xi)^{1/3} = (1 -\Xi \delta_\Xi)^{-1/3}$ :
 \begin{align}
 \label{eq:kin1D}
& J = J^E J^\Xi = \frac{F^E_{11} (F^E_{22})^2}{1 - \Xi \delta_\Xi}, \quad
F^E_{11} = \frac{J}{\bar{F}^P F^\Xi}, \quad F^E_{22} = \frac{(\bar{F}^P)^{1/2}}{F^\Xi}, \quad e^E_V = \ln J^E = \ln J - \ln J^\Xi;
\\
& \bar{e}^E_{11} =  -2 \bar{e}^E_{22} = \ln F^E_{11} - {\textstyle{\frac{1}{3}}} e^E_V,
\quad \bar{\bm e}^E = {\rm diag}(  \bar{e}^E_{11},  \bar{e}^E_{22},  \bar{e}^E_{22}),
\quad \gamma^P = -\sqrt{3} \ln {\bar F}^P, \quad {\bm{A}} = \bar{\bm F}^P (\bar{\bm F}^P)^{\mathsf T};
\\
& \lambda^P = \{ {\textstyle{\frac{1}{3}}} [ (\bar{F}^P)^2 + 2 / \bar{F}^P] \}^{1/2}, \quad \bar{\bm N}^P = - \sqrt{{\textstyle{\frac{2}{3}}}}  {\rm diag}(1,-{\textstyle{\frac{1}{2}}},-{\textstyle{\frac{1}{2}}}), \quad
\tau^A = \bar{\bm M}^A: \bar{\bm N}^P = -{\textstyle{\frac{\sqrt{3}}{2}}} \bar{M}^A_{11};
\\
& \bar{\bm M}^E = J \bar{\bm \sigma} = J {\rm diag} ( \bar{\sigma}_{11}, - {\textstyle{\frac{1}{2}}} \bar{\sigma}_{11},  - {\textstyle{\frac{1}{2}}} \bar{\sigma}_{11}), \quad \tau^E = \bar{\bm M}^E: \bar{\bm N}^P =
-{\textstyle{\frac{\sqrt{3}}{2}}} J \bar{\sigma}_{11} = {\textstyle{\frac{\sqrt{3}}{3}}} J \bar{\sigma}.
 \end{align}
 Secant bulk and shear moduli, Hugoniot specific heat, and Hugoniot Gr\"uneisen parameter are 
 \begin{align}
 \nonumber
 B_H = (1-\Xi) B^G_H + \Xi B^D_H  = &
(1-\Xi) B_0^{G,E} [ 1 - {\textstyle{\frac{1}{2}}}  B_1^{G,E} e^E_V + {\textstyle{\frac{1}{6}}} B_2^{G,E} (e^E_V)^2 - {\textstyle{\frac{1}{24}}} B_3^{G,E} (e^E_V)^3 ]
\\ & \quad +
\Xi B_0^{D,E} [ 1 - {\textstyle{\frac{1}{2}}}  B_1^{D,E} e^E_V + {\textstyle{\frac{1}{6}}} B_2^{D,E} (e^E_V)^2 - {\textstyle{\frac{1}{24}}} B_3^{D,E} (e^E_V)^3 ],
\label{eq:secantB}
 \end{align}
 \begin{align}
 \label{eq:secantG}
 G_H = G_0^{G,E} +{G^{G,E}_\theta}(\theta - \theta_0) 
-{B_0^{G,E}} ( G^{G,E}_p - G^{G,E}_2  e^E_V )e^E_V, \quad
c_V (\theta, \Xi) = (1-\Xi)c_V^G(\theta) + \Xi c_V^D (\theta),
 \end{align}
 \begin{align}
 \label{eq:secantGr}
 \Gamma = {\textstyle{\frac{1}{3}}} {\rm tr} {\bm \Gamma} = (1-\Xi) \frac{c_{V0}^G}{c_V} 
 [ \Gamma_0^{G,E} +  \Gamma_1^{G,E} e^E_V
+ {\textstyle{\frac{1}{2}}} \Gamma_2^{G,E} (e^E_V)^2]
+ \Xi \frac{c_{V0}^D}{c_V} [ \Gamma_0^{D,E} +  \Gamma_1^{D,E} e^E_V
+ {\textstyle{\frac{1}{2}}} \Gamma_2^{D,E} (e^E_V)^2].
 \end{align}
 
 Lagrangian longitudinal sound speed $C_L = C_E /J $ in the shocked state is useful for comparison
 to measured velocities of release or rarefaction waves, where $C_E$ is Eulerian sound speed
 \cite{davison2008}:
 \begin{align}
 \label{eq:CLdef}
 \rho_0 (C_L)^2 = -  \frac{\partial P}{\partial J} \biggr{\rvert}_{\eta,\gamma^P,\{ {\bm \alpha} \} } = 
 - \left[ \frac{\partial P}{\partial J} \biggr{\rvert}_{\theta,\gamma^P,\{ {\bm \alpha} \} } -
 \frac{\theta \Gamma } {J} 
 \frac{\partial P}{\partial \theta} \biggr{\rvert}_{J,\gamma^P,\{ {\bm \alpha} \} } \right].
 \end{align}
 Values of $C_L$ are calculated analytically via differentiation of
 the thermoelastic functions in Section~\ref{sec3.2} such as in \eqref{eq:secantB}--\eqref{eq:secantGr};
 $\partial \square / \partial J = (J^E/ J) \partial \square / \partial J^E$ when other variables are held fixed.
 
 In PMMA, plastic flow precedes shock decomposition as $J$ decreases and $P$ and $\theta$ increase on the Hugoniot. From Table~\ref{tableA1}, decomposition begins at $J^H_D = 0.58$, $P^H_D = 25.7$ GPa, and $\theta_D = 1289$ K. From Table~\ref{tableA2}, the Hugoniot elastic limit (HEL) for plastic flow occurs at $J^H_P = 0.937$,
 $P^H_P = 0.74$ GPa, and $\theta_P = 306$ K.
 For $J > J^H_P$, elastic conditions hold: $J = J^E$, $\bar{F}^P = 1$, $\gamma^P = \Xi = 0$.
 For $J \geq J^H_P$, plastic shear $\gamma^P = -\sqrt{3} \ln \bar{F}^P$ is determined by 
 static equilibrium conditions for each Hugoniot state \textit{behind} the shock front \cite{boettger1997,claytonCMT2022,claytonZAMM2024}. Inverting \eqref{eq:richeton} and using \eqref{eq:tauF},
 \begin{align}
 \label{eq:richeton0}
 \dot{\gamma}^P = 0 \quad \Rightarrow \quad \tau^F = 0  \quad \Rightarrow \quad \tau^E
 =  \{ \tau^A  + (1 - \Xi)   \langle \tau^\varsigma - \alpha^F_\theta (\theta - \theta_0) - \alpha^F_p  B_0^{G,E} e^E_V \rangle \} f^P_R,
 \end{align}
where $e^E_V \leq 0$. Conditions at the HEL $f^P_R (e^E_V = \ln J^H_P) = 1$ and $\tau^E ( \ln J^H_P)= \tau^E_P$ provide the constant
 \begin{align}
 \label{eq:alphaFP}
 \alpha^F_p = - [ \tau^E_P -\tau^\varsigma_0 + \alpha^F_\theta (\theta_P - \theta_0) ]/(B^{G,E}_0 \ln J^H_P).
 \end{align}
 Function $f^P_R(e^E_V, \theta(e^E_V))$ is a polynomial calibrated to match complex pressure and temperature dependent strength of PMMA on the Hugoniot presented later. With constants $f_i$ and $e^P_C$ in Table~\ref{tableA2},
 \begin{align}
 \label{eq:fRfit}
 f^P_R = \langle 1+ \sum_{i = 1}^6 f_i (e^P_V)^i \rangle {\mathsf H} (e^P_C - e^P_V), \qquad e^P_V = \langle \ln J^H_P - e^E_V \rangle. 
 \end{align}
 
 Kinetic equations for isotropic hardening and plastic free volume are given in rate form in \eqref{eq:tausig} and \eqref{eq:tausigdot}. In those equations, values of $\tau^\varsigma (t)$ and $\varphi (t) $ depend on the local history of
 $\dot{\gamma}^P(t)$. However, in the Hugoniot analysis of singular surfaces, the time history of $\gamma^P$ and its rate are unknown because transient flow processes in the shock front are unresolved. Therefore, substitute equations for $\varphi$, $\varsigma$, and $\tau^\varsigma$ that depend on $\gamma^P$ rather than its rate are posited for Hugoniot states:
 \begin{align}
 \label{eq:varphisub}
 & \varphi(\gamma^P) = [ 1 - \exp(-\gamma^P/a_\varphi) ] \varphi_\infty,
 \\
 \label{eq:tausub}
 & \tau^\varsigma(\gamma^P, \varphi(\gamma^P)) - \tau^\varsigma_0 =
 \mu^G_\varsigma  \varsigma(\gamma^P,\varphi) = \mu^G_\varsigma \{ 1 - \exp (-\gamma^P/b_\varphi)
 - [1- (\tau^\varsigma_\infty - \tau^\varsigma_0) / \mu^G_\varsigma] \varphi / \varphi_\infty \} .
 \end{align}
Constants $\tau^\varsigma_0$, $\mu^G_\varsigma$, $a_\varphi$, and $b_\varphi$ are listed in Table~\ref{tableA2}. As shown in Fig.~\ref{figA2}, \eqref{eq:varphisub} and \eqref{eq:tausub} give similar results for monotonic loading
 at room temperature to those for PMMA in Refs.~\cite{miehe2009,bouvard2010,dal2022}.
 Free volume $\varphi$ increases smoothly to its terminal value $\varphi_\infty$ with increasing $\gamma^P$.
 Isotropic stress $\tau^\varsigma$ increases rapidly to an initial peak, then decays more gradually to its
 terminal value that only modestly exceeds its initial value $\tau_0^\varsigma$.
 Back stress from anisotropic hardening $\tau^A$, in Fig.~\ref{figA2}(b) for reference, furnishes increasing
 shear strength at finite $\gamma^P$, as in Refs.~\cite{anand2003,miehe2009,bouvard2013,dal2022}.
 
 Metastable conditions for decomposition $\Xi = \tilde{\Xi}$ are assumed for Hugoniot states, similar to previous treatments of phase transformations in shock loading of metals \cite{boettger1997,claytonCMT2022,claytonZAMM2024}.
 For each decrement of $J = J^H $ on the principal Hugoniot, $\Xi(J^H)$ is obtained from \eqref{eq:Xifwd}.
 Threshold stress for decomposition onset $P^H_D$ is assigned from experiments quoted in Table~\ref{tableA1}, 
 with $J^H_D$ and $\theta_D$ corresponding calculated values at this shock pressure. Offset energy $\Psi^{D,\theta}_0$ and
 kinetic parameter $\beta_\Xi$ are obtained from methods discussed in Section~\ref{sec3.7}. Best agreement
 with experimental shock velocity and temperature data for shock pressures  $P > P^H_D$ was achieved by 
 setting thermodynamic latent heat $\lambda^D_\theta = 0$ and using a resistance function $R_\Xi (\theta)$ instead,
 where $\lambda^\Xi_\theta$ is a kinetic latent heat:
 \begin{align}
 \label{eq:lambdaXi}
 R_\Xi (\theta) = - \lambda^\Xi_\theta ( \theta / \theta_D - 1).
 \end{align}
 The positive value of $\lambda^\Xi_\theta$ in Table~\ref{tableA1} prevents room-temperature, pressure-induced decomposition 
 to hydrostatic pressures of nearly 800 GPa. Furthermore, $R_\Xi < 0$ for $\theta > \theta_D$, increasing the
 rate of decomposition versus $P$ with increasing Hugoniot temperature. Dissipation from $\dot{\Xi}$ in
 \eqref{eq:DXi} is unknown as kinetics inside the shock front are not resolved by the analysis, but results later confirm that total entropy production across every shock is positive in concurrence with the last of \eqref{eq:RHeqs}.
 
 Potential melting can be considered by analogous assumptions on metastability of Hugoniot states. From \eqref{eq:metfwd}, a minimum non-negative resistance function $R_\omega$ that prevents melting is determined by setting $F_\omega = R_\omega$ at each Hugoniot state $J = J^H$,
 where thermal and elastic Hugoniot properties of the melt are listed in Tables~\ref{tableA1} and \ref{tableA2}:
 \begin{align}
\nonumber
 & R_\omega(J^H) \geq F_\omega (J^H) = (1- \Xi) \langle \{ \Psi^G(J^H) - \Psi^M(J^H) \} \rangle
 \\ & 
 \nonumber 
 \approx (1-\Xi) \langle \{ \Psi^{G,E}_V (J^H)  [1 - B^{M,E}_0 / B^{G,E}_0]
 + \Psi^{G,E}_S (J^H)  [1 - G^{M,E}_0 / G^{G,E}_0] \\ & + \Psi^{G,E}_\beta (J^H) [1 - c^M_{V0}/c^G_{V0}]
 + \Psi^{G,\theta} (\theta(J^H)) - \Psi^{M,\theta} (\theta(J^H)) 
 + \Psi^{G,P}_A(J^H) - \Psi^{M,P}_A(J^H) \} \rangle.
  \label{eq:Romegamin}
 \end{align}
 
 A similar approach provides the minimum threshold energy $R_\xi$ needed to prevent fracture in metastable Hugoniot states far behind the wave front. 
  Under shock compression, $\phi = 0$ and $e^E_V < 0$, so $F^E_\xi = (1-\Xi) \Psi^{G,E}_S$, deviatoric elastic strain energy.
 With $\omega = 0$, \eqref{eq:Rxi} gives $R_\xi = (1-\Xi)R^G_\xi$. The minimum non-negative fracture resistance is determined
 from setting $\Phi = 0$ in \eqref{eq:Phifunc}:
 \begin{align}
 \label{eq:Rximin}
 R_\xi(J_H) \geq  (1-\Xi) \langle \{ \Psi^{G,E}_S (J^H) + c^P_\xi [\lambda^P(J^H) - 1]^2 \} \rangle.
 \end{align}
 
 \begin{figure}[hbp]
\centering
\subfigure[axial stress]{\includegraphics[width = 0.35\textwidth]{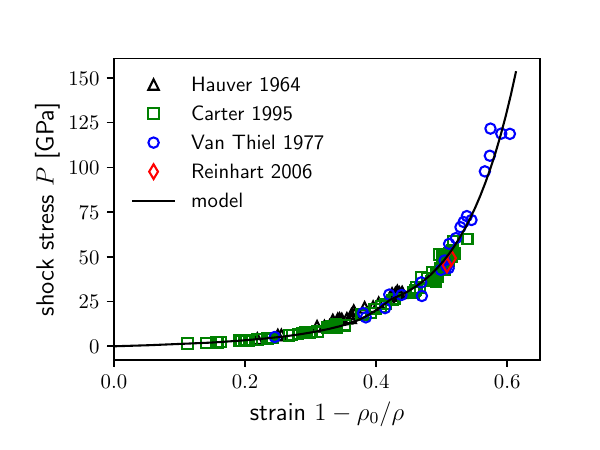} \label{fig1a}}
\subfigure[shock velocity]{\includegraphics[width = 0.35\textwidth]{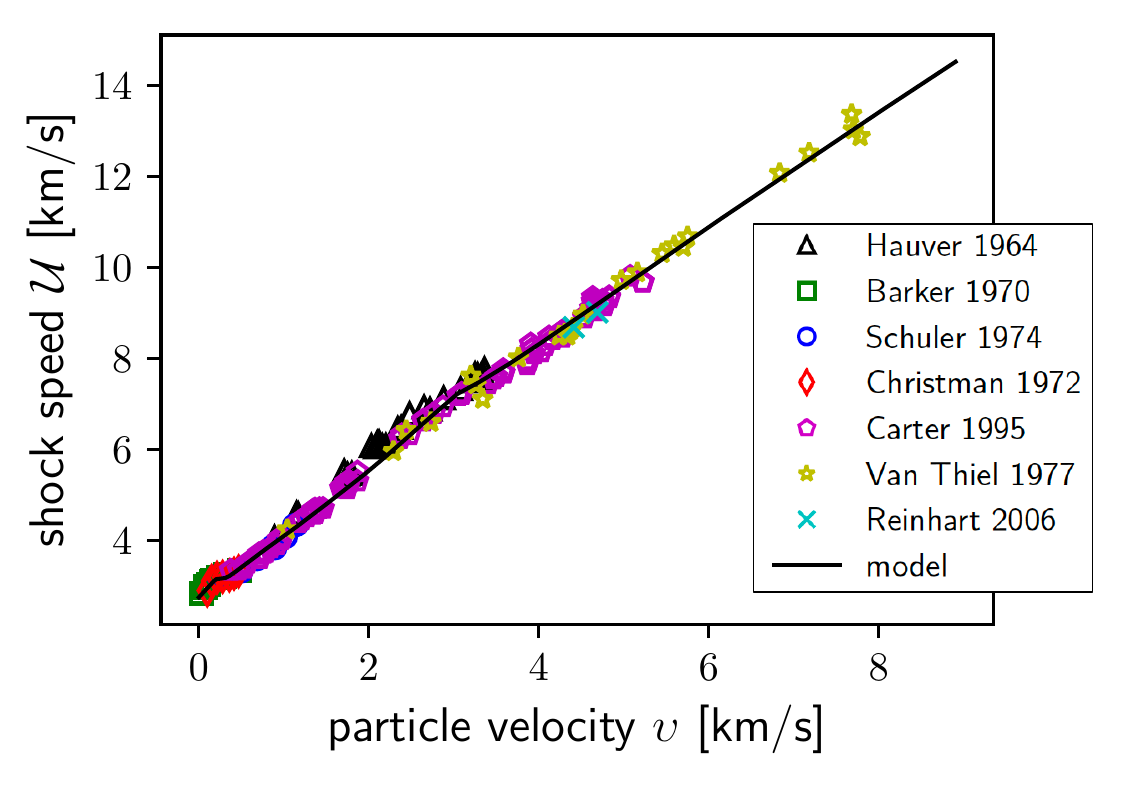} \label{fig1b}} \\
\subfigure[temperature]{\includegraphics[width = 0.35\textwidth]{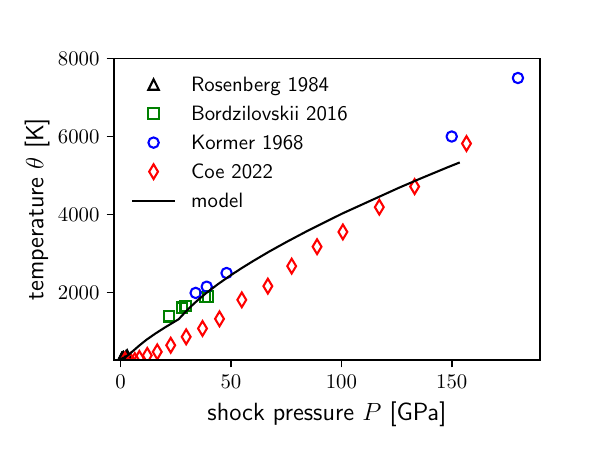} \label{fig1c}} 
\subfigure[wave speeds]{\includegraphics[width = 0.35\textwidth]{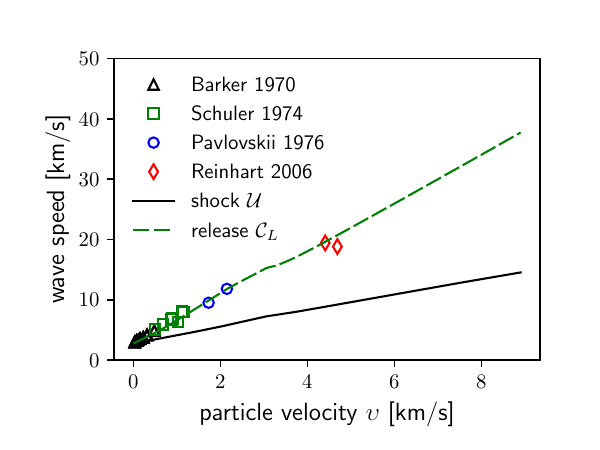} \label{fig1d}} 
\vspace{-0.2cm}
\caption{\label{fig1} Hugoniot response of PMMA to extreme pressure: (a) shock stress or shock pressure $P$ vs.~strain from model and experiments \cite{hauver1964,carter1995,vanthiel1977,reinhart2006} (b)
shock velocity $\mathcal U$ vs.~particle velocity $\upsilon$ from model and experiments \cite{hauver1964,barker1970,schuler1974,christman1972,carter1995,vanthiel1977,reinhart2006}
(c) temperature $\theta$ vs.~stress from model, experiments \cite{rosenberg1984,bordz2016,kormer1968},
and DFT \cite{coe2022} (d) Lagrangian sound speed $C_L$ from model and release experiments \cite{barker1970,schuler1974,pavlov1976,reinhart2006}
}
\end{figure}
 
 The high-pressure Hugoniot response is compared to experimental data in Fig.~\ref{fig1}.
 Experimental shock stress and shock velocity data \cite{hauver1964,barker1970,schuler1974,christman1972,carter1995,vanthiel1977,reinhart2006}
 are matched well to shock stresses $P$ evaluated to 120 GPa and shock speeds $\mathcal U$ evaluated to 13 km/s
 in respective Fig.~\ref{fig1}(a) and Fig.~\ref{fig1}(b).
 Experimental temperature data \cite{rosenberg1984,bordz2016,kormer1968}, which tend
 to exceed predictions from DFT \cite{coe2022}, are well captured to pressures exceeding 50 GPa.
 Extrapolation to pressures over 150 GPa suggests the model under-predicts
 temperature in this regime versus historical data \cite{kormer1968}. The assumed linear increase in
 specific heat with increasing temperature of the decomposed products is likely inaccurate at such high temperatures since
 $c_V^D(\theta)$ should plateau to an upper bound.
 Isentropic sound speed $C_L$ is shown with shock velocity $\mathcal U$ in Fig.~\ref{fig1}(d), along with
 release velocities from experiments \cite{barker1970,schuler1974,pavlov1976,reinhart2006}.
 Agreement is respectable, and both wave speeds increase monotonically with increasing particle velocity.
 The continuous increase in $\mathcal U$ precludes break-out of a precursor followed by a distinct plastic wave or transformation wave at the respective HEL or the shock decomposition threshold, for conditions that are not over-driven.
 
 \begin{figure}
\centering
\subfigure[shock velocity]{\includegraphics[width = 0.35\textwidth]{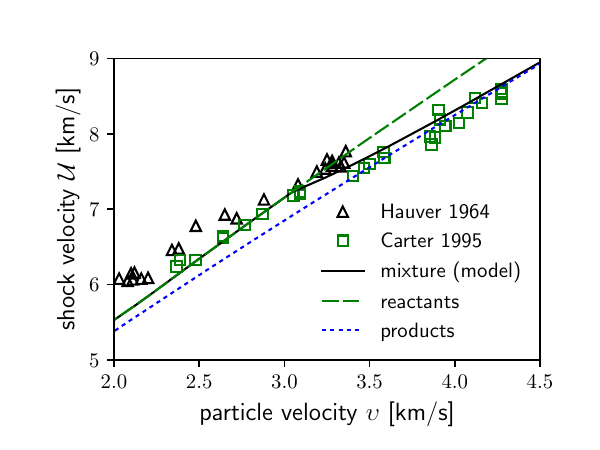} \label{fig2a}} 
\subfigure[decomposition fraction]{\includegraphics[width = 0.35\textwidth]{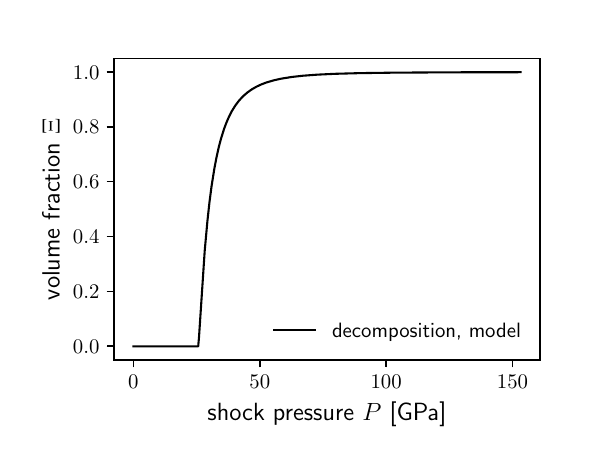} \label{fig2b}}
\vspace{-0.2cm}
\caption{\label{fig2} Hugoniot response of PMMA: (a) shock velocity $\mathcal U$ vs.~particle velocity $\upsilon$ from model (mixture, reactants, and decomposed products) and relevant experiments \cite{hauver1964,carter1995} in the vicinity of decomposition (b) predicted decomposition volume fraction $\Xi$ vs.~shock stress $P$ 
}
\end{figure}
 
 According to the model, shock decomposition commences at a pressure of 25.7 GPa and temperature of 1289 K, corresponding to a particle velocity of 3.0 km/s. The shock velocity-particle velocity Hugoniot is shown in this
 vicinity for the full model (mixture), isolated products, and isolated reactants in Fig.~\ref{fig2}(a), along with
 experimental data \cite{hauver1964,carter1995}. The predicted metastable decomposition fraction on the Hugoniot is
 reported in Fig.~\ref{fig2b}. The continuous decomposition process enables a smooth transition from
 the shock velocity Hugoniot of the reactants to the products, with a corresponding decrease in slope of $\mathcal U$ versus $\upsilon$. Transformation is 60\% complete at 31 GPa, 80\% complete at 36 GPa, and 90\% complete at 42 GPa where $\theta = 2124$ K.
 
\begin{figure}
\centering
\subfigure[axial stress]{\includegraphics[width = 0.35\textwidth]{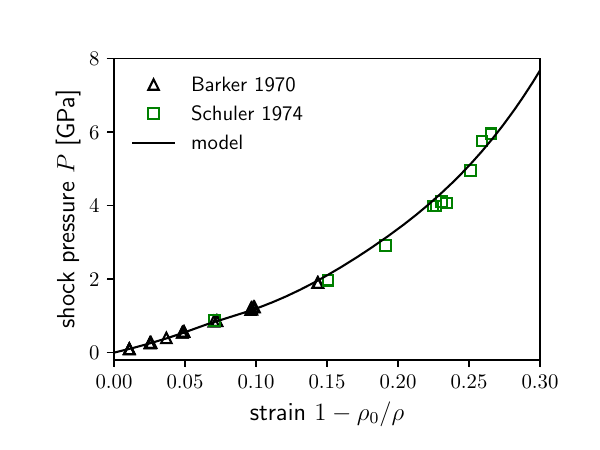} \label{fig3a}}
\subfigure[shock velocity]{\includegraphics[width = 0.35\textwidth]{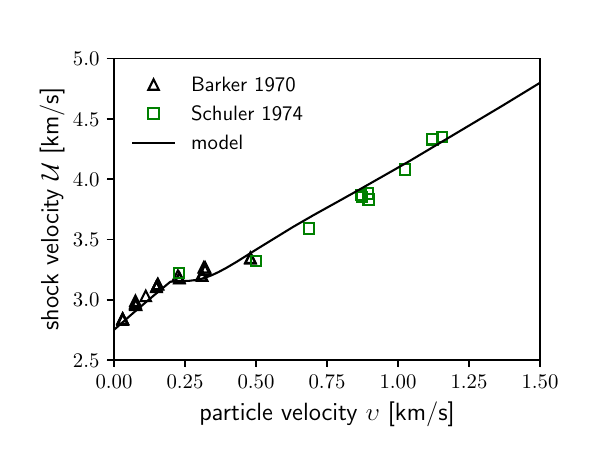} \label{fig3b}} \\
\subfigure[temperature]{\includegraphics[width = 0.35\textwidth]{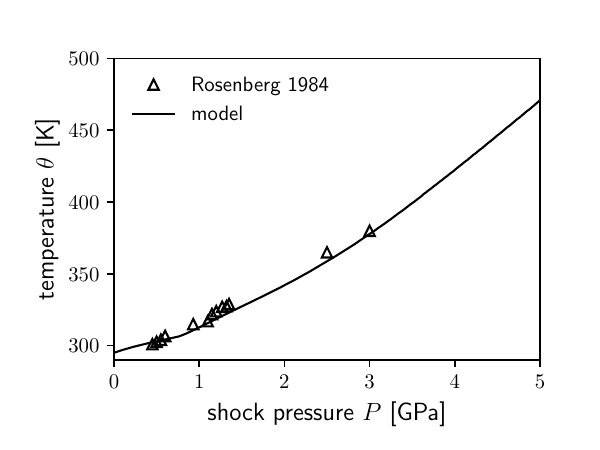} \label{fig3c}} 
\subfigure[wave speeds]{\includegraphics[width = 0.35\textwidth]{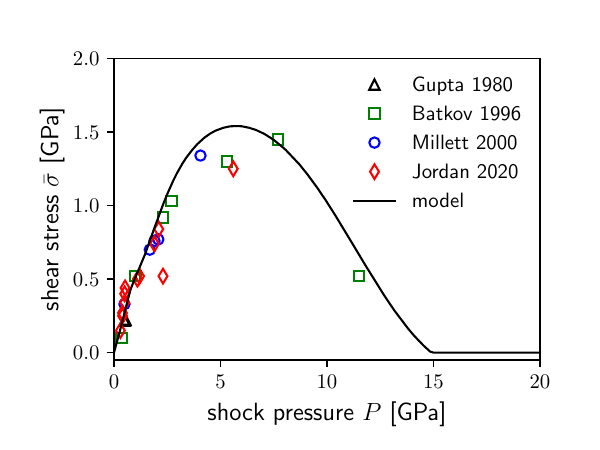} \label{fig3d}} 
\vspace{-0.2cm}
\caption{\label{fig3} Hugoniot response of PMMA to moderate pressure: (a) shock stress or shock pressure $P$ vs.~strain from model and experiments \cite{barker1970,schuler1974} (b)
shock velocity $\mathcal U$ vs.~particle velocity $\upsilon$ from model and experiments  \cite{barker1970,schuler1974}
(c) temperature $\theta$ vs.~stress from model and experiments \cite{rosenberg1984}
 (d) shear strength (Mises stress) $\bar{\sigma}$ from model and experiments \cite{gupta1980a,batkov1996,millett2000,jordan2020}
}
\end{figure}

 The Hugoniot response at more modest pressures is featured in Fig.~\ref{fig3}. 
 Data of Refs.~\cite{barker1970,schuler1974} on shock stress and shock velocity are captured in Fig.~\ref{fig3}(a) and
 Fig.~\ref{fig3}(b).
 Near and just above the HEL at $P^H_P = 0.74$ GPa, at the onset of plastic flow, shock velocity increases
 very slowly with particle velocity in the $\upsilon = 200 - 260$ m/s range. 
 Temperature closely follows experimental data \cite{rosenberg1984} in Fig.~\ref{fig3}(c).
 The model, in Fig.~\ref{fig3}(d), reflects the complex strength behavior of PMMA as measured by lateral stress gauges in experiments \cite{gupta1980a,batkov1996,millett2000,jordan2020}. Pressure hardening to a peak strength of 1.54 GPa between $P$ of 5 and 10 GPa is followed by a decrease in static strength to zero at  $P \gtrsim 15$ GPa as enabled by the function \eqref{eq:fRfit}. 
 The decay is attributed to thermal softening from shock temperature rise in Ref.~\cite{batkov1996}; according to the model, $\theta = 920$ K at $P = 15$ GPa. Although equilibrium strength far behind the wave front vanishes for $P \gtrsim 15$ GPa, the material has neither melted nor fractured (i.e., $\omega = \xi = 0$); dynamic strength $\tau^F > 0 $ for $\dot{\gamma}^P >0$ 
 persists in  \eqref{eq:richeton}.
 
\begin{figure}
\centering
\subfigure[thermodynamic coefficients]{\includegraphics[width = 0.32\textwidth]{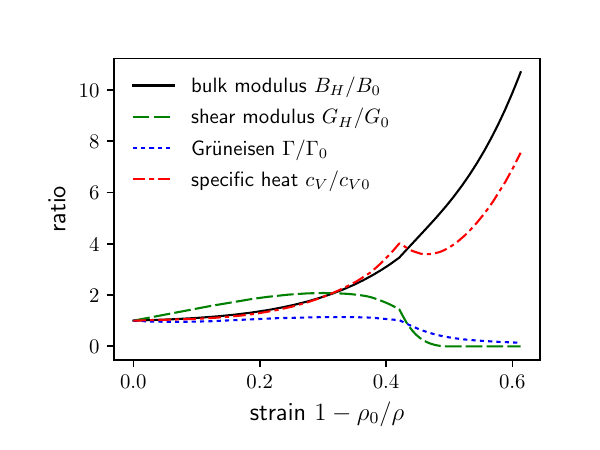} \label{fig4a}}
\subfigure[entropy]{\includegraphics[width = 0.32\textwidth]{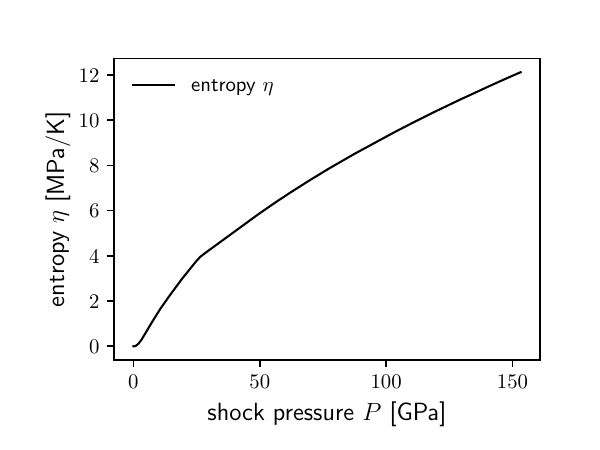} \label{fig4b}} 
\subfigure[kinetic resistance]{\includegraphics[width = 0.32\textwidth]{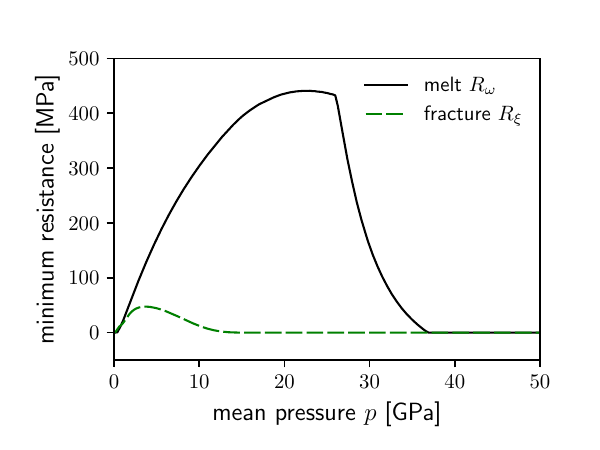} \label{fig4c}} 
\vspace{-0.2cm}
\caption{\label{fig4} Predicted Hugoniot response of PMMA: (a) ratio of Hugoniot to initial
bulk modulus, shear modulus, Gr\"uneisen parameter, and specific heat (b)
entropy per unit reference volume (c) minimum values of resistance functions $R_\omega$ and $R_\xi$
needed to suppress melting and fracture in high-pressure Hugoniot states}
\end{figure}

  Effective secant bulk modulus, shear modulus, Gr\"uneisen parameter, and specific heat on the Hugoniot, 
 defined in \eqref{eq:secantB}--\eqref{eq:secantGr} and all divided by their initial values, are reported in Fig.~\ref{fig4}(a). 
 The bulk modulus and specific heat increase with increasing compressive strain or pressure, with the exception
 of a momentary decrease in specific heat of the two-phase mixture of reactants and products during decomposition. 
 The shear modulus increases with pressure to a peak at 30\% compression, then decreases due to rapidly increasing shock temperature and later shock decomposition beginning at 42\% compression. Recall that decomposed products have null shear modulus. The Gr\"uneisen parameter remains relatively constant until shock decomposition, beyond which
 it decreases due to increasing specific heat. Entropy jump across the shock increases 
 monotonically with increasing pressure in Fig.~\ref{fig4}(b) and is always non-negative in accordance with the thermodynamic
 restriction in the last of \eqref{eq:RHeqs}. The change in slope at $P \approx 25$ GPa arises from differences in thermodynamic properties of the reactants and products.
 
 Respective resistance functions $R_\omega$ and $R_\xi$ needed to prohibit melting and fracture in Hugoniot states, calculated in \eqref{eq:Romegamin} and \eqref{eq:Rximin}, are shown versus mean pressure $p = P - \frac{2}{3} \bar{\sigma}$ in Fig.~\ref{fig4}(d). Melting is promoted by temperature above $\theta_G$ due to specific heat differences and the lower bulk and shear moduli of the melt relative to the glass (Table~\ref{tableA2}). Function $R_\omega$ provides
 the resistance to melt from high pressure noted elsewhere \cite{olabisi1975,saeki1992,menikoff2004,clements2012a}.
 Under compressive loading, fracture is driven only by shearing energy from elasticity and extreme plastic flow.
 The latter is negligible, and the former increases initially then reduces to zero at $P \gtrsim 15$ GPa when shear
 strength, and thus deviatoric strain energy density, vanish. 
 Modest resistance $R_\xi > 0 $ is required at lower pressures to reflect experimental tendencies \cite{satapathy2000,rittel2008}.
 The maximum value of $R_\xi (J^H)$ calculated this way for compression is $R_\xi = 95 \Gamma_\xi / l_\xi$, 
 reassuringly less than the value of $R^G_\xi = 115 \Gamma_\xi / l_\xi$ in Table~\ref{tableA2} used to model spall of the glass in Section~\ref{sec4.3}. 
 The minimum melt resistance is notably larger than the 
 fracture resistance but decreases rapidly at pressures exceeding the decomposition threshold since the products do not melt.
  These functions do not consider off-Hugoniot states within shock fronts that require a more in-depth computational study beyond the present scope.
 Furthermore, for comparison with experimental data, release waves have been deemed isentropic and thermoelastic
 per usual assumptions \cite{menikoff2004,davison2008,dattelbaum2019} and \eqref{eq:CLdef}.
 Depending on thermodynamic properties, melting and/or decomposition might be possible on shock
 release if temperatures remain high enough while pressure reduces.
 Shock-release melting has been studied elsewhere for metals \cite{williams2025a}, as have shear localization- and fracture-induced melting that could arise at lower pressures \cite{rittel2000,bjerke2002,claytonARX2025}.

\subsection{Low-pressure shock propagation}
\label{sec4.2}

At low impact stresses $P \lesssim P^H_P = 0.74$ GPa, structured steady waves are observed in planar shock compression experiments on PMMA after sufficient run distance.
These wave forms, as inferred from free-surface velocity histories, comprise an initial elastic shock resulting from instantaneous glassy response, followed
by a rounded rise to a terminal particle velocity nearly identical to half of the impact velocity for symmetric loading configurations \cite{barker1970,schuler1970,nunziato1973p,schuler1973}.
Such rounding, over finite rise times, originates from transient viscoelastic relaxation to equilibrium Hugoniot states.
Constitutive parameters are representative of $\mu$s time scales in the latter states, as discussed at the beginning of Section~\ref{sec4} (e.g., equilibrium Hugoniot stiffness exceeds quasi-static elastic stiffness). 

Steady wave behavior in this regime is analyzed by adapting methods of Refs.~\cite{claytonJMPS2021,claytonPRE2024}. For $J > J^H_P = 0.937$, the response in
\eqref{eq:Funi} is thermoelastic, and none of fracture, melting, or decomposition occur: ${\bm F} = {\bm F}^E$, $J = J^E$, $\bar{F}^P = \lambda^P = 1, \gamma^P = \xi = \omega = \Xi = 0$. 
A single volumetric ${\bm \Upsilon}_V$ and single shear ${\bm \Upsilon}_S$ tensor state variable suffice here for viscoelasticity,
so $\{ {\bm \alpha} \} \rightarrow ({\bm \Upsilon}_V, {\bm \Upsilon}_S)$.
Quantities corresponding to the head of the steady wave
are labeled by $\square_i = \square^-$, where $\square^-$ is the state immediately following the initial elastic
shock front treated as a singular surface \cite{schuler1970,nunziato1973p,schuler1973,schuler1974}. The $\square^+$
state ahead of the steady wave, and ahead of the initial shock, obeys \eqref{eq:plusstate}.
The state $\square_i$ is determined using \eqref{eq:RHeqs2} and procedures of Section~\ref{sec4.1}, albeit with instantaneous, rather than relaxed Hugoniot, thermoelastic properties defined in what follows. 
Quantities for final state at the tail of the steady wave are labeled $\square_f$. Even though the wave is of finite width (i.e., a ``structured'' shock), \eqref{eq:RHeqs} apply because the wave is steady \cite{claytonNEIM2019,davison2008}.
Taking the $\square^+$ state for the entire wave form as \eqref{eq:plusstate} and $\square_f$ as the end state, then \eqref{eq:RHeqs2} provide the equilibrium Hugoniot quantities at $\square_f$ \cite{nunziato1973p}, where suitable relaxed thermodynamic properties are given later. The arbitrary datum time, at the head of the steady wave, is $t_i = 0$.
The steady Lagrangian wave speed is $\mathcal U$.

For uniaxial strain with $J(X,t) = \partial x(X,t) / \partial X$, material and momentum conservation laws are
\begin{align}
\label{eq:1Dcont}
\dot{J} = \partial \upsilon / \partial X, \qquad \partial P / \partial X = - \rho_0 \, \partial \upsilon / \partial t.
\end{align}
Define the moving coordinate systems $Y(X,t)$ and resulting differential relations for function $f(Y)$:
\begin{align}
\label{eq:Ydef}
Y = X - {\mathcal U} t, \quad \partial f / \partial X = {\d } f / {\d } Y,  \quad \partial f / \partial t = - \mathcal{U} \d f / \d Y.
\end{align}
In a steady wave form, $J = J(Y), \upsilon = \upsilon(Y), P = P(Y)$, etc. Then \eqref{eq:1Dcont} and \eqref{eq:Ydef} produce
\begin{align}
\label{eq:SW1}
\frac{\d J}{\d Y} = -\frac{1}{\mathcal U} \frac{\d \upsilon}{\d Y}, \quad \frac{\d P}{\d Y} = \rho_0 {\mathcal U} \frac{\d \upsilon}{\d Y} \quad \Rightarrow \quad \frac{\d P}{\d Y} = - \rho_0 {\mathcal U}^2 \frac{\d J}{\d Y}.
\end{align}
The rightmost equality in \eqref{eq:SW1} is the differential stress-strain equation along the Rayleigh line
with slope $ \d P / \d J = - \rho_0 {\mathcal U}^2$. In the present application, the domain over which \eqref{eq:SW1} is used
is the Rayleigh line connecting the instantaneous Hugoniot at $\square_i$ to the relaxed Hugoniot at $\square_f$ \cite{schuler1973}. Integrating \eqref{eq:SW1} between any two points in the steady wave form gives
\begin{align}
\label{eq:RHSW}
 {\mathcal U} \llbracket J \rrbracket = - \llbracket \upsilon \rrbracket, \qquad \llbracket P \rrbracket =
 \rho_0 {\mathcal U}  \llbracket \upsilon \rrbracket = - \rho_0 {\mathcal U}^2  \llbracket J \rrbracket ,
\end{align}
where $\llbracket \square \rrbracket$ is the difference between the two points. Noting $\rho_0 = \rho J$, these agree with \eqref{eq:RHeqs}.

The longitudinal compressive stress $P$ and its derivative obey the constitutive relations
\begin{align}
\label{eq:Ptan} 
P(J,\eta,{\bm \Upsilon}_V, {\bm \Upsilon}_S) = - \frac{\partial U ({\bm F}(J),\eta,{\bm \Upsilon}_V, {\bm \Upsilon}_S)}{\partial J},
\quad \frac{ \d P}{ \d Y } = \frac{ \partial P}{ \partial J} \frac{ \d J}{ \d Y} + \frac{ \partial P}{ \partial \eta} \frac{ \d \eta}{ \d Y}
+ \frac{ \partial P}{ \partial {\bm \Upsilon}_V } : \frac{ \d {\bm \Upsilon}_V}{ \d Y} + \frac{ \partial P}{ \partial {\bm \Upsilon}_S}: \frac{ \d {\bm \Upsilon}_S}{ \d Y}.
\end{align}
Equating the second of \eqref{eq:Ptan} with \eqref{eq:SW1} and using the isentropic sound speed definition \eqref{eq:CLdef},
\begin{align}
\label{eq:dJdY}
\frac{\d J}{ \d Y}  = \frac{1}{\rho_0 [ (C_L)^2 - {\mathcal U}^2 ] } \left[ \frac{ \partial P}{ \partial \eta} \frac{ \d \eta}{ \d Y}
+ \frac{ \partial P}{ \partial {\bm \Upsilon}_V }:  \frac{ \d {\bm \Upsilon}_V}{ \d Y} + \frac{ \partial P}{ \partial {\bm \Upsilon}_S}: \frac{ \d {\bm \Upsilon}_S}{ \d Y} \right] = - \frac{1}{\mathcal U} \frac{\partial J}{ \partial t} \geq 0.
\end{align}
Both $Y$ and $J$  decrease from the head to the tail of the steady compression wave, and $C_L > {\mathcal U}$.

Equation \eqref{eq:dJdY} is integrated numerically for $J(Y)$ along the Rayleigh line from $J_i$ to $J_f$,
noting $\d Y = - {\mathcal U} \d t$ at fixed $X$.
For initial conditions $X_i = Y_i = 0 \leftrightarrow t_i = 0$,  volume ratio $J_i$ is the solution of \eqref{eq:RHeqs2} corresponding to $\mathcal U$ using instantaneous constitutive parameters. Complementary initial conditions on $P_i$, $\theta_i$, $\eta_i$, and $\upsilon_i$ are obtained from this solution, and $({\bm \Upsilon}_V)_i = ({\bm \Upsilon}_S)_i = {\bf 0}$.
End conditions for $J_f$ are obtained from solution of \eqref{eq:RHeqs2}
using relaxed Hugoniot constitutive parameters and the same value of $\mathcal U$.
The steady wave analysis terminates at this volume, when the Rayleigh line intersects the relaxed Hugoniot at ($J_f,P_f$) at time $t_f = - Y_f / {\mathcal U}$, whereby the constant end state is
\begin{align}
\label{eq:SWend}
J = J_f, \quad P = P_f; \quad \theta \rightarrow \theta_f, 
\quad  {\eta} \rightarrow \eta_f,
\quad \dot{\bm \Upsilon}_V \rightarrow {\bf 0},
 \quad \dot{\bm \Upsilon}_S \rightarrow {\bf 0}, \qquad [ \forall \, t > t_f ].
\end{align}
If the relaxed Hugoniot pair ($J_f,P_f$) is intersected prior to attainment of the latter four conditions in \eqref{eq:SWend} (e.g., total viscoelastic relaxation may require infinite time), then the latter are enforced
as jump conditions on temperature, entropy, and internal variable rates at $t = t_f$ for consistency.

The thermoelastic model and thermoelastic properties, namely elastic constants and Gr\"uneisen parameters, used in Section~\ref{sec4.1} are insufficient in the present setting, as the former depict a compromise between
instantaneous and relaxed Hugoniot responses with transient viscoelastic effects excluded. 
Here in Section~\ref{sec4.2}, 
distinct thermoelastic constants for instantaneous and relaxed Hugoniot states are defined 
following Refs.~\cite{schuler1970,nunziato1973p,schuler1973,schuler1974},
and free energy $\Psi^G$ transitions from instantaneous to relaxed response via
evolution of internal configurational variables ${\bm \Upsilon}_V$ and ${\bm \Upsilon}_S$.
Neither the instantaneous nor relaxed Hugoniot should be extrapolated to plastic states $P \gg P^H_P$.

According to the framework of Section~\ref{sec3.3}, configurational energies $\Psi^\Upsilon_V,\Psi^\Upsilon_S$
are not prescribed in analytical form, but can be obtained via numerical integration, for example \eqref{eq:Qsm} and
\eqref{eq:QVl}, where internal variable rates are given by \eqref{eq:pivisc} and \eqref{eq:piviscV}.
In the current analysis, an analytical form of $P(J,\theta,{\bm \Upsilon}_V, {\bm \Upsilon}_S)$, and therefore
of $\Psi^\Upsilon_V$ and $\Psi^\Upsilon_S$, is required to accurately calculate
$\partial P / \partial {\bm \Upsilon}_V$, $ \partial P / \partial {\bm \Upsilon}_S$, and $\d \eta / \d Y$ entering \eqref{eq:dJdY}. Strain-like internal variables referred to Lagrangian coordinates are
related to scalar volumetric strain-like measure $e^\Upsilon_V$ and spatial deviatoric measure $ \bar{\bm e}^\Upsilon$:
\begin{align}
\label{eq:Upsdef}
{\bm \Upsilon}_V =({\bm R}^E)^{\mathsf T} e^\Upsilon_V  {\bm R}^E, \quad
{\bm \Upsilon}_S =({\bm R}^E)^{\mathsf T} \bar{\bm e}^\Upsilon  {\bm R}^E, \quad
 \bar{\bm e}^\Upsilon = ( \bar{\bm e}^\Upsilon)^{\mathsf T}, \quad
{\rm tr} \, \bar{\bm e}^\Upsilon = 0.
\end{align}

For isotropic viscoelastic response, $\Psi^\Upsilon_V = \Psi^\Upsilon_V(e^\Upsilon_V,e^E_V,\theta)$
and $\Psi^\Upsilon_S = \Psi^\Upsilon_S (\bar{\bm e}^\Upsilon,\bar{\bm e}^E,\theta)$. These
are not needed individually, but rather are combined into a total viscoelastic free energy that, similar to the 
example in Ref.~\cite{holzapfel1996a}, depends only on the differences $\Delta e_V = e^E_V - e^\Upsilon_V$ and
$\Delta \bar{\bm e} = \bar{\bm e}^E - \bar{\bm e}^\Upsilon$:
\begin{align}
\label{eq:psiviscdef}
\Psi^\Upsilon  = \Psi^\Upsilon  (e^E_V,\bar{\bm e}^E;
e^\Upsilon_V, \bar{\bm e}^V; \theta) =  \Psi^\Upsilon (\Delta e_V,\Delta \bar{\bm e},\theta), \qquad
\hat{\Psi}^\Upsilon = \Psi^\Upsilon (\Delta e_V = e^E_V, \Delta \bar{\bm e} = \bar{\bm e}^E, \theta).
\end{align}
Function $\hat{\Psi}^\Upsilon$ produces instantaneous response, wherein configurational variables in \eqref{eq:Upsdef} vanish. Viscous stresses ${\bm Q}^V_l, {\bm Q}^S_m$ and
their spatial counterparts  ${\bm q}^V_l, {\bm q}^S_m$ are defined similarly but not identically to \eqref{eq:Qsm}, \eqref{eq:dotQS}, 
\eqref{eq:QVl}, and \eqref{eq:dotQV}, where subscripts $l = 1$ and $m = 1$ are not needed:
\begin{align}
\label{eq:QVdef}
& {\bm Q}^V =  J ({\bm R}^E)^{\mathsf T} {\bm q}^V {\bm R}^E, \quad
{\bm q}^V = q^V {\bf 1} = \frac{1}{J} \frac{\partial \Psi^\Upsilon}{\partial e^E_V} {\bf 1} = 
-  \frac{1}{J} \frac{\partial \Psi^\Upsilon}{\partial e^\Upsilon_V} {\bf 1}=  - (p^\Upsilon_V + p^\Upsilon_S) {\bf 1},  \quad
\hat{q}^V = \frac{1}{J} \frac{\partial \hat{\Psi}^\Upsilon}{\partial e^E_V};
\\
\label{eq:QSdef}
& {\bm Q}^S =  J ({\bm R}^E)^{\mathsf T} {\bm q}^S {\bm R}^E, \quad
{\bm q}^S = \frac{1}{J} \frac{\partial \Psi^\Upsilon}{\partial \bar{\bm e}^E}  = 
-  \frac{1}{J} \frac{\partial \Psi^\Upsilon}{\partial \bar{\bm e}^\Upsilon} = \bar{\bm \sigma}^\Upsilon_S,  \qquad
 \hat{\bm q}^S = \frac{1}{J}  \frac{\partial \hat{\Psi}^\Upsilon}{\partial \bar{\bm e}^E}.
 \end{align}
Two relaxation times $\tau^V_l \rightarrow \tau^V = {\rm const}$ and $\tau^S_m \rightarrow \tau^S= {\rm const}$ are sufficient for the current problem.
Analogs of \eqref{eq:dotQS}, \eqref{eq:convint}, \eqref{eq:dotQV}, and \eqref{eq:convintV} are as follows, with initial conditions at $t_0 \rightarrow t_i = 0$:
\begin{align}
\label{eq:dotQVuni}
 &  \dot{q}^V + \frac{q ^V} {\tau^V}  =  D_t {\hat{q}}^V , 
\quad
q^V (t) = 
q^V_0 \exp \biggr{[} \frac{-t}{ \tau^V} \biggr{]}
 + 
 \int_{0^+}^t \exp \biggr{[} \frac{s-t}{ \tau^V} \biggr{]}
 D_s  \hat{q}^V {\d } s, 
 \quad q^V_0 
  = \frac{1}{J} \frac{\partial \hat{\Psi}^\Upsilon}{\partial e^E_V} \bigr{\rvert}_{t = 0};
\\
\label{eq:dotQSuni}
&  \dot{\bm q}^S + \frac{{\bm q}^S}{ \tau^S}
 =  D_t  \hat{\bm q}^S, \quad 
 {\bm q}^S (t) = 
 {\bm q}^S_0 \exp \biggr{[} \frac{-t}{ \tau^S} \biggr{]} + 
 \int_{0^+}^t \exp \biggr{[} \frac{s-t}{ \tau^S} \biggr{]}
 D_s  \hat{\bm q}^S   {\d } s, \quad
{\bm q}^S_{0} =   \frac{1}{J}  \frac{\partial \hat{\Psi}^\Upsilon}{\partial \bar{\bm e}^E} \bigr{\rvert}_{t =0}.
\end{align}
The spatial tensor rate equation \eqref{eq:dotQSuni} is valid generally only if continuum spin $\dot{\bm R}{\bm R}^{\mathsf T} = {\bf 0}$; otherwise an objective rate of ${\bm q}^S$ should be used. The present form is acceptable for uniaxial strain wherein ${\bm R} = {\bm R}^E = {\bf 1}$. Also in the uniaxial setting,
${\bm q}^S = {\rm diag}(q^S_{11},-\frac{1}{2} q^S_{11},-\frac{1}{2} q^S_{11})$, and $\hat{\bm q}^S$ is of similar form, so \eqref{eq:dotQSuni} need only be integrated for  a single component (e.g., $q^S_{11}$).

Combining \eqref{eq:QVdef} with \eqref{eq:dotQVuni}, combining \eqref{eq:QSdef} with \eqref{eq:dotQSuni}, and
transforming the integration variable from $t$ to $Y$, two zero-valued functions $Z^V(Y)$ and $Z^S(Y)$ are defined:
\begin{align}
\label{eq:ZV}
& Z^V(Y) = \frac{1}{J} \frac{\partial \Psi^\Upsilon(e^\Upsilon_V(Y), \bar{\bm e}^\Upsilon (Y), \ldots)}{\partial e^\Upsilon_V} +
q^V_0 \exp \biggr{[} \frac{Y}{ {\mathcal U} \tau^V} \biggr{]}
 + 
 \int_{0^+}^Y \exp \biggr{[} \frac{Y-s}{ {\mathcal U} \tau^V} \biggr{]}
 D_s  \hat{q}^V {\d } s = 0,
 \\
 \label{eq:ZS}
  & Z^S(Y) =  \frac{1}{J} \frac{\partial \Psi^\Upsilon(e^\Upsilon_V(Y), \bar{\bm e}^\Upsilon (Y), \ldots)}{\partial \bar{e}^\Upsilon_{11}} +
(q^S_0)_{11} \exp \biggr{[} \frac{Y}{ {\mathcal U} \tau^V} \biggr{]}
 + 
 \int_{0^+}^Y \exp \biggr{[} \frac{Y-s}{ {\mathcal U} \tau^S} \biggr{]}
 D_s  \hat{q}^S_{11} {\d } s = 0.
\end{align}
Equations \eqref{eq:ZV} and \eqref{eq:ZS} are solved simultaneously, by iteration, for $e^\Upsilon_V(Y)$ and
$\bar{e}^\Upsilon_{11}(Y)$ at each increment $\d Y$ in the wave. Other potentially nonzero components
of $\bar{\bm e}^\Upsilon$ are $\bar{e}^\Upsilon_{22} = \bar{e}^\Upsilon_{33} = -\frac{1}{2} \bar{e}^\Upsilon_{11}$.
Derivatives are obtained via standard difference approximations taken at each increment:
\begin{align}
\label{eq:dZ}
\frac{ \d e^\Upsilon_V}{\d Y} = \lim_{\d Y \rightarrow 0}
\frac{{\rm arg}0(Z^V(Y+ \d Y)) - {\rm arg}0(Z^V(Y) )}{ \d Y}, 
\quad
\frac{ \d \bar{e}^\Upsilon_{11}}{\d Y} = \lim_{\d Y \rightarrow 0}
\frac{ {\rm arg}0(Z^S(Y+ \d Y)) - {\rm arg}0(Z^S(Y) )}{ \d Y},
\end{align}
where arguments of $Z^V = 0 $ and $Z^S = 0 $ are $e^\Upsilon_V(Y)$ and $\bar{e}^\Upsilon_{11}(Y)$, respectively,
from \eqref{eq:ZV} and \eqref{eq:ZS}.

Let $\square^{G,E}$ and $\square^{G,I}$ label coefficients of respective relaxed and instantaneous
thermoelastic responses. Specifically, these comprise $B^{G,*}_i$ and $\Gamma^{G,*}_i$ where $i=0,1,2,3$
and $i = 0,1,2$
for bulk moduli and Gr\"uneisen parameters, and shear modulus parameters $G^{G,*}_0$ and $G^{G,*}_p$.
Values are in Table~\ref{tableA3}.
Specific heat parameters $c_{V0}^G$ and $c_{\theta}^G$, and $G^{G,*}_\theta$, are the same for relaxed and instantaneous states, all unchanged from Table~\ref{tableA1}.
Higher-order bulk moduli $B^{G,E}_i =  B^{G,I}_i$ for $i > 0$. 
The viscoelastic free energy function combining volumetric and shear response is, valid for $e^E_V \leq 0$,
\begin{align}
\nonumber
\Psi^\Upsilon   = &  %(\Delta e_V,\Delta \bar{\bm e},\theta) 
(B_0^{G,I} - B_0^{G,E}) (\Delta e_V)^2 \left[ {\textstyle{\frac{1}{2}}} - {\textstyle{\frac{1}{6}}}  B_1^{G,E} \Delta e_V + {\textstyle{\frac{1}{24}}} B_2^{G,E} (\Delta e_V)^2 - {\textstyle{\frac{1}{120}}} B_3^{G,E} (\Delta e_V)^3 \right]
\\
\nonumber 
&  + [ (G^{G,I}_0 - G^{G,E}_0) - (B_0^{G,I} G_p^{G,I} - B_0^{G,E} G_p^{G,E}) \Delta e_V 
+  (B_0^{G,I} G_2^{G,I} - B_0^{G,E} G_2^{G,E}) (\Delta e_V)^2 ] \Delta \bar{\bm e}: \Delta \bar{\bm e}
\\
& \!  -c^G_{V0} (\theta - \theta_0) \Delta e_V [ ( \Gamma_0^{G,I}- \Gamma_0^{G,E}) + {\textstyle{\frac{1}{2}}} 
(\Gamma_1^{G,I} - \Gamma_1^{G,E}) \Delta e_V
+ {\textstyle{\frac{1}{6}}}(\Gamma_2^{G,I} - \Gamma_2^{G,E})(\Delta e_V)^2].
\label{eq:PsiUpsTot}
\end{align}
The total free energy $\Psi = \Psi^{G,E}_V + \Psi^{G,E}_S + \Psi^{G,E}_\beta + \Psi^{G,\theta} + \Psi^{\Upsilon}$ furnishes the complete instantaneous thermoelastic response when $e^\Upsilon_V = 0 \Rightarrow \Delta e_V = e^E_V$ and
$\bar{\bm e}^\Upsilon = {\bf 0} \Rightarrow \Delta \bar{\bm e} = \bar{\bm e}^E$. The relaxed Hugoniot energy, but not
all of its strain derivatives,
is recovered when  $\Delta e_V = 0$ and $\Delta \bar{\bm e} = {\bf 0}$, giving $\Psi^{\Upsilon} = 0$.

Quantities on the right side of the shock structure differential equation \eqref{eq:dJdY} can be found analytically using \eqref{eq:QVdef}, \eqref{eq:QSdef}, \eqref{eq:PsiUpsTot}, and entropy production from ${\mathfrak D} = {\mathfrak D}^\Upsilon_V + {\mathfrak D}^\Upsilon_S$:
\begin{align}
\label{eq:dPdUVS}
& 
\frac{ \partial P}{ \partial {\bm \Upsilon}_V }:  \frac{ \d {\bm \Upsilon}_V}{ \d Y} = 
- \{ J \biggr{[} \frac{\partial p^\Upsilon}{\partial J} \biggr{\rvert}_{\theta, \bar{\bm e}^\Upsilon} -
 \frac{\theta \Gamma } {J} 
 \frac{\partial p^\Upsilon}{\partial \theta} \biggr{\rvert}_{J,\bar{\bm e}^\Upsilon } \biggr{]} - q^V \} \frac{\d e^\Upsilon_V}{\d Y} ,
 \qquad
 \frac{ \partial P}{ \partial {\bm \Upsilon}_S }:  \frac{ \d {\bm \Upsilon}_S}{ \d Y} =
\frac{ \d q^S_{11}}{\d \bar{e}^E_{11}} \biggr{\rvert}_{\theta, e^\Upsilon_V} \frac{\d \bar{e}^\Upsilon_{11}}{\d Y},
 \\
 \label{eq:dPdeta}
 & 
\frac{ \partial P}{ \partial \eta }\biggr{\rvert}_{J, e^\Upsilon_V,\bar{\bm e}^\Upsilon}   \frac{ \d \eta}{ \d Y} =
% \frac{\theta}{c_V^G} \frac{ \partial P}{ \partial \theta }\biggr{\rvert}_{J, e^\Upsilon_V,\bar{\bm e}^\Upsilon}  \frac{ \d \eta}{ \d Y} =
   \frac{- ( {\mathfrak D}^\Upsilon_V + {\mathfrak D}^\Upsilon_S) }{{\mathcal U} c_V^G} \frac{ \partial P}{ \partial \theta }\biggr{\rvert}_{J, e^\Upsilon_V,\bar{\bm e}^\Upsilon}
= 
 \frac{J}{c_V^G} \biggr{[} q^V  \frac{\d e^\Upsilon_V}{\d Y}  + \frac{3}{2} q^S_{11}  \frac{\d \bar{e}^\Upsilon_{11}}{\d Y}\biggr{]}
  \frac{ \partial P}{ \partial \theta }\biggr{\rvert}_{J, e^\Upsilon_V,\bar{\bm e}^\Upsilon} .
\end{align}
Bulk shock velocity $\kappa^V$ of Section~\ref{sec3.8} is not needed in the current analytical-numerical method, and shear 
viscosity $\mu^V$ vanishes by definition in the glass phase.
Adiabatic temperature rise in the wave is obtained by integrating the following form of energy balance \eqref{eq:tempratefin} with respect to $Y$:
\begin{align}
\label{eq:dthetadY}
\frac{\d \theta}{ \d Y} = -\frac{ \theta}{J} \biggr{[} \Gamma - \frac{4}{3} \frac{G^{G,E}}{c_V^G} \ln J \biggr{]} \frac{ \d J} {\d Y}
+  \frac{J}{c_V^G} \biggr{[} q^V - \theta \frac{\partial q^V}{\partial \theta} \biggr{]}  \frac{\d e^\Upsilon_V}{\d Y}
+  \frac{3}{2} \frac{J}{c_V^G} \biggr{[} q^S_{11} - \theta \frac{\partial q^S_{11}}{\partial \theta} \biggr{]}  \frac{\d \bar{e}^\Upsilon_{11}}{\d Y}.
\end{align}
Given the solution $J= J(Y)$ from \eqref{eq:dJdY}, particle velocity from \eqref{eq:RHSW} is 
$\upsilon(Y) = \upsilon_i +{\mathcal U} [J_i - J(Y)]$.

The ``critical strain gradient'' $(\partial J / \partial X)^-$ immediately behind the initial instantaneous shock,
for which a steady wave exists, can be derived in closed form using various thermodynamic frameworks \cite{chen1970a,chen1971,chen1972,bowen1974,claytonIJES2022,claytonPRE2024}.
For the present viscoelastic framework and initial conditions, the analytical solution derived in Ref.~\cite{claytonPRE2024}
applies, here specialized to the simpler case of a single-phase solid rather than solid-fluid mixture.
Denote by $\delta_t \square$ the time derivative of a quantity $\square (X,t)$
with respect to an observer moving with the shock at Lagrangian speed $\mathcal U$:
 $\delta_t \square = \dot{\square} +{ \mathcal U} \partial \square / \partial X$.
 The shock evolution equation for strain amplitude $\llbracket J(X,t) \rrbracket = J^-(X,t) - 1 $ is \cite{claytonPRE2024}:
 \begin{align}
 \label{eq:deltaJ}
 & \delta_t \llbracket J \rrbracket = 
{\mathcal U}  \frac{(1- \hat{\xi}) (2 - \hat{\zeta})  \{ \Lambda - (\partial  J / \partial X)^{-} \} }
{ (3 \hat{\xi} +1 ) - \hat{\zeta} (3 \hat{\xi} -1 ) },
\quad
\Lambda = 
\frac{ 1 + \llbracket J \rrbracket } {(1- \hat{\xi})  \hat{\mathsf C }^{-} }
\biggr{\{}
\frac{\rho_0}{J^{-}} [ {\bm{L}}^{-} :  \dot{\bm{a}}^{-} ] \biggr{\}}.
\end{align}
For the current application, the requisite mechanical and thermodynamic quantities are
\begin{align}
\nonumber 
& \hat{\mathsf C} = \rho_0 (C_L)^2, \quad \hat{\xi} = ({\mathcal U}/C_L^-)^2,
\quad \hat{\mathsf G} = -\rho \theta \Gamma, \quad \hat{\zeta} = \hat{\mathsf G}^- \llbracket J \rrbracket / (\rho_0 \theta^-), \quad {\bm \pi} = - \{ q^V, {\bm q}^S \},
\\
\label{eq:dJ2}
&  {\bm { a}} = \{ e^\Upsilon_V, \bar{\bm{e}}^\Upsilon \}, \quad
\dot{\bm { a}} = -  {\mathcal U} \biggr{ \{ }  \frac{ \d e^\Upsilon_V }{ \d Y},  \frac{ \d \bar{\bm{e}}^\Upsilon} { \d Y} \biggr{\}}, \quad
 \bm{ A} = - \frac{\partial P}{\partial {\bm a} }\biggr{\rvert}_{J,\eta}, 
\quad {\bm L} = \frac{1}{\rho_0 {\mathcal U} }  \biggr{[} {\bm A} - \frac{ \hat{\mathsf G} }{\rho \theta} {\bm \pi} \biggr{]}.
\end{align}
When $\Lambda = \Lambda_c = (\partial J / \partial X)^{-}$, the initial elastic shock is steady
with $J^-(t) = J_i = {\rm const}$, $P^-(t) = P_i = {\rm const}$, and ${\mathcal U} = \rm{const}$,
consistent with steady-wave assumptions of Section~\ref{sec4.2}. Calculations verified that this closed-form solution agrees with the numerical outcome of \eqref{eq:dJdY} at $Y = Y_i = 0$.

\begin{figure}
\centering
\subfigure[Hugoniot stress]{\includegraphics[width = 0.32\textwidth]{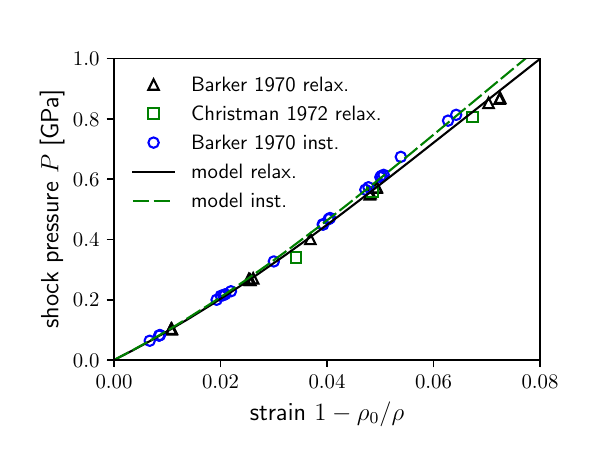} \label{fig5a}}
\subfigure[shock velocity]{\includegraphics[width = 0.32\textwidth]{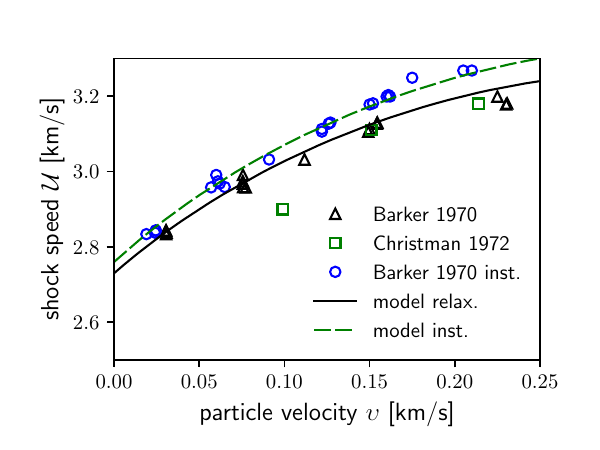} \label{fig5b}} 
\subfigure[Hugoniot temperature]{\includegraphics[width = 0.32\textwidth]{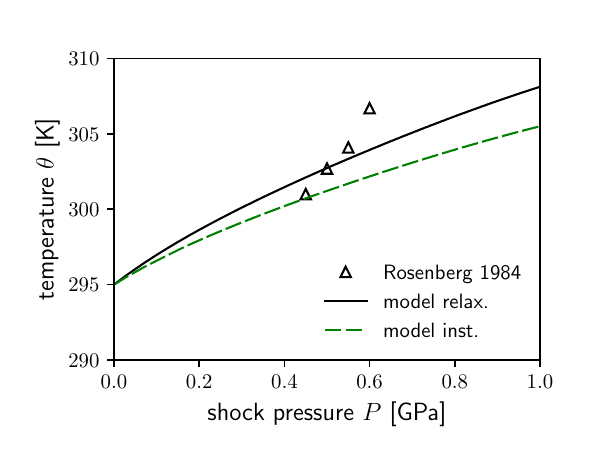} \label{fig5c}} \\
\vspace{-0.2cm}
\subfigure[Hugoniot wave speeds]{\includegraphics[width = 0.32\textwidth]{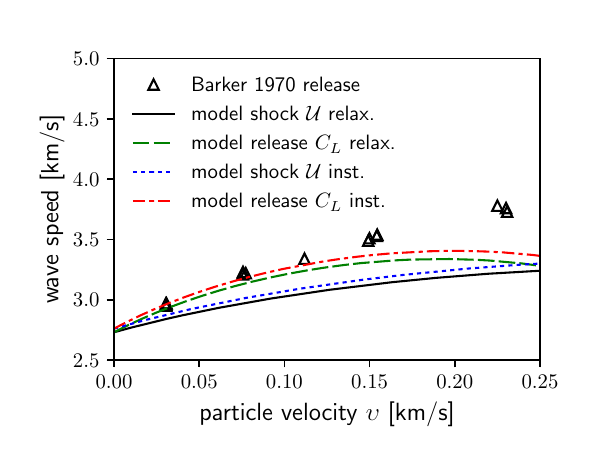} \label{fig5d}}
\subfigure[steady velocity profiles]{\includegraphics[width = 0.32\textwidth]{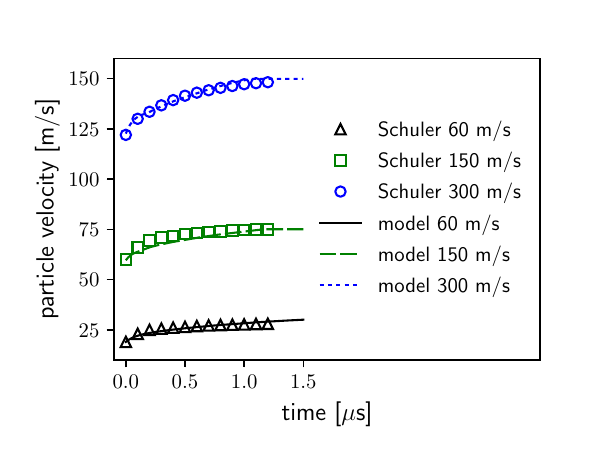} \label{fig5e}} 
\subfigure[sound speed in steady wave]{\includegraphics[width = 0.32\textwidth]{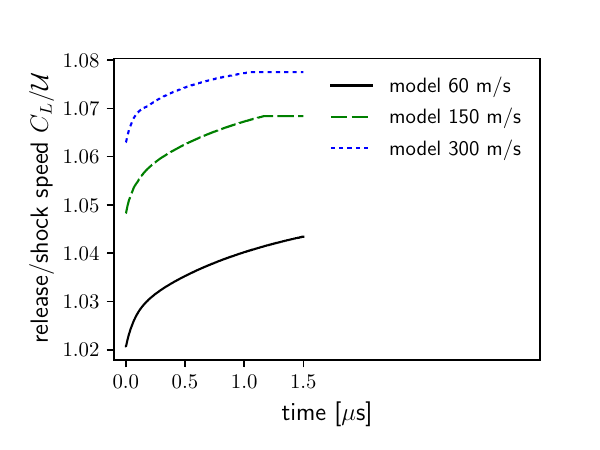} \label{fig5f}} 
\vspace{-0.2cm}
\caption{\label{fig5} Low-pressure shock response of PMMA: (a) instantaneous and relaxed shock pressure $P$
from model and experiments \cite{barker1970,christman1972} (b)
 instantaneous and relaxed shock velocity $\mathcal U$
from model and experiments \cite{barker1970,christman1972}
(c) instantaneous and relaxed Hugoniot temperature $\theta$ from model and experiments \cite{rosenberg1984}
(d) shock $\mathcal U$ and sound $C_L$ velocity from model and experiments \cite{barker1970}
(e) steady profiles of particle velocity $\upsilon$ vs.~time from model and experiments \cite{schuler1970}
(f) ratio of local sound speed $C_L$ to steady wave velocity $\mathcal U$ (model)}
\end{figure}

Outcomes of the analysis are shown in Fig.~\ref{fig5}.
Model depictions of instantaneous and relaxed stress-volume and shock velocity-particle velocity
Hugoniots are compared with experimental data \cite{barker1970,schuler1974} in Fig.~\ref{fig5}(a) and Fig.~\ref{fig5}(b). These data have been used to calibrate \eqref{eq:PsiUpsTot}.
Instantaneous curves exceed relaxed counterparts.
Temperature from the relaxed rather than instantaneous model more closely matches experiment \cite{rosenberg1984} in Fig.~\ref{fig5}(c). Conversely, instantaneous sound speed $C_L$ more closely agrees with experimental
release velocities \cite{barker1970} in Fig.~\ref{fig5}(d). Failure of the model to capture release wave data for $\upsilon \gtrsim 150 $ m/s is not unexpected. At higher velocities, plastic deformation occurs in PMMA and \eqref{eq:PsiUpsTot} is decreasingly  accurate, showing that this energy function and its parameter values should not be extrapolated to higher pressures. Similar issues are reported in other studies \cite{schuler1970,nunziato1973p,schuler1973}. The current steady wave analysis is limited to particle velocities $\upsilon_i \lesssim 125$ m/s for the initial instantaneous elastic shock front, for which accuracy of $C_L$ is confirmed. For $0 < \upsilon  \lesssim 170 $ m/s, $C_L$ and $\mathcal U$ increase monotonically with increasing $\upsilon$, and $C_L / {\mathcal U} > 1$.

\begin{table}
\caption{Steady wave loading conditions (model); $\square_i, \square_f$ denote initial, final states}
\centering
\footnotesize
\label{table1}
\begin{tabular}{cccc}
\hline
\bf{Nominal impact [m/s]} & \bf{Wave speed $\mathcal{U}$ [m/s]} & \bf{Volume $J_i,J_f$} & \bf{Temperature $\theta_i,\theta_f$} \\
\hline
60 & 2834 & 0.9933, 0.9892 & 296.06, 297.03 \\
150 & 2966 & 0.9799, 0.9747 & 298.05, 299.59 \\
300 & 3120 & 0.9607, 0.9520 & 300.75, 303.41 \\
\hline
 \end{tabular}
\end{table}

\begin{table}
\caption{Critical strain gradient $\Lambda_c = (\partial J / \partial X)^-$ for steady waves in PMMA }
\centering
\footnotesize
\label{table2}
\begin{tabular}{cccc}
\hline
\bf{Nominal impact [m/s]} & \bf{Strain $1 - J^-$} & $\Lambda_c$  [1/m] (experiment \cite{nunziato1973am}) & $\Lambda_c$  [1/m] (model, eq.~\eqref{eq:deltaJ}) \\
\hline
60 &  0.0067 & $-$ & 6.4 \\
150 & 0.0201 & $10.0 \pm 3$ & 9.4 \\
220 & 0.0300 & $13.7$ & $-$ \\
300 & 0.0393 & $16.3 \pm 6$ & 16.3  \\
\hline
 \end{tabular}
\end{table}

The steady wave analysis is applied to three experiments of Ref.~\cite{schuler1970}, where loading protocols are listed in Table~\ref{table1}. Nominal impact velocity is very close to twice the final particle velocity in the tail of the steady wave \cite{schuler1970,schuler1973}. Particle velocity histories computed via \eqref{eq:RHSW} with numerical integration of
\eqref{eq:dJdY} are compared with data extracted from free surface profiles \cite{schuler1970} in Fig.~\ref{fig5}(e).
A best fit to all three experimental profiles is obtained via adjustment of relaxation times $\tau^V$ and $\tau^S$,
where the same two values in Table~\ref{tableA3} are applied simultaneously to all three profiles.
Closest agreement is obtained for $\tau^V / \tau^S \approx 18.5$, meaning slower relaxation for volumetric versus shear viscoelasticity. Axial stress $P(t)$ in the steady wave at $X$ follows identical trends because its change is $\rho_0 {\mathcal U} \times$ the particle velocity change.  Ratio $C_L/ {\mathcal U} > 1$ increases monotonically from its initial value at the head to a constant at tail of the wave---attained when the Rayleigh line for each imposed $\mathcal U$ intersects the relaxed $P$ versus $\rho$ Hugoniot---for all three cases in Fig.~\ref{fig5}(f), a necessary condition for steady wave propagation in
 nonlinear viscoelastic materials with memory \cite{schuler1973}. The critical strain gradient $\Lambda = \Lambda_c$ for steady wave propagation predicted by the second of \eqref{eq:deltaJ} is compared with experimental data \cite{nunziato1973am} in Table~\ref{table2}. Although experimental data demonstrate wide bounds on precision, close agreement of the model with median experimental values is encouraging.

\subsection{Spall fracture}
\label{sec4.3}

The spall process in PMMA is modeled by uniaxial tension calculations.
The objective is determination of properties and parameters for crazing and dynamic tensile fracture
at very high rates with concurrent evaluation of model capabilities. This is facilitated by comparison
with experimental spall data across a range of strain rates and starting temperatures
\cite{bartowski2000,geraskin2009,zaretsky2019}.
Notably, Ref.~\cite{geraskin2009} reports an increase in spall strength with increasing rate,
and Ref.~\cite{zaretsky2019} reports a decrease in spall strength with increasing initial temperature
spanning the glass transition. When strain rates are not reported explicitly as in Refs.~\cite{bartowski2000,zaretsky2019}, free surface velocity profiles and sound speeds are used to calculate strain rates during the spall process
\cite{diamond2025} to be used as loading conditions for the model.

For initial temperatures $\theta_i = \theta_0$, viscoelastic relaxation is not incorporated explicitly,
but calculations using instantaneous (e.g., $G^{G,I}_0, B^{G,I}_0$) and relaxed Hugoniot
(e.g., $G^{G,E}_0, B^{G,E}_0$) properties bound the predicted response. Results later show
that differences among bounded spall strengths are negligible. For $\theta_i > \theta_0$, viscoelastic properties
are not well known: $\tau^V$ and $\tau^S$ should include temperature dependence. Such temperature
dependence has been safely omitted in Section~\ref{sec4.2} where $| \theta - \theta_0 | \lesssim 8.4$ K. In Section~\ref{sec4.3},
like Section~\ref{sec4.1}, viscoelastic response is implicitly included in the
Hugoniot properties of the glass and melt.  Bulk and shear moduli of the melt phase are calculated
using sound speeds and shock velocity reported in Ref.~\cite{zaretsky2019} for $\theta_i = 413$ K. 
Specific heats differ among glass and melt (Table~\ref{tableA1}); Gr\"uneisen parameters $\Gamma^{*,E}_0$ are
assumed the same.

The applied axial strain rate is $\dot{J}^F > 0$. Isochoric plastic strain from shear yielding $\gamma^P(t)$,
free volume change $\varphi(t)$, porosity from crazing or chain scission $\phi(t)$, and damage from
fracture $\xi(t)$ can all be nonzero. Impact stresses and temperatures are far below decomposition thresholds,
so $\Xi(t) = 0$.  Experimental data for impacts below and above the HEL show little or no difference
in spall strength \cite{bartowski2000}, so nonzero initial plastic deformation is excluded for simplicity.
Subsequent calculations demonstrate isochoric plastic deformation during tensile extension is small. Adiabatic conditions are assumed, and bulk shock viscosity of Section~\ref{sec3.7} is inactive for tensile loading. Shear viscosity is not impossible
for the melt phase at high temperature. However, data in Ref.~\cite{martinez1980} show a logarithmic
decrease in $\mu^V$ with increasing strain rate. Extrapolating such data to the present regime (i.e., $\dot{J}^F$ on order of $10^4$ to $10^5$/s) gives viscous shear stresses of order 10 kPa, negligible relative
to spall strengths of order 100 MPa and therefore omitted without important consequence.
Current 1-D calculations are restricted to a single material point at $X$ undergoing
axial strain rate $\dot{J}^F = J_i^{-1/3} \partial \upsilon / \partial X$. Stress wave interactions and the width
of the spall-damaged zone are not captured explicitly; these require advanced numerical methods
with space-time discretization beyond the current scope.
Initial conditions are $J^F(0) = 1$, $\gamma^P(0) = \varphi(0) = \phi(0) = \xi(0) = 0$. Initial temperatures
vary, where $\theta_i \geq \theta_0$. If $\theta_i > \theta_0$, then an initial spherical thermal expansion
$J_i \approx \exp ( \Gamma_0^{*,E} c^*_{V0} (\theta_i - \theta_0) / B^{*,E}_0)$ alleviates initial thermal pressure. 

Consistent with \eqref{eq:Dpexp}, material symmetry, and above assumptions, the deformation gradient is
 \begin{align}
 \label{eq:Fspall}
 {\bm F} = 
  \begin{bmatrix}
J^F & 0 & 0 \\
 0 & 1 & 0 \\
 0 & 0 & 1
 \end{bmatrix} (J_i)^{1/3}= 
 {\bm F}^E \bar{\bm F}^P {\bm F}^\phi = 
 \begin{bmatrix}
 F^E_{11} & 0 & 0 \\
 0 & F^E_{22} & 0 \\
 0 & 0 & F^E_{22} 
 \end{bmatrix}
 \begin{bmatrix}
 \bar{F}^P & 0 & 0 \\
 0 & \bar{F}^{P-1/2} & 0 \\
 0 & 0 & \bar{F}^{P-1/2}
 \end{bmatrix}
 \begin{bmatrix}
F^\phi & 0 & 0 \\
 0 & 1 & 0 \\
 0 & 0 & 1
 \end{bmatrix}.
 \end{align}
 The following kinematic and stress relations hold, where $J^\phi = F^\phi = 1/(1 -\phi)$ and ${\bm F}^E$
 contains $J_i$:
 \begin{align}
 \label{eq:kin1Dspall}
& J = J_i J^F = J^E J^\phi = \frac{F^E_{11} (F^E_{22})^2}{1 - \phi}, \quad
F^E_{11} = \frac{J_i^{1/3} J^F }{\bar{F}^P F^\phi}, \quad F^E_{22} = J_i^{1/3} (\bar{F}^P)^{1/2}, \quad e^E_V = \ln J^E = \ln J - \ln J^\phi;
\\
& \bar{e}^E_{11} =  -2 \bar{e}^E_{22} = \ln F^E_{11} - {\textstyle{\frac{1}{3}}} e^E_V,
\quad \bar{\bm e}^E = {\rm diag}(  \bar{e}^E_{11},  \bar{e}^E_{22},  \bar{e}^E_{22}),
\quad \gamma^P = \sqrt{3} \ln {\bar F}^P, \quad {\bm{A}} = \bar{\bm F}^P (\bar{\bm F}^P)^{\mathsf T};
\\
& \lambda^P = \{ {\textstyle{\frac{1}{3}}} [ (\bar{F}^P)^2 + 2 / \bar{F}^P] \}^{1/2}, \quad \bar{\bm N}^P = \sqrt{{\textstyle{\frac{2}{3}}}}  {\rm diag}(1,-{\textstyle{\frac{1}{2}}},-{\textstyle{\frac{1}{2}}}), \quad
\tau^A = \bar{\bm M}^A: \bar{\bm N}^P =  {\textstyle{\frac{\sqrt{3}}{2}}} \bar{M}^A_{11};
\\
& \bar{\bm M}^E = J \bar{\bm \sigma} = J {\rm diag} ( \bar{\sigma}_{11}, - {\textstyle{\frac{1}{2}}} \bar{\sigma}_{11},  - {\textstyle{\frac{1}{2}}} \bar{\sigma}_{11}), \quad \tau^E = \bar{\bm M}^E: \bar{\bm N}^P =
{\textstyle{\frac{\sqrt{3}}{2}}} J \bar{\sigma}_{11} = {\textstyle{\frac{\sqrt{3}}{3}}} J \bar{\sigma};
\\ 
& {\bm N}_\phi = {\rm diag}(1,0,0), \qquad M^E_\phi = M^E_{11} = J \sigma_{11} = J(\bar{\sigma}_{11} - p) = J P, \qquad P = \sigma_{11} \geq 0.
 \end{align}
 Axial stress, positive in tension, is $P$. Spall strength is identified as $P_c = \max_{t \in [0,\infty)} P(t)$.
 
 Unlike static equilibrium conditions for Hugoniot response in Section~\ref{sec4.1}, $\dot{\gamma}^P(t) \geq 0$.
 Equation \eqref{eq:richeton} is used in dynamic form, with literature parameters in Table~\ref{tableA2}.
 In \eqref{eq:tauF}, $f^P_R = 1$ in tensile loading in accordance with \eqref{eq:fRfit}. 
 Equations \eqref{eq:varphisub} and \eqref{eq:tausub} are used for isotropic hardening/softening and free volume, consistent with parameters for the shock-loading regime in Table~\ref{tableA2}.
 Porosity rate $\dot {\phi}$ from crazing in the glass or from scission in the melt is found from \eqref{eq:epsphidot}.
 Fracture rate $\dot {\xi}$ is given by \eqref{eq:ratedep}, with $\Phi$ in \eqref{eq:Phifunc}. 
 Surface energies $\Gamma^G_\xi$ and $\Gamma^M_\xi$ are constants in Table~\ref{tableA2}.
 A constant resistance $R_\xi^G$ is sufficient for the glass phase, while linear temperature dependence of $R_\xi^M$ is
 needed to capture experimental data:
 \begin{align}
 \label{eq:Rxispall}
 R^G_\xi = R^G_{\xi 0}, \qquad R^M_\xi = R^M_{\xi 0} [ 1 + \alpha^M_{\xi \theta} (\theta / \theta_G - 1) ].
 \end{align}
 A recent study \cite{williams2025b} similarly modeled a change in spall properties across a solid-solid (phase) transition, there for metals.
  Dissipation from ${\mathfrak D}^P = \tau^P \dot{\gamma}^P$, ${\mathfrak D}^\phi = M^E_\phi \dot{\phi}/(1-\phi)$, and ${\mathfrak D}^\xi = \zeta \dot{\xi}$ contributes to temperature rate calculated via \eqref{eq:tempratefin},
  as do thermoelastic coupling and quantities \eqref{eq:DAtheta} and \eqref{eq:Dxitheta}.
  For comparison with elevated temperature data, $\omega(t) = 0$ for $\theta_i \leq 391$ K and $\omega(t) = 1$ for
  $\theta_i \geq 403$ K, as suggested by observations on the PMMA in Ref.~\cite{zaretsky2019}
  that appears to have a slightly higher glass transition temperature $\theta_G$.
  Neither melting nor freezing are assumed to occur during dynamic deformation (i.e., $\dot {\omega} = 0$), as assumed likewise in Section~\ref{sec4.1}. Pressures and time scales are assumed to inhibit such transitions. 
  Degradation of $\Psi^{M,E}_\beta$ by $f^E_V$ is omitted for the melt phase to
  avoid net negative dissipation from release of this (negative) energy.  Magnitude of the latter is large due
  to initial thermal expansion and high specific heat of the melt phase.
  
  Spanning rates in Refs.~\cite{bartowski2000,geraskin2009}, for room-temperature calculations (i.e., $\theta_i = \theta_0$), imposed values of $\dot{J}^F$ range from $10^4$ to $10^7$/s.
 For elevated temperature calculations, from velocity profiles in Ref.~\cite{zaretsky2019}, the following
  loading conditions ($\theta_i,\dot{J}^F$) are modeled in units of (K, $10^4$/s): (298, 5.48), (331, 4.84), (361, 3.78), (380, 3.19), (391, 2.95), (403, 1.33), (413, 1.31). Strain rate during spall tends to decrease with increasing temperature
  as the material becomes more compliant.
     
\begin{figure}
\centering
\subfigure[spall strength]{\includegraphics[width = 0.32\textwidth]{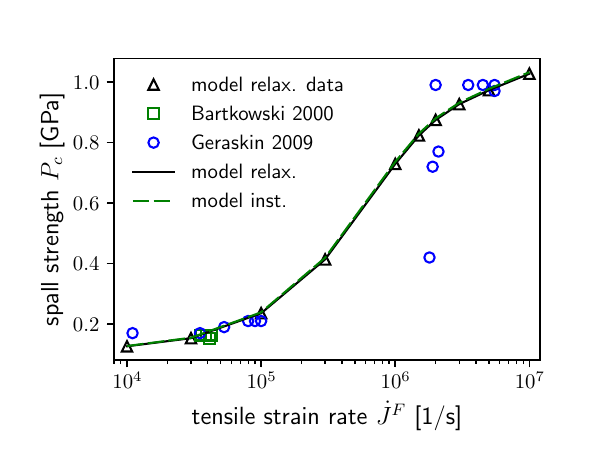} \label{fig6a}}
\subfigure[axial stress]{\includegraphics[width = 0.32\textwidth]{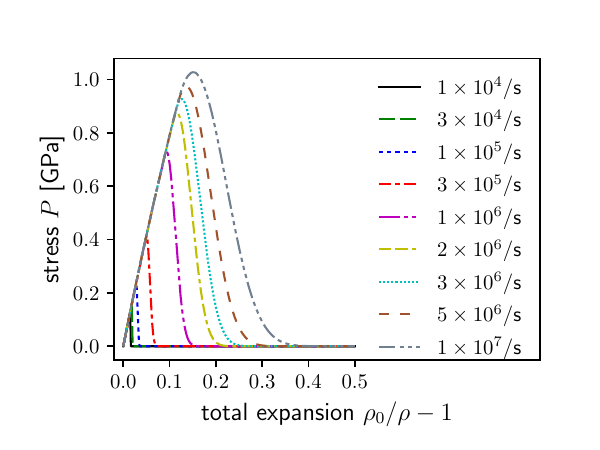} \label{fig6b}} 
\subfigure[fracture variable]{\includegraphics[width = 0.32\textwidth]{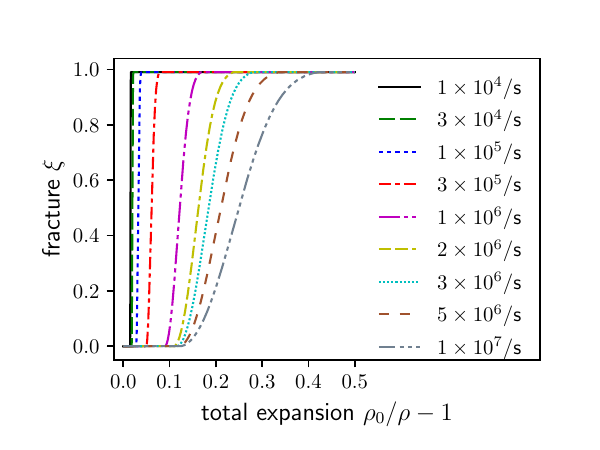} \label{fig6c}} \\
\subfigure[porosity]{\includegraphics[width = 0.32\textwidth]{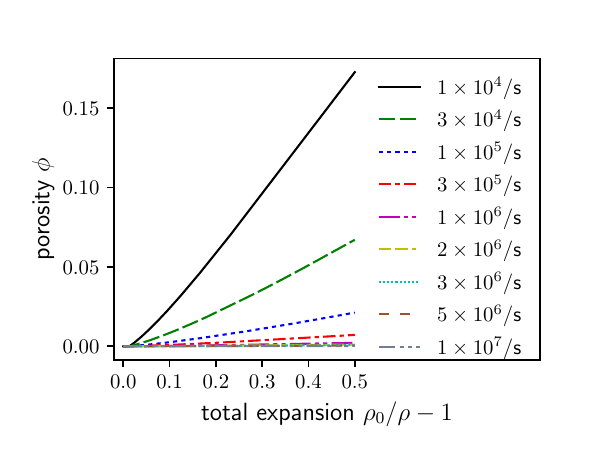} \label{fig6d}}
\subfigure[plastic strain]{\includegraphics[width = 0.32\textwidth]{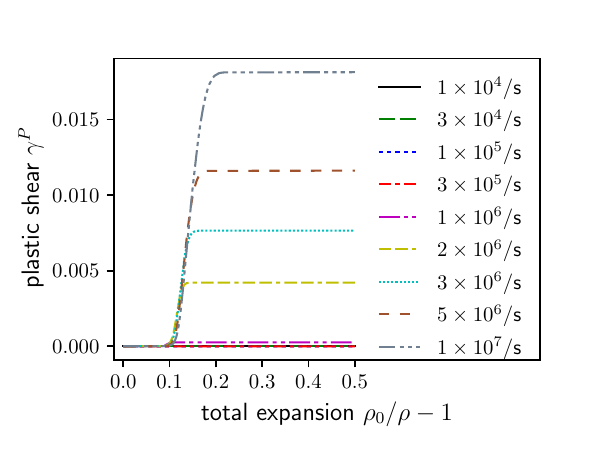} \label{fig6e}} 
\subfigure[temperature]{\includegraphics[width = 0.32\textwidth]{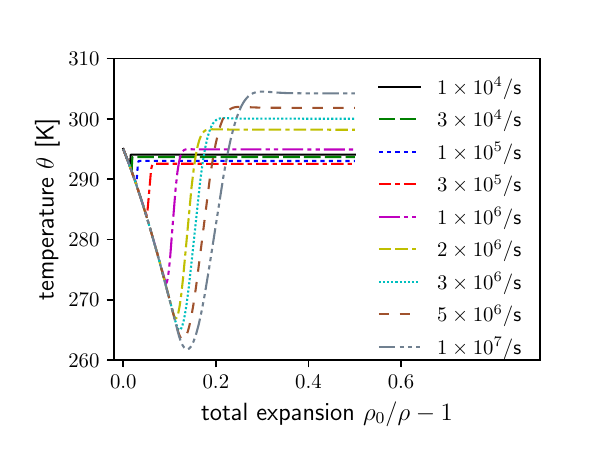} \label{fig6f}} 
\vspace{-0.2cm}
\caption{\label{fig6} Room-temperature spall response of PMMA: (a) spall stress $P_c$
vs.~strain rate $\dot{J}^F$ from model (relaxed Hugoniot and instantaneous elastic properties) and experiments \cite{bartowski2000,geraskin2009} 
(b) axial stress $P$ vs.~expansive strain $J-1$
(c) order parameter $\xi$ 
(d) void fraction $\phi$ 
(e) isochoric plastic strain $\gamma^P$ 
(f) temperature $\theta$}
\end{figure}
  
  Results for $\theta_i = \theta_0 = 295$ K are reported in Fig.~\ref{fig6}.
  Spall strength $P_c$ is compared with experimental data \cite{bartowski2000,geraskin2009} in Fig.~\ref{fig6}(a).
  Many trends are similar; agreement is modest. An outlying experimental data point corresponding to $P_c = 0.42$ GPa at $\dot{J}^F = 1.8 \times 10^6$/s is noted, and spall strength tends to attain a flatter plateau at the highest rates in experiments.  According to the model, spall strength increases relatively slowly with strain rate up to $\approx 10^5$/s, increases rapidly between $10^5$ and $10^6$/s, then increases slowly again at even higher rates.
Results have been achieved by adjusting five parameters in Table~\ref{tableA2}: $\dot{\epsilon}_0$, $n$, and $c^\phi_\xi$ for crazing along with $R^G_{\xi 0}$ and $\beta_\xi$ for fracture. Spall strength $P_c$ is highly sensitive to craze-induced porosity that initiates brittle fracture in the glass. Outcomes in Fig.~\ref{fig6}(a) are nearly indistinguishable whether instantaneous or relaxed Hugoniot elastic constants are used, suggesting transient effects of viscoelasticity should be small. 

Predictions for axial stress $P$, fracture order parameter $\xi$, void volume fraction $\phi$, plastic shear strain $\gamma^P$, and temperature $\theta$ are shown versus total  (including thermal) expansion $J - 1$ in the remaining five parts of Fig.~\ref{fig6}, where each curve corresponds to a strain rate $\dot{J}^F$ listed in the legend. 
The slower the rate of increase of $\xi$ with $J$ in Fig.~\ref{fig6}(c), the larger the peak stress attained in Fig.~\ref{fig6}(b).
 Greater terminal porosities are achieved in Fig.~\ref{fig6}(d) for lower expansion rates because more time is available for
 rate-dependent void growth; time needed for spall decreases with increasing $\dot{J}^F$.
 The highest plastic shear strains are attained for highest strain rates in Fig.~\ref{fig6}(e) since a larger driving shear stress is achieved prior to load drop, and plastic strain-rate sensitivity is small compared to that for craze strain: $m/n \approx 0.05$. Temperature in Fig.~\ref{fig6}(f) initially decreases with increasing strain and tensile pressure due to thermoelastic coupling. After the onset of fracture, temperature increases due to dissipation and reduction of the tangent bulk modulus from damage.
 
 \begin{figure}
\centering
\subfigure[spall strength]{\includegraphics[width = 0.32\textwidth]{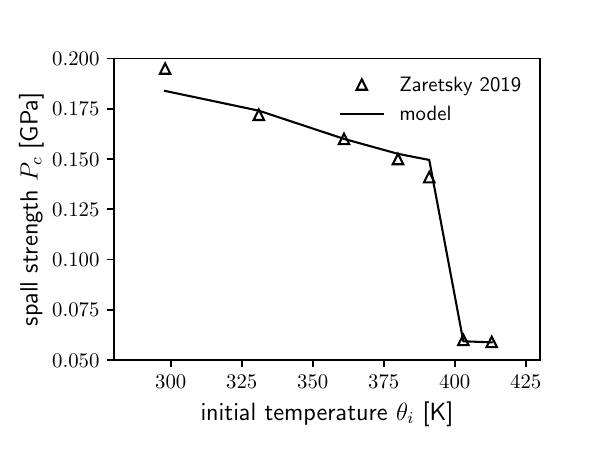} \label{fig7a}}
\subfigure[fracture variable]{\includegraphics[width = 0.32\textwidth]{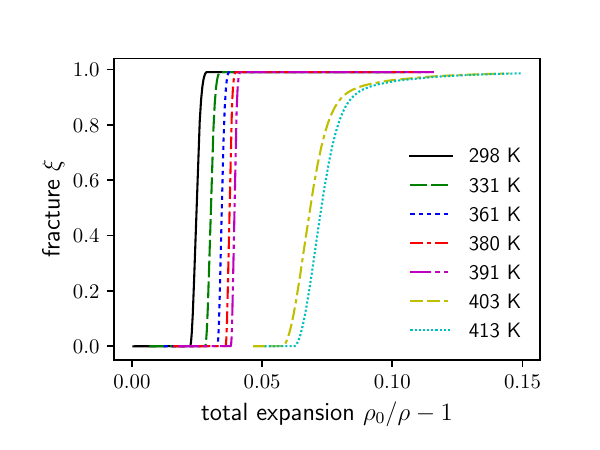} \label{fig7b}}\
\subfigure[porosity]{\includegraphics[width = 0.32\textwidth]{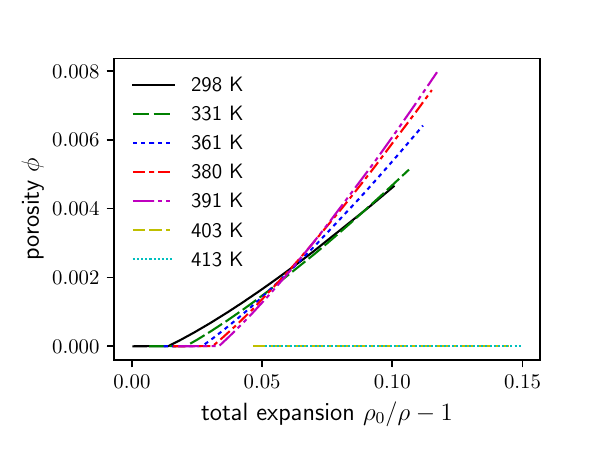} \label{fig7c}}
\vspace{-0.2cm}
\caption{\label{fig7} Elevated-temperature spall response of PMMA: (a) spall stress $P_c$
vs.~strain rate $\dot{J}^F$ from model and experiments \cite{zaretsky2019} 
(b) order parameter $\xi$ vs.~expansive strain $J-1$
(c) void fraction $\phi$ }
\end{figure}
 
 Outcomes for higher $\theta_i \in [298,413]$ K are shown in Fig.~\ref{fig7}. Close agreement with
 experimental values obtained from pull-back signals \cite{zaretsky2019} is witnessed
 in Fig.~\ref{fig7}(a).  The severe drop in $P_c$ above the glass transition is evident for $\theta_i > 391$ K.
 Results are obtained by calibration of $W^M_{\xi 0}$ and $\alpha^M_{\xi \theta}$ in Table~\ref{tableA2} without adjusting
  any properties assigned previously to achieve Fig.~\ref{fig6}. 
 Evolution of $\xi$ and $\phi$ are shown in respective FIg.~\ref{fig6}(b) and Fig.~\ref{fig6}(c), where initial temperatures
 for each curve are given in the legend. Total expansion $J$ contains thermal expansion $J_i$. Thus, $J(0) > 1$ and $J_i$ increases with $\theta_i$. Initial expansion is higher in the melt due to higher coefficient of thermal expansion, manifesting from lower bulk modulus and higher specific heat \cite{richeton2007,clements2012a}.
 Below the glass transition, spall strength decreases with
 increasing temperature because strain rate decreases concurrently. 
 Similar to the case for room-temperature results in Fig.~\ref{fig6}, lower strain rates here at higher starting temperatures correlate with higher porosity from crazing below the glass transition in Fig.~\ref{fig7}(c), leading to smaller peak stresses and lower spall strength. Plastic strain $\lambda^P$ is too small (i.e., $\lambda^P(t) \ll \lambda^P_\phi = 1.39$) for cavitation and pore growth by chain scission in the melt. Thus $\phi = 0$ for $\theta_i \geq  403$ K in Fig.~\ref{fig7}(c),
 and spall fracture occurs in the melt by strain-energy driven damage to rupture in the absence of crazing or void growth.
 Although the rate of increase in $\xi$ with $J$ is more gradual in melt than the glass in Fig.~\ref{fig7}(b),
 strength $P_c$ is lower in the former. This is a net product of low melt-to-glass ratios of threshold fracture energy,
 Hugoniot bulk modulus, and Hugoniot shear modulus in Tables~\ref{tableA2} and \ref{tableA3}:  $W^M_{\xi 0}/W^G_{\xi 0} \approx 0.06$, $B^{M,E}_0/ B^{G,E}_0 \approx 0.8 $, and $G^{M,E}_0/G^{G,E}_0 \approx 0.65 $.

\section{Conclusions}
\label{sec5}
A constitutive framework has been established for the dynamic thermomechanical response
of amorphous polymers applicable to extreme pressures and loading rates. The framework accounts for nonlinear thermoelasticity, isochoric plasticity, crazing, and rupture.
Fracture, melting across the glass transition, and shock decomposition across the high-pressure transition
are captured by distinct order parameters enabling a smooth, thermodynamically consistent material response.

The model has been applied toward PMMA. Experimental Hugoniot data are well-represented to shock pressures up to 120 GPa where temperatures exceed 4200 K.  On the principal Hugoniot, decomposition begins at respective shock pressure and
temperature of 25.7 GPa and 1289 K and is 90\% complete at 42 GPa and 2124 K.
A continuous decomposition process enables shock and release velocities to increase continuously with particle velocity.
Kinetic barriers preclude fracture and melting at high pressure; the latter exceed the former in magnitude.
Low-pressure steady waves have been adequately represented using a single relaxation time for each of bulk and shear viscoelasticity. Shear relaxation time is smaller than bulk relaxation time.
Experimental spall strengths are reasonably captured (with a few exceptions): spall resistance increases with increasing expansion rate and decreasing temperature. Spall fracture is dominated by crazing at temperatures below the glass transition; rupture occurs without crazing above the glass transition.

Calculations in the present study have been limited to one spatial dimension. Implementation in a 3-D numerical scheme (e.g., finite element or hydrocode) is foreseen as a necessary future step to address more complicated loading conditions.

\clearpage
\setcounter{section}{0} \renewcommand{\thesection}{A.\arabic{section}}

\appendix

\section{Appendix: PMMA properties and parameters}
\setcounter{figure}{0} \renewcommand{\thefigure}{A.\arabic{figure}}
\setcounter{table}{0} \renewcommand{\thetable}{A.\arabic{table}}
\setcounter{section}{0} \renewcommand{\thesection}{A.\arabic{section}}
\label{secA}

\begin{figure}[hbp]
\centering
\subfigure[melt fraction]{\includegraphics[width = 0.32\textwidth]{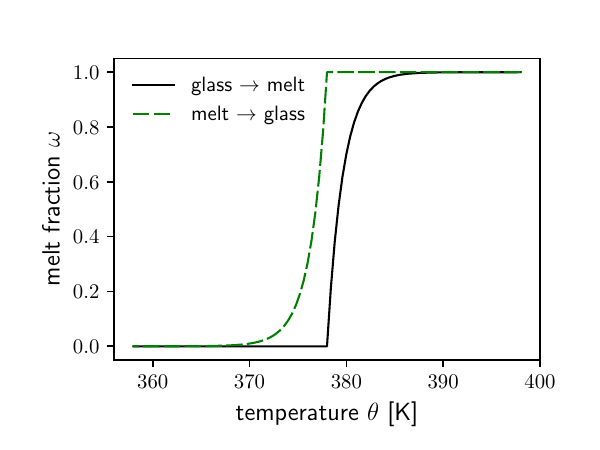} \label{figA1a}}
\subfigure[free energy]{\includegraphics[width = 0.32\textwidth]{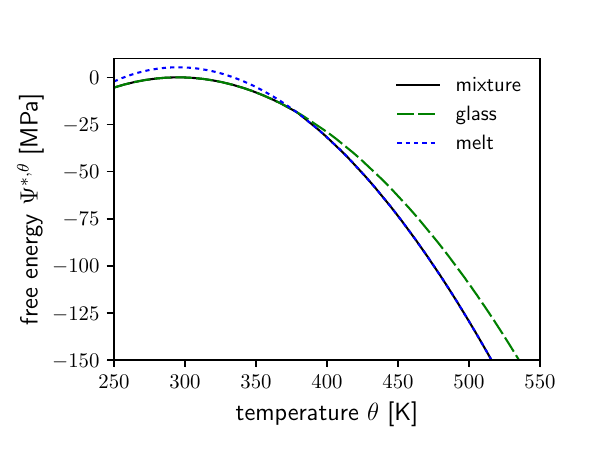} \label{figA1b}} 
\subfigure[specific heat]{\includegraphics[width = 0.32\textwidth]{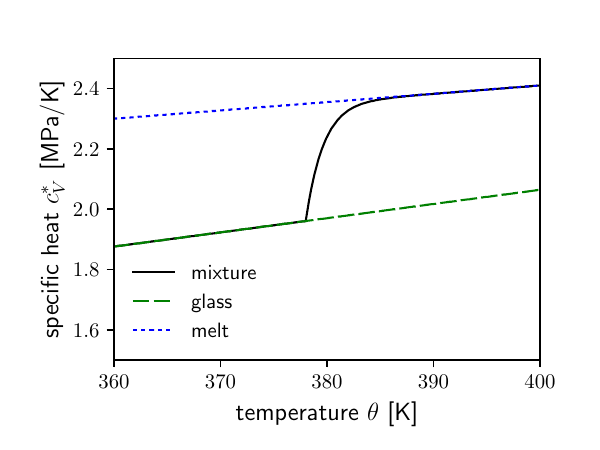} \label{figA1c}} 
\vspace{-0.2cm}
\caption{\label{figA1} Metastable melt and freezing processes at null pressure or strain: (a) melt fraction
for melting and freezing (b) free energies and (c) specific heats of phases and mixture undergoing melting. Glass transition is $\theta_G = 378$ K.}
\end{figure}

\begin{figure}[hbp]
\centering
\subfigure[isotropic stress and free volume]{\includegraphics[width = 0.32\textwidth]{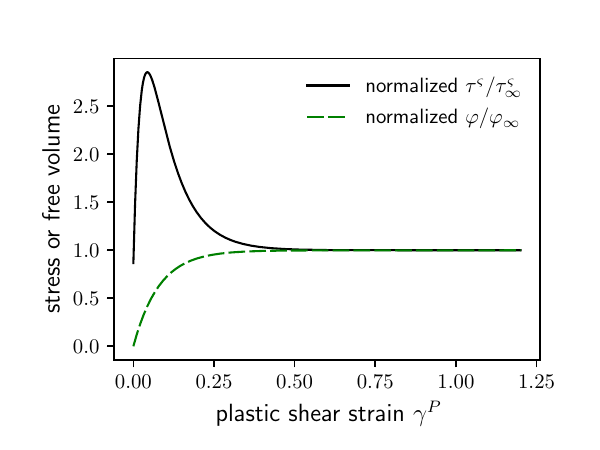} \label{figA2a}}
\subfigure[yield stress components]{\includegraphics[width = 0.32\textwidth]{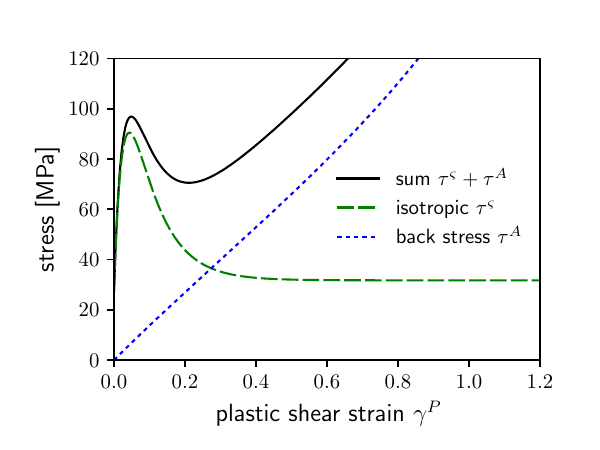} \label{figA2b}} 
\vspace{-0.2cm}
\caption{\label{figA2} Evolution of plastic free volume and shear strength versus monotonic plastic shear strain $\gamma^P$: (a) isotropic stress $\tau^\varsigma$ and free volume $\varphi$ normalized by $\tau^\varsigma_\infty = 31.7$ MPa and $\varphi_\infty = 3 \times 10^{-4}$ (b) isotropic stress $\tau^\varsigma$, anisotropic hardening stress $\tau^A$, and their sum (no pressure hardening or thermal softening)}
\end{figure}

 \begin{table}[hbp]
\caption{Physical properties and model parameters for Hugoniot analysis of PMMA, $\theta_0 = 295$ K}
\centering
\footnotesize
\label{tableA1}
\begin{tabular}{llrl}
\hline
\bf{Property [units]} & \bf{Definition} & \bf{Value} & \bf{Reference or source} \\
\hline
$\rho_0$ [g/cm$^3$]  & mass density at $\theta = \theta_0$ & 1.185 & \cite{barker1970,schuler1970}   \\
$\theta_G$ [K] & static, atmospheric glass transition temperature & 378 & \cite{menikoff2004} \\
$c_{V0}^G$ [MPa/K] & glass specific heat per unit volume at $\theta = \theta_0$ & 1.570 & \cite{richeton2007} \\
$c_{V0}^M$ [MPa/K] & melt specific heat per unit volume at $\theta = \theta_0$ & 2.121 & \cite{richeton2007} \\
$c_{V0}^D$ [MPa/K] & product specific heat per unit volume at $\theta = \theta_0$ & 1.570 & assumed \\
$c_{\theta}^G/c_{V0}^G$ [1/K] & temperature derivative of glass specific heat & $3.0 \times 10^{-3}$ & \cite{richeton2007} \\
$c_{\theta}^M/c_{V0}^M$ [1/K] & temperature derivative of melt specific heat & $1.3 \times 10^{-3}$ & \cite{richeton2007} \\
$c_{\theta}^D/c_{V0}^D$ [1/K] & temperature derivative of product specific heat & $1.3 \times 10^{-3}$ & assumed \\
$\psi^{M,\theta}_0$ [MPa] & datum free energy of melt phase & 5.35 & calculated \\
$\beta_\omega$ [MPa] & kinetic factor for melt transition width & 0.195 & calculated \\
$P^H_D$ [GPa] & Shock pressure threshold for decomposition onset & 25.7 & \cite{hauver1964,carter1995} \\
$J^H_D$ [-] & Hugoniot volume ratio at decomposition onset & 0.58 & calculated \\
$ \theta_D$ [K] & Hugoniot temperature at onset of decomposition & 1289 & calculated \\
$ \delta_\Xi$ [-] & relative volume change under shock decomposition & -0.034 & \cite{carter1995,dattelbaum2019} \\
$\psi^{D,\theta}_0$ [GPa] & datum free energy of decomposed product phase & -0.103 & calculated \\
$ \lambda^\Xi_\theta$ [GPa] & kinetic latent heat of decomposed products & 6.69 & calculated \\
$\beta_\Xi$ [GPa] & kinetic factor for decomposition transition width & 1.5 & calculated \\
$B^{G,E}_0$ [GPa] & isothermal bulk modulus of glass & 5.71 & from $C_L = 2.75$ km/s \cite{menikoff2004} \\
$B^{G,E}_1$ [-] & linear volume stiffening of bulk modulus of glass & 1.3 & calibrated \\
$B^{G,E}_2$ [-] & quadratic volume stiffening of bulk modulus of glass  & 15.0 & calibrated \\
$B^{G,E}_3$ [-] & cubic volume stiffening of bulk modulus of glass & 200 & calibrated \\
$B^{D,E}_0$ [GPa] & isothermal bulk modulus of products & 8.0 & calibrated \\
$B^{D,E}_1$ [-] & linear volume stiffening of bulk modulus of products & 1.5 & calibrated \\
$B^{D,E}_2$ [-] & quadratic volume stiffening of bulk modulus of products & 15.0 & calibrated \\
$B^{D,E}_3$ [-] & cubic volume stiffening of bulk modulus of products & 120 & calibrated \\
$G^{G,E}_0$ [GPa] & elastic shear modulus of glass & 2.29 & \cite{christman1972} \\
$G^{G,E}_\theta$ [MPa/K] & linear thermal softening of shear modulus & -5.2 & \cite{christman1972} \\
$G^{G,E}_p$ [-] & linear pressure stiffening of shear modulus  & 2.0 & calibrated \\
$G^{G,E}_2$ [-] & quadratic pressure stiffening of shear modulus & 0 & not used \\
$\Gamma^{G,E}_0$ [-] & ambient pressure Gr\"uneisen parameter of glass & 0.657 & \cite{nunziato1973g} \\
$\Gamma^{G,E}_1$ [-] & linear volume dependence of Gr\"uneisen parameter & 0.657 & \cite{menikoff2004} \\
$\Gamma^{G,E}_2$ [-] & quadratic volume dependence of Gr\"uneisen parameter & 16.0 & \cite{menikoff2004} \\
$\Gamma^{D,E}_0$ [-] & ambient pressure Gr\"uneisen parameter of products & 0.667 & ideal gas limit \cite{coe2022} \\
 $\Gamma^{D,E}_1 = \Gamma^{D,E}_2$ [-] & volume dependence of Gr\"uneisen parameter of products & 0 & not used \\
 \hline
 \end{tabular}
\end{table}

\begin{table}
\caption{Plasticity, fracture, and melt parameters for shock and spall analysis of PMMA}
\centering
\footnotesize
\label{tableA2}
\begin{tabular}{llrl}
\hline
\bf{Property [units]} & \bf{Definition} & \bf{Value} & \bf{Reference or source} \\
\hline
$P^H_P$ [GPa] & Hugoniot elastic limit (HEL) &  0.740 & \cite{barker1970,schuler1970,schuler1974,rosenberg1994} \\
$J^H_P$ [-] & volume ratio at HEL & 0.937 & calculated \\
$\theta_P$ [-] & temperature at HEL & 306 & calculated \\
$N^P_0$ [-] & number of links per polymer chain & 2.2525 & \cite{richeton2007} \\
$N^P_\theta $ [1/K] & thermal scaling of number of links & $2.5 \times 10^{-3}$ &  \cite{richeton2007} \\
$\mu_{A0} $ [MPa] & modulus for anisotropic hardening & 10.2 & \cite{dal2022} \\
$\mu_{A \theta} $  [MPa/K] & thermal scaling of plastic modulus & 0.0942 & \cite{richeton2007} \\ 
$\tau^{\varsigma}_0 $ [MPa] & initial isotropic plastic strength & 27.5 & \cite{dal2022} \\
$\tau^{\varsigma}_\infty $ [MPa] & terminal isotropic plastic strength & 31.7 & \cite{dal2022} \\
$\varphi_\infty $ [-] & terminal plastic free volume change & $3 \times 10^{-4}$ & \cite{dal2022} \\
$a_\varphi $ [-] & kinetic parameter for free volume change & 0.08 & calibrated \\
$b_\varphi $ [-] & kinetic parameter for isotropic strength &  0.025 & calibrated \\
$\mu^G_\varsigma$ [MPa] & modulus for isotropic hardening & 151.2 & calibrated \\
$\alpha^F_\theta$ [-] & thermal softening of plastic flow & 0.331 & calculated \\
$\alpha^F_p$ [-] & pressure hardening of plastic flow & 0.538 & from HEL strength \\
$ \dot{\gamma}_0$ [1/s] & reference plastic shearing rate & $1.16 \times 10^{16}$ & \cite{richeton2007} \\
$H_\beta / k_B $ [K] & activation energy of secondary relaxation & $ 1.082 \times 10^4 $ & \cite{richeton2007} \\
$c^G_1$ [-] & shear yielding kinetic parameter & 32.58 &  \cite{richeton2007} \\
$c^G_2$ [K] & shear yielding kinetic parameter & 83.5 &  \cite{richeton2007} \\
$m$ [-] & rate sensitivity exponent for shear flow & 0.157 &  \cite{richeton2007} \\
$V_F / k_B $ [K/MPa] & activation energy of secondary relaxation &  7.062  & \cite{richeton2007} \\
$f_1,f_2,f_3$ [-] & fitting parameters for Hugoniot strength & -14.5, 262.3, -1979 & calibrated \\
$f_4,f_5,f_6$ [-] & fitting parameters for Hugoniot strength & 7369, -13732, 10119 & calibrated \\
$e^c_V$ [-] & inelastic strain above which $f^P_R$ vanishes & 0.41 & calibrated \\
$c^\phi_1$ [MPa] & craze initiation parameter & 45.6 & \cite{dal2022} \\
$c^\phi_2 $ [MPa$^2$] & craze initiation parameter & 1130 & \cite{dal2022} \\
$ \dot{\epsilon}_0$ [1/s] & reference craze strain rate & $2 \times 10^3$ & calibrated for spall \\
$n$ [-] & rate sensitivity exponent for craze strain & 3.33  &  calibrated for spall \\
$\lambda^P_\phi$ [-] & chain scission initiation stretch & 1.39 &  \cite{gearing2004} \\
$\Gamma_\xi$ [J/m$^2$]  & low temperature dynamic fracture energy & 150 & \cite{atkins1975} \\
$l_\xi$ [$\mu$m] & phase-field length constant & 300 & \cite{narayan2021} \\
$c^P_\xi l_\xi / \Gamma_\xi$ [-] & ductile fracture driving force & 25 & \cite{dal2022} \\
$c^\phi_\xi l_\xi / \Gamma_\xi $  [-] &  craze fracture driving force & $10^9$ & calibrated for spall \\ 
$R_{\xi 0}^G  l_\xi / \Gamma_\xi $ [-] & glass fracture resistance & 115 & calibrated for spall \\
$R_{\xi 0}^M  l_\xi / \Gamma_\xi $ [-] & melt fracture resistance & 6.96 & calibrated for spall \\
$\alpha^M_{\xi \theta}$  [-] & temperature dependence of melt fracture & 9 & calibrated for spall \\
$\beta_\xi$ [$\mu$s] & phase-field fracture kinetic time & 5 & calibrated for spall \\
$B^{M,E}_0$ [GPa] & isothermal bulk modulus of melt & 3.31 & from shock speed \cite{zaretsky2019} \\
$G^{M,E}_0$ [GPa] & dynamic shear modulus of melt & 1.50 & from shear wave speed \cite{zaretsky2019} \\
$G^{M,E}_\theta$ [MPa/K] & linear thermal softening of shear modulus & -$3 \times 10^{-3}$ & \cite{srivastava2010} \\
\hline
 \end{tabular}
\end{table}

\clearpage

\begin{table}
\caption{Refined elastic and viscoelastic parameters for steady wave analysis of PMMA, $\theta_0 = 295$ K}
\centering
\footnotesize
\label{tableA3}
\begin{tabular}{llrl}
\hline
\bf{Property [units]} & \bf{Definition} & \bf{Value} & \bf{Reference or source} \\
\hline
$B^{G,E}_0$ [GPa] & relaxed isothermal bulk modulus & 5.622 & from $C_L = 2.73$ km/s \cite{nunziato1973ac,nunziato1973p,schuler1974} \\
$B^{G,I}_0$ [GPa] & instantaneous isothermal bulk modulus & 5.817 & from $C_L = 2.76$ km/s \cite{asay1969,nunziato1973p,schuler1974} \\
$B^{G,E}_1 = B^{G,I}_1$ [-] & linear volume stiffening of  bulk modulus  & 15.0 & calibrated \\
$B^{G,E}_2 = B^{G,I}_2 $ [-] & quadratic volume stiffening of  bulk modulus  & -400 & calibrated \\
$B^{G,E}_3 = B^{G,I}_3 $ [-] & cubic volume stiffening of  bulk modulus& 0 & not used \\
$G^{G,E}_0$ [GPa] & relaxed elastic shear modulus  & 2.257 & calibrated \\
$G^{G,I}_0$ [GPa] & instantaneous elastic shear modulus  & 2.302 & \cite{christman1972,stephens1972} \\
$G^{G,E}_p$ [-] & linear pressure stiffening of relaxed modulus  & 1.50 & calibrated \\
$G^{G,I}_p$ [-] & linear pressure stiffening of instant modulus  & 1.65 & calibrated \\
$\Gamma^{G,I}_0$ [-] & ambient pressure instant Gr\"uneisen parameter& 0.55 & \cite{nunziato1973g} \\
$\Gamma^{G,I}_1$ [-] & linear volume dependence of $\Gamma^{G,I}$ & 0.55 & \cite{menikoff2004} \\
$\Gamma^{G,I}_2$ [-] & quadratic volume dependence of $\Gamma^{G,I}$ & 16.0 & \cite{menikoff2004} \\
$\tau^V = \tau^V_1$ [$\mu$s] & bulk viscoelastic relaxation time  & 1.200 & calibrated \\
$\tau^S = \tau^S_1$ [$\mu$s] & shear viscoelastic relaxation time & 0.065 & calibrated \\
\hline
 \end{tabular}
\end{table}

\bibliography{refs}

\begin{thebibliography}{180}
\providecommand{\natexlab}[1]{#1}
\providecommand{\url}[1]{\texttt{#1}}
\expandafter\ifx\csname urlstyle\endcsname\relax
  \providecommand{\doi}[1]{doi: #1}\else
  \providecommand{\doi}{doi: \begingroup \urlstyle{rm}\Url}\fi

\bibitem[Saeki et~al.(1992)Saeki, Tsubokawa, Yamanaka, and
  Yamaguchi]{saeki1992}
S.~Saeki, M.~Tsubokawa, J.~Yamanaka, and T.~Yamaguchi.
\newblock Correlation between the equation of state and the pressure dependence
  of glass transition and melting temperatures in polymers and rare-gas solids.
\newblock \emph{Polymer}, 33:\penalty0 577--584, 1992.

\bibitem[Richeton et~al.(2007)Richeton, Ahzi, Vecchio, Jiang, and
  Makradi]{richeton2007}
J.~Richeton, S.~Ahzi, K.S. Vecchio, F.C. Jiang, and A.~Makradi.
\newblock Modeling and validation of the large deformation inelastic response
  of amorphous polymers over a wide range of temperatures and strain rates.
\newblock \emph{International Journal of Solids and Structures}, 44:\penalty0
  7938--7954, 2007.

\bibitem[Srivastava et~al.(2010)Srivastava, Chester, Ames, and
  Anand]{srivastava2010}
V.~Srivastava, S.A. Chester, N.M. Ames, and L.~Anand.
\newblock A thermo-mechanically-coupled large-deformation theory for amorphous
  polymers in a temperature range which spans their glass transition.
\newblock \emph{International Journal of Plasticity}, 26:\penalty0 1138--1182,
  2010.

\bibitem[Clements(2012{\natexlab{a}})]{clements2012a}
B.E. Clements.
\newblock A continuum glassy polymer model applicable to dynamic loading.
\newblock \emph{Journal of Applied Physics}, 112:\penalty0 083511,
  2012{\natexlab{a}}.

\bibitem[Beyler and Hirschler(2002)]{beyler2002}
C.L. Beyler and M.M. Hirschler.
\newblock Thermal decomposition of polymers.
\newblock In \emph{{SFPE} Handbook of Fire Protection Engineering}, volume~2,
  chapter~7, pages 111--131. National Fire Protection Association, Quincy (MA),
  2002.

\bibitem[Grujicic et~al.(2012)Grujicic, Bell, and Pandurangan]{grujicic2012}
M.~Grujicic, W.C. Bell, and B.~Pandurangan.
\newblock Design and material selection guidelines and strategies for
  transparent armor systems.
\newblock \emph{Materials and Design}, 34:\penalty0 808--819, 2012.

\bibitem[Barker and Hollenbach(1970)]{barker1970}
L.M. Barker and R.E. Hollenbach.
\newblock Shock-wave studies of {PMMA}, fused silica, and sapphire.
\newblock \emph{Journal of Applied Physics}, 41:\penalty0 4208--4226, 1970.

\bibitem[Argon(1975)]{argon1975}
A.S. Argon.
\newblock Role of heterogeneities in the crazing of glassy polymers.
\newblock \emph{Pure and Applied Chemistry}, 43:\penalty0 247--272, 1975.

\bibitem[Arruda et~al.(1995)Arruda, Boyce, and Jayachandran]{arruda1995}
E.M. Arruda, M.C. Boyce, and R.~Jayachandran.
\newblock Effects of strain rate, temperature and thermomechanical coupling on
  the finite strain deformation of glassy polymers.
\newblock \emph{Mechanics of Materials}, 19:\penalty0 193--212, 1995.

\bibitem[Fleck et~al.(1990)Fleck, Stronge, and Liu]{fleck1990}
N.A. Fleck, W.J. Stronge, and J.H. Liu.
\newblock High strain-rate shear response of polycarbonate and polymethyl
  methacrylate.
\newblock \emph{Proceedings of the Royal Society of London A}, 429\penalty0
  (1877):\penalty0 459--479, 1990.

\bibitem[Mulliken and Boyce(2006)]{mulliken2006}
A.D. Mulliken and M.C. Boyce.
\newblock Mechanics of the rate-dependent elastic--plastic deformation of
  glassy polymers from low to high strain rates.
\newblock \emph{International Journal of Solids and Structures}, 43:\penalty0
  1331--1356, 2006.

\bibitem[Richeton et~al.(2006{\natexlab{a}})Richeton, Ahzi, Vecchio, Jiang, and
  Adharapurapu]{richeton2006a}
J.~Richeton, S.~Ahzi, K.S. Vecchio, F.C. Jiang, and R.R. Adharapurapu.
\newblock Influence of temperature and strain rate on the mechanical behavior
  of three amorphous polymers: characterization and modeling of the compressive
  yield stress.
\newblock \emph{International Journal of Solids and Structures}, 43:\penalty0
  2318--2335, 2006{\natexlab{a}}.

\bibitem[Gearing and Anand(2004)]{gearing2004}
B.P. Gearing and L.~Anand.
\newblock On modeling the deformation and fracture response of glassy polymers
  due to shear-yielding and crazing.
\newblock \emph{International Journal of Solids and Structures}, 41:\penalty0
  3125--3150, 2004.

\bibitem[Narayan and Anand(2021)]{narayan2021}
S.~Narayan and L.~Anand.
\newblock Fracture of amorphous polymers: a gradient-damage theory.
\newblock \emph{Journal of the Mechanics and Physics of Solids}, 146:\penalty0
  104164, 2021.

\bibitem[Williams et~al.(1955)Williams, Landel, and Ferry]{williams1955}
M.L. Williams, R.F. Landel, and J.D. Ferry.
\newblock The temperature dependence of relaxation mechanisms in amorphous
  polymers and other glass-forming liquids.
\newblock \emph{Journal of the American Chemical Society}, 7:\penalty0
  3701--3707, 1955.

\bibitem[Anand and Gurtin(2003)]{anand2003}
L.~Anand and M.E. Gurtin.
\newblock A theory of amorphous solids undergoing large deformations, with
  application to polymeric glasses.
\newblock \emph{International Journal of Solids and Structures}, 40:\penalty0
  1465--1487, 2003.

\bibitem[Jatin and Basu(2014)]{jatin2014}
V.S. Jatin and S.~Basu.
\newblock Investigations into the origins of plastic flow and strain hardening
  in amorphous glassy polymers.
\newblock \emph{International Journal of Plasticity}, 56:\penalty0 139--155,
  2014.

\bibitem[White and Lipson(2016)]{white2016}
R.P. White and J.E.G. Lipson.
\newblock Polymer free volume and its connection to the glass transition.
\newblock \emph{Macromolecules}, 49:\penalty0 3987--4007, 2016.

\bibitem[Dal et~al.(2022)Dal, Gultekin, Basdemir, and Acan]{dal2022}
H.~Dal, O.~Gultekin, S.~Basdemir, and A.~Acan.
\newblock Ductile--brittle failure of amorphous glassy polymers: a phase-field
  approach.
\newblock \emph{Computer Methods in Applied Mechanics and Engineering},
  401:\penalty0 115639, 2022.

\bibitem[Menikoff(2004)]{menikoff2004}
R.~Menikoff.
\newblock Constitutive model for polymethyl methacrylate at high pressure.
\newblock \emph{Journal of Applied Physics}, 96:\penalty0 7696--7704, 2004.

\bibitem[Asay et~al.(1969)Asay, Lamberson, and Guenther]{asay1969}
J.R. Asay, D.L. Lamberson, and A.H. Guenther.
\newblock Pressure and temperature dependence of the acoustic velocities in
  polymethylmethacrylate.
\newblock \emph{Journal of Applied Physics}, 40:\penalty0 1768--1783, 1969.

\bibitem[Kono(1960)]{kono1960}
R.~Kono.
\newblock The dynamic bulk viscosity of polystyrene and polymethyl
  methacrylate.
\newblock \emph{Journal of the Physical Society of Japan}, 15:\penalty0
  718--725, 1960.

\bibitem[Sane and Knauss(2001)]{sane2001}
S.B. Sane and W.G. Knauss.
\newblock The time-dependent bulk response of poly (methyl methacrylate).
\newblock \emph{Mechanics of Time-Dependent Materials}, 5:\penalty0 293--324,
  2001.

\bibitem[Clements(2009)]{clements2009}
B.E. Clements.
\newblock Nonequilibrium volumetric response of shocked polymers.
\newblock \emph{AIP Conference Proceedings}, 1195:\penalty0 1223--1228, 2009.

\bibitem[Anand et~al.(2009)Anand, Ames, Srivastava, and Chester]{anand2009}
L.~Anand, N.M. Ames, V.~Srivastava, and S.A. Chester.
\newblock A thermo-mechanically coupled theory for large deformations of
  amorphous polymers. {Part I}: formulation.
\newblock \emph{International Journal of Plasticity}, 25:\penalty0 1474--1494,
  2009.

\bibitem[Bouvard et~al.(2010)Bouvard, Ward, Hossain, Marin, Bammann, and
  Horstemeyer]{bouvard2010}
J.-L. Bouvard, D.K. Ward, D.~Hossain, E.B. Marin, D.J. Bammann, and M.F.
  Horstemeyer.
\newblock A general inelastic internal state variable model for amorphous
  glassy polymers.
\newblock \emph{Acta Mechanica}, 213:\penalty0 71--96, 2010.

\bibitem[Bouvard et~al.(2013)Bouvard, Francis, Tschopp, Marin, Bammann, and
  Horstemeyer]{bouvard2013}
J.-L. Bouvard, D.K. Francis, M.A. Tschopp, E.B. Marin, D.J. Bammann, and M.F.
  Horstemeyer.
\newblock An internal state variable material model for predicting the time,
  thermomechanical, and stress state dependence of amorphous glassy polymers
  under large deformation.
\newblock \emph{International Journal of Plasticity}, 42:\penalty0 168--193,
  2013.

\bibitem[Coleman and Noll(1961)]{coleman1961}
B.D. Coleman and W.~Noll.
\newblock Foundations of linear viscoelasticity.
\newblock \emph{Reviews of Modern Physics}, 33:\penalty0 239--249, 1961.

\bibitem[Schuler(1970)]{schuler1970}
K.W. Schuler.
\newblock Propagation of steady shock waves in polymethyl methacrylate.
\newblock \emph{Journal of the Mechanics and Physics of Solids}, 18:\penalty0
  277--293, 1970.

\bibitem[Schuler et~al.(1973)Schuler, Nunziato, and Walsh]{schuler1973}
K.W. Schuler, J.W. Nunziato, and E.K. Walsh.
\newblock Recent results in nonlinear viscoelastic wave propagation.
\newblock \emph{International Journal of Solids and Structures}, 9:\penalty0
  1237--1281, 1973.

\bibitem[Nunziato and Walsh(1973{\natexlab{a}})]{nunziato1973p}
J.W. Nunziato and E.K. Walsh.
\newblock Propagation of steady shock waves in non-linear thermoviscoelastic
  solids.
\newblock \emph{Journal of the Mechanics and Physics of Solids}, 21:\penalty0
  317--335, 1973{\natexlab{a}}.

\bibitem[Jinaga et~al.(2025)Jinaga, Zulueta, Burgoa, Cobian, Freitas, Lackner,
  Major, and Noels]{jinaga2025}
U.K. Jinaga, K.~Zulueta, A.~Burgoa, L.~Cobian, U.~Freitas, M.~Lackner,
  Z.~Major, and L.~Noels.
\newblock A consistent finite-strain thermomechanical
  quasi-nonlinear-viscoelastic viscoplastic constitutive model for
  thermoplastic polymers.
\newblock \emph{International Journal of Solids and Structures}, page 113517,
  2025.

\bibitem[Holzapfel and Simo(1996)]{holzapfel1996a}
G.A. Holzapfel and J.C. Simo.
\newblock A new viscoelastic constitutive model for continuous media at finite
  thermomechanical changes.
\newblock \emph{International Journal of Solids and Structures}, 33:\penalty0
  3019--3034, 1996.

\bibitem[Holzapfel(1996)]{holzapfel1996b}
G.A. Holzapfel.
\newblock On large strain viscoelasticity: continuum formulation and finite
  element applications to elastomeric structures.
\newblock \emph{International Journal for Numerical Methods in Engineering},
  39:\penalty0 3903--3926, 1996.

\bibitem[Sadik and Yavari(2024)]{sadik2024}
S.~Sadik and A.~Yavari.
\newblock Nonlinear anisotropic viscoelasticity.
\newblock \emph{Journal of the Mechanics and Physics of Solids}, 182:\penalty0
  105461, 2024.

\bibitem[Sidoroff(1974)]{sidoroff1974}
F.~Sidoroff.
\newblock Nonlinear viscoelastic model with intermediate configuration.
\newblock \emph{Journal de Mecanique}, 13:\penalty0 679--713, 1974.

\bibitem[Reese and Govindjee(1998)]{reese1998}
S.~Reese and S.~Govindjee.
\newblock A theory of finite viscoelasticity and numerical aspects.
\newblock \emph{International Journal of Solids and Structures}, 35:\penalty0
  3455--3482, 1998.

\bibitem[Clayton(2011)]{claytonNCM2011}
J.D. Clayton.
\newblock \emph{Nonlinear Mechanics of Crystals}.
\newblock Springer, Dordrecht, 2011.

\bibitem[Francis et~al.(2014)Francis, Bouvard, Hammi, and
  Horstemeyer]{francis2014}
D.K. Francis, J.-L. Bouvard, Y.~Hammi, and M.F. Horstemeyer.
\newblock Formulation of a damage internal state variable model for amorphous
  glassy polymers.
\newblock \emph{International Journal of Solids and Structures}, 51:\penalty0
  2765--2776, 2014.

\bibitem[Boyce et~al.(1988)Boyce, Parks, and Argon]{boyce1988}
M.C. Boyce, D.M. Parks, and A.S. Argon.
\newblock Large inelastic deformation of glassy polymers. {Part I}: rate
  dependent constitutive model.
\newblock \emph{Mechanics of Materials}, 7:\penalty0 15--33, 1988.

\bibitem[Argon(1973)]{argon1973}
A.S. Argon.
\newblock A theory for the low-temperature plastic deformation of glassy
  polymers.
\newblock \emph{Philosophical Magazine}, 28:\penalty0 839--865, 1973.

\bibitem[Arruda et~al.(1993)Arruda, Boyce, and Quintus-Bosz]{arruda1993b}
E.M. Arruda, M.C. Boyce, and H.~Quintus-Bosz.
\newblock Effects of initial anisotropy on the finite strain deformation
  behavior of glassy polymers.
\newblock \emph{International Journal of Plasticity}, 9:\penalty0 783--811,
  1993.

\bibitem[Dupaix and Boyce(2007)]{dupaix2007}
R.B. Dupaix and M.C. Boyce.
\newblock Constitutive modeling of the finite strain behavior of amorphous
  polymers in and above the glass transition.
\newblock \emph{Mechanics of Materials}, 39:\penalty0 39--52, 2007.

\bibitem[Ames et~al.(2009)Ames, Srivastava, Chester, and Anand]{ames2009}
N.M. Ames, V.~Srivastava, S.A. Chester, and L.~Anand.
\newblock A thermo-mechanically coupled theory for large deformations of
  amorphous polymers. {Part II}: applications.
\newblock \emph{International Journal of Plasticity}, 25:\penalty0 1495--1539,
  2009.

\bibitem[Richeton et~al.(2005)Richeton, Ahzi, Daridon, and
  Remond]{richeton2005}
J.~Richeton, S.~Ahzi, L.~Daridon, and Y.~Remond.
\newblock A formulation of the cooperative model for the yield stress of
  amorphous polymers for a wide range of strain rates and temperatures.
\newblock \emph{Polymer}, 46:\penalty0 6035--6043, 2005.

\bibitem[Richeton et~al.(2006{\natexlab{b}})Richeton, Ahzi, Makradi, and
  Vecchio]{richeton2006b}
J.~Richeton, S.~Ahzi, A.~Makradi, and K.S. Vecchio.
\newblock Constitutive modeling of polymer materials at impact loading rates.
\newblock \emph{Journal de Physique IV}, 134:\penalty0 103--107,
  2006{\natexlab{b}}.

\bibitem[Wu and Van Der~Giessen(1993)]{wu1993}
P.D. Wu and E.~Van Der~Giessen.
\newblock On improved network models for rubber elasticity and their
  applications to orientation hardening in glassy polymers.
\newblock \emph{Journal of the Mechanics and Physics of Solids}, 41:\penalty0
  427--456, 1993.

\bibitem[Arruda and Boyce(1993)]{arruda1993a}
E.M. Arruda and M.C. Boyce.
\newblock A three-dimensional constitutive model for the large stretch behavior
  of rubber elastic materials.
\newblock \emph{Journal of the Mechanics and Physics of Solids}, 41:\penalty0
  389--412, 1993.

\bibitem[Khandagale et~al.(2023)Khandagale, Breitzman, Majidi, and
  Dayal]{khand2023}
P.~Khandagale, T.~Breitzman, C.~Majidi, and K.~Dayal.
\newblock Statistical field theory for nonlinear elasticity of polymer networks
  with excluded volume interactions.
\newblock \emph{Physical Review E}, 107:\penalty0 064501, 2023.

\bibitem[Palm et~al.(2006)Palm, Dupaix, and Castro]{palm2006}
G.~Palm, R.B. Dupaix, and J.~Castro.
\newblock Large strain mechanical behavior of poly (methyl
  methacrylate)({PMMA}) near the glass transition temperature.
\newblock \emph{Journal of Engineering Materials and Technology}, 128:\penalty0
  559--563, 2006.

\bibitem[Schuler and Nunziato(1974)]{schuler1974}
K.W. Schuler and J.W. Nunziato.
\newblock The dynamic mechanical behavior of polymethyl methacrylate.
\newblock \emph{Rheologica Acta}, 13:\penalty0 265--273, 1974.

\bibitem[Merzhievskii and Voronin(2012)]{merz2012}
L.A. Merzhievskii and M.S. Voronin.
\newblock Modeling of shock-wave deformation of polymethyl methacrylate.
\newblock \emph{Combustion, Explosion, and Shock Waves}, 48:\penalty0 226--235,
  2012.

\bibitem[Popova et~al.(2015)Popova, Mayer, and Khishchenko]{popova2015}
T.V. Popova, A.E. Mayer, and K.V. Khishchenko.
\newblock Numerical investigations of shock wave propagation in
  polymethylmethacrylate.
\newblock \emph{Journal of Physics: Conference Series}, 653:\penalty0 012045,
  2015.

\bibitem[Popova et~al.(2018)Popova, Mayer, and Khishchenko]{popova2018}
T.V. Popova, A.E. Mayer, and K.V. Khishchenko.
\newblock Evolution of shock compression pulses in polymethylmethacrylate and
  aluminum.
\newblock \emph{Journal of Applied Physics}, 123:\penalty0 235902, 2018.

\bibitem[Dorogoy et~al.(2010)Dorogoy, Rittel, and Brill]{dorogoy2010}
A.~Dorogoy, D.~Rittel, and A.~Brill.
\newblock A study of inclined impact in polymethylmethacrylate plates.
\newblock \emph{International Journal of Impact Engineering}, 37:\penalty0
  285--294, 2010.

\bibitem[Holmquist et~al.(2016)Holmquist, Bradley, Dwivedi, and
  Casem]{holmquist2016}
T.J. Holmquist, J.~Bradley, A.~Dwivedi, and D.~Casem.
\newblock The response of polymethyl methacrylate ({PMMA}) subjected to large
  strains, high strain rates, high pressures, a range in temperatures, and
  variations in the intermediate principal stress.
\newblock \emph{European Physical Journal Special Topics}, 225:\penalty0
  343--354, 2016.

\bibitem[Lawlor et~al.(2024)Lawlor, Gandhi, and Ravichandran]{lawlor2024}
B.P. Lawlor, V.~Gandhi, and G.~Ravichandran.
\newblock Full-field quantitative visualization of shock-driven pore collapse
  and failure modes in {PMMA}.
\newblock \emph{Journal of Applied Physics}, 136:\penalty0 225901, 2024.

\bibitem[Coleman and Noll(1963)]{coleman1963}
B.D. Coleman and W.~Noll.
\newblock The thermodynamics of elastic materials with heat conduction and
  viscosity.
\newblock \emph{Archive for Rational Mechanics and Analysis}, 13:\penalty0
  167--178, 1963.

\bibitem[Coleman and Gurtin(1967)]{coleman1967}
B.D. Coleman and M.E. Gurtin.
\newblock Thermodynamics with internal state variables.
\newblock \emph{Journal of Chemical Physics}, 47:\penalty0 597--613, 1967.

\bibitem[Christman(1972)]{christman1972}
D.R. Christman.
\newblock Dynamic properties of polymethymethacrylate ({PMMA}) ({Plexiglas}).
\newblock Technical Report DNA-2810F, Defense Nuclear Agency, Washington, D.C.,
  1972.

\bibitem[Washabaugh and Knauss(1993)]{washabaugh1993}
P.D. Washabaugh and W.G. Knauss.
\newblock Non-steady, periodic behavior in the dynamic fracture of {PMMA}.
\newblock \emph{International Journal of Fracture}, 59:\penalty0 189--197,
  1993.

\bibitem[Argon and Salama(1977)]{argon1977}
A.S. Argon and M.M. Salama.
\newblock Growth of crazes in glassy polymers.
\newblock \emph{Philosophical Magazine}, 36:\penalty0 1217--1234, 1977.

\bibitem[G'Sell et~al.(2002)G'Sell, Hiver, and Dahoun]{gsell2002}
C.~G'Sell, J.M. Hiver, and A.~Dahoun.
\newblock Experimental characterization of deformation damage in solid polymers
  under tension, and its interrelation with necking.
\newblock \emph{International Journal of Solids and Structures}, 39:\penalty0
  3857--3872, 2002.

\bibitem[Satapathy and Bless(2000)]{satapathy2000}
S.~Satapathy and S.~Bless.
\newblock Deep punching {PMMA}.
\newblock \emph{Experimental Mechanics}, 40:\penalty0 31--37, 2000.

\bibitem[Rittel and Brill(2008)]{rittel2008}
D.~Rittel and A.~Brill.
\newblock Dynamic flow and failure of confined polymethylmethacrylate.
\newblock \emph{Journal of the Mechanics and Physics of Solids}, 56:\penalty0
  1401--1416, 2008.

\bibitem[Zhang et~al.(2021)Zhang, Townsend, Petrinic, and
  Pellegrino]{zhang2021}
L.~Zhang, D.~Townsend, N.~Petrinic, and A.~Pellegrino.
\newblock Pressure and temperature dependent dynamic flow and failure behavior
  of {PMMA} at intermediate strain rates.
\newblock \emph{International Journal of Impact Engineering}, 158:\penalty0
  104026, 2021.

\bibitem[Archer and Lesser(2010)]{archer2010}
J.S. Archer and A.J. Lesser.
\newblock Shear band formation and mode {II} fracture of polymeric glasses.
\newblock \emph{Journal of Polymer Science B: Polymer Physics}, 49:\penalty0
  103--114, 2010.

\bibitem[Bjerke and Lambros(2002)]{bjerke2002}
T.~Bjerke and J.~Lambros.
\newblock Heating during shearing and opening dominated dynamic fracture of
  polymers.
\newblock \emph{Experimental Mechanics}, 42:\penalty0 107--114, 2002.

\bibitem[Estevez et~al.(2000)Estevez, Tijssens, and Van~der
  Giessen]{estevez2000}
R.~Estevez, M.G.A. Tijssens, and E.~Van~der Giessen.
\newblock Modeling of the competition between shear yielding and crazing in
  glassy polymers.
\newblock \emph{Journal of the Mechanics and Physics of Solids}, 48:\penalty0
  2585--2617, 2000.

\bibitem[Arias et~al.(2007)Arias, Knap, Chalivendra, Hong, Ortiz, and
  Rosakis]{arias2007}
I.~Arias, J.~Knap, V.B. Chalivendra, S.~Hong, M.~Ortiz, and A.J. Rosakis.
\newblock Numerical modelling and experimental validation of dynamic fracture
  events along weak planes.
\newblock \emph{Computer Methods in Applied Mechanics and Engineering},
  196:\penalty0 3833--3840, 2007.

\bibitem[Schluter et~al.(2016)Schluter, Kuhn, Muller, and Gross]{schluter2016}
A.~Schluter, C.~Kuhn, R.~Muller, and D.~Gross.
\newblock An investigation of intersonic fracture using a phase field model.
\newblock \emph{Archive of Applied Mechanics}, 86:\penalty0 321--333, 2016.

\bibitem[Clayton(2005{\natexlab{a}})]{claytonJMPS2005}
J.D. Clayton.
\newblock {Dynamic plasticity and fracture in high density polycrystals:
  constitutive modeling and numerical simulation}.
\newblock \emph{Journal of the Mechanics and Physics of Solids}, 53:\penalty0
  261--301, 2005{\natexlab{a}}.

\bibitem[Clayton(2005{\natexlab{b}})]{claytonIJSS2005}
J.D. Clayton.
\newblock {Modeling dynamic plasticity and spall fracture in high density
  polycrystalline alloys}.
\newblock \emph{International Journal of Solids and Structures}, 42:\penalty0
  4613--4640, 2005{\natexlab{b}}.

\bibitem[Foulk and Vogler(2010)]{foulk2010}
J.W. Foulk and T.J. Vogler.
\newblock A grain-scale study of spall in brittle materials.
\newblock \emph{International Journal of Fracture}, 163:\penalty0 225--242,
  2010.

\bibitem[Miehe et~al.(2015{\natexlab{a}})Miehe, Hofacker, Schanzel, and
  Aldakheel]{miehe2015b}
C.~Miehe, M.~Hofacker, L.M. Schanzel, and F.~Aldakheel.
\newblock Phase field modeling of fracture in multi-physics problems. {Part II.
  C}oupled brittle-to-ductile failure criteria and crack propagation in
  thermo-elastic-plastic solids.
\newblock \emph{Computer Methods in Applied Mechanics and Engineering},
  294:\penalty0 486--522, 2015{\natexlab{a}}.

\bibitem[Li et~al.(2024)Li, Deng, Xu, and Liu]{li2024}
K.~Li, H.~Deng, W.~Xu, and Y.~Liu.
\newblock Modelling of fracture-involved large strain behaviors of amorphous
  glassy polymers via a unified physically-based constitutive model coupled
  with phase field method.
\newblock \emph{Engineering Fracture Mechanics}, 311:\penalty0 110546, 2024.

\bibitem[Geraskin et~al.(2009)Geraskin, Khishchenko, Krasyuk, Pashinin,
  Semenov, and Vovchenko]{geraskin2009}
A.A. Geraskin, K.V. Khishchenko, I.K. Krasyuk, P.P. Pashinin, A.Y. Semenov, and
  V.I. Vovchenko.
\newblock Specific features of spallation processes in polymethyl methacrylate
  under high strain rate.
\newblock \emph{Contributions in Plasma Physics}, 49:\penalty0 451--454, 2009.

\bibitem[Diamond and Ramesh(2025)]{diamond2025}
J.M. Diamond and K.T. Ramesh.
\newblock Spallation of polycarbonate on nanosecond timescales.
\newblock \emph{Physical Review E}, 111:\penalty0 025503, 2025.

\bibitem[Zaretsky and Kanel(2019)]{zaretsky2019}
E.B. Zaretsky and G.I. Kanel.
\newblock Response of poly (methyl methacrylate) to shock-wave loading at
  elevated temperatures.
\newblock \emph{Journal of Applied Physics}, 126:\penalty0 085902, 2019.

\bibitem[Cherepanov et~al.(2024)Cherepanov, Savinykh, Garkushin, and
  Razorenov]{cherepanov2024}
I.A. Cherepanov, A.S. Savinykh, G.V. Garkushin, and S.V. Razorenov.
\newblock Spall strength of polycarbonate at a temperature of 20--185 {C}.
\newblock \emph{Technical Physics}, 69:\penalty0 1938--1944, 2024.

\bibitem[Curran et~al.(2012)Curran, Shockey, and Seaman]{curran1973}
D.R. Curran, D.A. Shockey, and L.~Seaman.
\newblock Dynamic fracture criteria for a polycarbonate.
\newblock \emph{Journal of Applied Physics}, 44:\penalty0 4025--4038, 2012.

\bibitem[Dattlebaum et~al.(2026)Dattlebaum, Schilling, Clements, Jordan, Welch,
  and Stull]{dattelbaum2024}
D.M. Dattlebaum, B.F. Schilling, B.E. Clements, J.L. Jordan, C.F. Welch, and
  J.A. Stull.
\newblock Shock response and dynamic failure of high-density ({HDPE}) and
  ultra-high molecular weight polyethylene ({UHMWPE}).
\newblock \emph{Journal of Dynamic Behavior of Materials}, 12:\penalty0 37--48,
  2026.

\bibitem[Dewapriya and Miller(2022)]{dewapriya2022}
M.A.N. Dewapriya and R.E. Miller.
\newblock Molecular dynamics study on the shock induced spallation of
  polyethylene.
\newblock \emph{Journal of Applied Physics}, 131:\penalty0 025102, 2022.

\bibitem[Clayton(2026)]{claytonAMECH2025}
J.D. Clayton.
\newblock Phase-field theory of adiabatic shear.
\newblock \emph{Acta Mechanica}, 237:\penalty0 239--273, 2026.

\bibitem[Carter and Marsh(1995)]{carter1995}
W.J. Carter and S.P. Marsh.
\newblock Hugoniot equation of state of polymers.
\newblock Technical Report LA-13006-MS, Los Alamos National Laboratory, Los
  Alamos (NM), 1995.

\bibitem[Dattelbaum and Coe(2019)]{dattelbaum2019}
D.M. Dattelbaum and J.D. Coe.
\newblock Shock-driven decomposition of polymers and polymeric foams.
\newblock \emph{Polymers}, 11:\penalty0 493, 2019.

\bibitem[Hauver and Melani(1964)]{hauver1964}
G.E. Hauver and A.~Melani.
\newblock Shock compression of plexiglas and polystyrene.
\newblock Technical Report BRL-R-1259, Ballistic Research Laboratory, Aberdeen
  Proving Ground (MD), 1964.

\bibitem[Hauver(1965)]{hauver1965}
G.E. Hauver.
\newblock Shock-induced polarization in plastics. {II. Experimental} study of
  plexiglas and polystyrene.
\newblock \emph{Journal of Applied Physics}, 36:\penalty0 2113--2118, 1965.

\bibitem[Lentz et~al.(2020)Lentz, Coe, and Velizhanin]{lentz2020}
M.K. Lentz, J.D. Coe, and K.A. Velizhanin.
\newblock Reshock analysis for {PMMA} driven above the threshold for chemical
  decomposition.
\newblock \emph{AIP Conference Proceedings}, 2272:\penalty0 070027, 2020.

\bibitem[Bordzilovskii et~al.(2021)Bordzilovskii, Voronin, and
  Karakhanov]{bordz2021}
S.A. Bordzilovskii, M.S. Voronin, and S.M. Karakhanov.
\newblock Temperature of polymethyl methacrylate in a secondary shock wave.
\newblock \emph{Combustion, Explosion, and Shock Waves}, 57:\penalty0 736--745,
  2021.

\bibitem[Coe et~al.(2022)Coe, Lentz, Velizhanin, Gammel, Kaushagen, Jones, and
  Cochrane]{coe2022}
J.D. Coe, M.~Lentz, K.A. Velizhanin, J.T. Gammel, J.~Kaushagen, K.~Jones, and
  K.R. Cochrane.
\newblock The equation of state and shock-driven decomposition of
  polymethylmethacrylate ({PMMA}).
\newblock \emph{Journal of Applied Physics}, 131, 2022.

\bibitem[Huber et~al.(2023)Huber, Dattelbaum, Lang, Coe, Peterson, Bartram, and
  Gibson]{huber2023}
R.C. Huber, D.M. Dattelbaum, J.M. Lang, J.D. Coe, J.H. Peterson, B.~Bartram,
  and L.L. Gibson.
\newblock Polyimide dynamically compressed to decomposition pressures: two-wave
  structures captured by velocimetry and modeling.
\newblock \emph{Journal of Applied Physics}, 133:\penalty0 035106, 2023.

\bibitem[Graham(1979)]{graham1979}
R.A. Graham.
\newblock Shock-induced electrical activity in polymeric solids. {A}
  mechanically induced bond scission model.
\newblock \emph{Journal of Physical Chemistry}, 83:\penalty0 3048--3056, 1979.

\bibitem[Chan and Ruoff(1981)]{chan1981}
K.-S. Chan and A.L. Ruoff.
\newblock Static compression of polymethyl methacrylate to 100 {GPa}.
\newblock \emph{Journal of Applied Physics}, 52:\penalty0 5395--5396, 1981.

\bibitem[Maerzke et~al.(2019)Maerzke, Coe, Ticknor, Leiding, Tinka~Gammel, and
  Welch]{maerzke2019}
K.A. Maerzke, J.D. Coe, C.~Ticknor, J.A. Leiding, J.~Tinka~Gammel, and C.F.
  Welch.
\newblock Equations of state for polyethylene and its shock-driven
  decomposition products.
\newblock \emph{Journal of Applied Physics}, 126:\penalty0 045902, 2019.

\bibitem[Kashiwagi et~al.(1985)Kashiwagi, Hirata, and Brown]{kashiwagi1985}
T.~Kashiwagi, T.~Hirata, and J.E. Brown.
\newblock Thermal and oxidative degradation of poly (methyl methacrylate)
  molecular weight.
\newblock \emph{Macromolecules}, 18:\penalty0 131--138, 1985.

\bibitem[Stoliarov et~al.(2003)Stoliarov, Westmoreland, Nyden, and
  Forney]{stoliarov2003}
S.I. Stoliarov, P.R. Westmoreland, M.R. Nyden, and G.P. Forney.
\newblock A reactive molecular dynamics model of thermal decomposition in
  polymers: {I. Poly} (methyl methacrylate).
\newblock \emph{Polymer}, 44:\penalty0 883--894, 2003.

\bibitem[Korobeinichev et~al.(2019)Korobeinichev, Paletsky, Gonchikzhapov,
  Glaznev, Gerasimov, Naganovsky, Shundrina, Snegirev, and Vinu]{korob2019}
O.P. Korobeinichev, A.A. Paletsky, M.B. Gonchikzhapov, R.K. Glaznev, I.E.
  Gerasimov, Y.K. Naganovsky, I.K. Shundrina, A.Y. Snegirev, and R.~Vinu.
\newblock Kinetics of thermal decomposition of {PMMA} at different heating
  rates and in a wide temperature range.
\newblock \emph{Thermochimica Acta}, 671:\penalty0 17--25, 2019.

\bibitem[Van~Thiel et~al.(1977)Van~Thiel, Shaner, and
  Salinas~(editors)]{vanthiel1977}
M.~Van~Thiel, J.~Shaner, and E.~Salinas~(editors).
\newblock Compendium of shock wave data. {V}olume 3.
\newblock Technical Report UCRL-50108-3, Lawrence Livermore Laboratory,
  LIvermore (CA)), 1977.

\bibitem[Scheidler(2009)]{scheidler2009}
M.~Scheidler.
\newblock Viscoelastic models for nearly incompressible materials.
\newblock Technical Report ARL-TR-4992, Army Research Laboratory, Aberdeen
  Proving Ground (MD), 2009.

\bibitem[Clayton(2014{\natexlab{a}})]{claytonIJES2014}
J.D. Clayton.
\newblock Analysis of shock compression of strong single crystals with
  logarithmic thermoelastic-plastic theory.
\newblock \emph{International Journal of Engineering Science}, 79:\penalty0
  1--20, 2014{\natexlab{a}}.

\bibitem[Clayton(2019{\natexlab{a}})]{claytonNEIM2019}
J.D. Clayton.
\newblock \emph{Nonlinear Elastic and Inelastic Models for Shock Compression of
  Crystalline Solids}.
\newblock Springer, Cham, 2019{\natexlab{a}}.

\bibitem[Levitas and Samani(2011)]{levitas2011b}
V.I. Levitas and K.~Samani.
\newblock Coherent solid/liquid interface with stress relaxation in a
  phase-field approach to the melting/solidification transition.
\newblock \emph{Physical Review B}, 84:\penalty0 140103, 2011.

\bibitem[Bahloul et~al.(2020)Bahloul, Doghri, and Adam]{bahloul2020}
A.~Bahloul, I.~Doghri, and L.~Adam.
\newblock An enhanced phase field model for the numerical simulation of polymer
  crystallization.
\newblock \emph{Polymer Crystallization}, 3:\penalty0 e10144, 2020.

\bibitem[Levitas et~al.(2011)Levitas, Idesman, and Palakala]{levitas2011a}
V.I. Levitas, A.V. Idesman, and A.K. Palakala.
\newblock Phase-field modeling of fracture in liquid.
\newblock \emph{Journal of Applied Physics}, 110:\penalty0 033531, 2011.

\bibitem[Clayton(2024)]{claytonPRE2024}
J.D. Clayton.
\newblock Universal phase-field mixture representation of thermodynamics and
  shock wave mechanics in porous soft biologic continua.
\newblock \emph{Physical Review E}, 110:\penalty0 035001, 2024.

\bibitem[Clayton(2025)]{claytonARX2025}
J.D. Clayton.
\newblock Analysis of adiabatic strain localization coupled to ductile fracture
  and melting, with application and verification for simple shear.
\newblock \emph{AppliedMath}, 5:\penalty0 169, 2025.

\bibitem[Gibbs and DiMarzio(1958)]{gibbs1958}
J.H. Gibbs and E.A. DiMarzio.
\newblock Nature of the glass transition and the glassy state.
\newblock \emph{Journal of Chemical Physics}, 28:\penalty0 373--383, 1958.

\bibitem[Wu(1999)]{wu1999}
J.~Wu.
\newblock The glassy state, ideal glass transition, and second-order phase
  transition.
\newblock \emph{Journal of Applied Polymer Science}, 71:\penalty0 143--150,
  1999.

\bibitem[Millett and Bourne(2000)]{millett2000}
J.C.F. Millett and N.K. Bourne.
\newblock The deviatoric response of polymethylmethacrylate to one-dimensional
  shock loading.
\newblock \emph{Journal of Applied Physics}, 88:\penalty0 7037--7040, 2000.

\bibitem[Gupta et~al.(1980)Gupta, Keough, Henley, and Walter]{gupta1980a}
Y.M. Gupta, D.D. Keough, D.~Henley, and D.F. Walter.
\newblock Measurement of lateral compressive stresses under shock loading.
\newblock \emph{Applied Physics Letters}, 37:\penalty0 395--397, 1980.

\bibitem[Gupta(1980)]{gupta1980b}
Y.M. Gupta.
\newblock Determination of the impact response of {PMMA} using combined
  compression and shear loading.
\newblock \emph{Journal of Applied Physics}, 51:\penalty0 5352--5361, 1980.

\bibitem[Jordan et~al.(2002)Jordan, Casem, Sturtevant, and
  Sutherland]{jordan2020}
J.L. Jordan, D.T. Casem, B.T. Sturtevant, and G.~Sutherland.
\newblock Dynamic strength in polymethylmethacrylate.
\newblock \emph{AIP Conference Proceedings}, 2272:\penalty0 040006, 2002.

\bibitem[Batkov et~al.(1996)Batkov, Novikov, and Fishman]{batkov1996}
Y.V. Batkov, S.A. Novikov, and N.D. Fishman.
\newblock Shear stresses in polymers under shock compression.
\newblock \emph{AIP Conference Proceedings}, 370:\penalty0 577--580, 1996.

\bibitem[Rosenberg and Partom(1994)]{rosenberg1994}
Z.~Rosenberg and Y.~Partom.
\newblock Accounting for the {Hugoniot} elastic limits of polymers by using
  pressure-dependent yield criterion.
\newblock \emph{Journal of Applied Physics}, 76:\penalty0 1935--1936, 1994.

\bibitem[Nunziato and Schuler(1973)]{nunziato1973e}
J.W. Nunziato and K.W. Schuler.
\newblock Evolution of steady shock waves in polymethyl methacrylate.
\newblock \emph{Journal of Applied Physics}, 44:\penalty0 4774--4775, 1973.

\bibitem[Nunziato and Walsh(1973{\natexlab{b}})]{nunziato1973am}
J.W. Nunziato and E.K. Walsh.
\newblock Amplitude behavior of shock waves in a thermoviscoelastic solid.
\newblock \emph{International Journal of Solids and Structures}, 9:\penalty0
  1373--1383, 1973{\natexlab{b}}.

\bibitem[Chen and Gurtin(1970)]{chen1970a}
P.J. Chen and M.E. Gurtin.
\newblock On the growth of one-dimensional shock waves in materials with
  memory.
\newblock \emph{Archive for Rational Mechanics and Analysis}, 36:\penalty0
  33--46, 1970.

\bibitem[Chen and Gurtin(1972)]{chen1972}
P.J. Chen and M.E. Gurtin.
\newblock Thermodynamic influences on the growth of one-dimensional shock waves
  in materials with memory.
\newblock \emph{Zeitschrift fur Angewandte Mathematik und Physik (ZAMP)},
  23:\penalty0 69--79, 1972.

\bibitem[Chen and Gurtin(1971)]{chen1971}
P.J. Chen and M.E. Gurtin.
\newblock Growth and decay of one‐dimensional shock waves in fluids with
  internal state variables.
\newblock \emph{Physics of Fluids}, 14:\penalty0 1091--1094, 1971.

\bibitem[Clayton(2021)]{claytonJMPS2021}
J.D. Clayton.
\newblock Nonlinear thermodynamic phase field theory with application to
  fracture and dynamic inelastic phenomena in ceramic polycrystals.
\newblock \emph{Journal of the Mechanics and Physics of Solids}, 157:\penalty0
  104633, 2021.

\bibitem[Clayton and Williams(2022)]{claytonPROCA2022}
J.D. Clayton and C.L. Williams.
\newblock Modelling the anomalous shock response of titanium diboride.
\newblock \emph{Proceedings of the Royal Society of London A}, 478:\penalty0
  20220253, 2022.

\bibitem[Malvern(1969)]{malvern1969}
L.E. Malvern.
\newblock \emph{{Introduction to the Mechanics of a Continuous Medium}}.
\newblock Prentice-Hall, Englewood Cliffs NJ, 1969.

\bibitem[Marsden and Hughes(1983)]{marsden1983}
J.E. Marsden and T.J.R. Hughes.
\newblock \emph{{Mathematical Foundations of Elasticity}}.
\newblock Prentice-Hall, Englewood Cliffs NJ, 1983.

\bibitem[Gurtin(1996)]{gurtin1996}
M.E. Gurtin.
\newblock Generalized {Ginzburg-Landau and Cahn-Hilliard} equations based on a
  microforce balance.
\newblock \emph{Physica D}, 92:\penalty0 178--192, 1996.

\bibitem[Borden et~al.(2016)Borden, Hughes, Landis, Anvari, and
  Lee]{borden2016}
M.J. Borden, T.J.R. Hughes, C.M. Landis, A.~Anvari, and I.J. Lee.
\newblock A phase-field formulation for fracture in ductile materials: finite
  deformation balance law derivation, plastic degradation, and stress
  triaxiality effects.
\newblock \emph{Computer Methods in Applied Mechanics and Engineering},
  312:\penalty0 130--166, 2016.

\bibitem[Miehe et~al.(2015{\natexlab{b}})Miehe, Schaenzel, and
  Ulmer]{miehe2015a}
C.~Miehe, L.-M. Schaenzel, and H.~Ulmer.
\newblock Phase field modeling of fracture in multi-physics problems. {Part I.
  Balance} of crack surface and failure criteria for brittle crack propagation
  in thermo-elastic solids.
\newblock \emph{Computer Methods in Applied Mechanics and Engineering},
  294:\penalty0 449--485, 2015{\natexlab{b}}.

\bibitem[Gurtin and Anand(2005)]{gurtin2005}
M.E. Gurtin and L.~Anand.
\newblock A theory of strain-gradient plasticity for isotropic, plastically
  irrotational materials. {P}art {II}: finite deformations.
\newblock \emph{International Journal of Plasticity}, 21:\penalty0 2297--2318,
  2005.

\bibitem[Clayton(2014{\natexlab{b}})]{claytonDGKC2014}
J.D. Clayton.
\newblock \emph{{Differential Geometry and Kinematics of Continua}}.
\newblock World Scientific, Singapore, 2014{\natexlab{b}}.

\bibitem[Anand(1979)]{anand1979}
L.~Anand.
\newblock On {H. Hencky’s} approximate strain-energy function for moderate
  deformations.
\newblock \emph{Journal of Applied Mechanics}, 46:\penalty0 78--82, 1979.

\bibitem[Fitzgerald(1980)]{fitzgerald1980}
J.E. Fitzgerald.
\newblock A tensorial {H}encky measure of strain and strain rate for finite
  deformations.
\newblock \emph{Journal of Applied Physics}, 51:\penalty0 5111--5115, 1980.

\bibitem[Criscione et~al.(2000)Criscione, Humphrey, Douglas, and
  Hunter]{criscione2000}
J.C. Criscione, J.D. Humphrey, A.S. Douglas, and W.C. Hunter.
\newblock An invariant basis for natural strain which yields orthogonal stress
  response terms in isotropic hyperelasticity.
\newblock \emph{Journal of the Mechanics and Physics of Solids}, 48:\penalty0
  2445--2465, 2000.

\bibitem[Clayton(2019{\natexlab{b}})]{claytonJMPS2019}
J.D. Clayton.
\newblock Nonlinear thermomechanics for analysis of weak shock profile data in
  ductile polycrystals.
\newblock \emph{Journal of the Mechanics and Physics of Solids}, 124:\penalty0
  714--757, 2019{\natexlab{b}}.

\bibitem[Schanzel et~al.(2012)Schanzel, Dal, and Miehe]{schanzel2012}
L.~Schanzel, H.~Dal, and C.~Miehe.
\newblock A new continuum approach to the coupling of shear yielding and
  crazing with fracture in glassy polymers.
\newblock \emph{Proceedings of Applied Mathematics and Mechanics (PAMM)},
  12:\penalty0 337--338, 2012.

\bibitem[Basdemir et~al.(2022)Basdemir, Gultekin, and Dal]{bas2022}
S.~Basdemir, O.~Gultekin, and H.~Dal.
\newblock Coupled thermo-viscoplastic fracture model for ductile-brittle
  failure of amorphous glassy polymers with phase-field approach.
\newblock \emph{Proceedings of Applied Mathematics and Mechanics (PAMM)},
  22:\penalty0 e202200280, 2022.

\bibitem[Miehe et~al.(2010)Miehe, Welschinger, and Hofacker]{miehe2010}
C.~Miehe, F.~Welschinger, and M.~Hofacker.
\newblock Thermodynamically consistent phase-field models of fracture:
  Variational principles and multi-field {FE} implementations.
\newblock \emph{International Journal for Numerical Methods in Engineering},
  83:\penalty0 1273--1311, 2010.

\bibitem[Clayton et~al.(2023)Clayton, Leavy, and Knap]{claytonMRC2023}
J.D. Clayton, R.B. Leavy, and J.~Knap.
\newblock Phase field theory for pressure-dependent strength in brittle solids
  with dissipative kinetics.
\newblock \emph{Mechanics Research Communications}, 129:\penalty0 104097, 2023.

\bibitem[Loo et~al.(2001)Loo, Register, Ryan, and Dee]{loo2001}
Y.-L. Loo, R.~A. Register, A.J. Ryan, and G.T. Dee.
\newblock Polymer crystallization confined in one, two, or three dimensions.
\newblock \emph{Macromolecules}, 34:\penalty0 8968--8977, 2001.

\bibitem[Rittel(1999)]{rittel1999}
D.~Rittel.
\newblock On the conversion of plastic work to heat during high strain rate
  deformation of glassy polymers.
\newblock \emph{Mechanics of Materials}, 31:\penalty0 131--139, 1999.

\bibitem[Shao et~al.(2017)Shao, Zhu, Wang, and Zhao]{shao2017}
G.~Shao, S.~Zhu, Y.~Wang, and Q.~Zhao.
\newblock An internal state variable thermodynamic model for determining the
  {Taylor-Quinney} coefficient of glassy polymers.
\newblock \emph{International Journal of Mechanical Sciences}, 126:\penalty0
  261--269, 2017.

\bibitem[Clayton and Knap(2014)]{claytonIJF2014}
J.D. Clayton and J.~Knap.
\newblock A geometrically nonlinear phase field theory of brittle fracture.
\newblock \emph{International Journal of Fracture}, 189:\penalty0 139--148,
  2014.

\bibitem[Davison(2008)]{davison2008}
L.~Davison.
\newblock \emph{Fundamentals of Shock Wave Propagation in Solids}.
\newblock Springer, Berlin, 2008.

\bibitem[Born(1939)]{born1939}
M.~Born.
\newblock Thermodynamics of crystals and melting.
\newblock \emph{Journal of Chemical Physics}, 7:\penalty0 591--603, 1939.

\bibitem[Clayton and Lloyd(2022)]{claytonCMT2022}
J.D. Clayton and J.T. Lloyd.
\newblock Finite strain continuum theory for phase transformations in
  ferromagnetic elastic-plastic solids.
\newblock \emph{Continuum Mechanics and Thermodynamics}, 34:\penalty0
  1579--1620, 2022.

\bibitem[Nunziato and Walsh(1973{\natexlab{c}})]{nunziato1973g}
J.W. Nunziato and E.K. Walsh.
\newblock Instantaneous and equilibrium {Gr\"uneisen} parameters for a
  nonlinear viscoelastic polymer.
\newblock \emph{Journal of Applied Physics}, 44:\penalty0 1207--1211,
  1973{\natexlab{c}}.

\bibitem[Holzapfel and Gasser(2001)]{holzapfel2001}
G.A. Holzapfel and T.C. Gasser.
\newblock A viscoelastic model for fiber-reinforced composites at finite
  strains: continuum basis, computational aspects and applications.
\newblock \emph{Computer Methods in Applied Mechanics and Engineering},
  190:\penalty0 4379--4403, 2001.

\bibitem[Holzapfel et~al.(2002)Holzapfel, Gasser, and Stadler]{holzapfel2002}
G.A. Holzapfel, T.C. Gasser, and M.~Stadler.
\newblock A structural model for the viscoelastic behavior of arterial walls:
  continuum formulation and finite element analysis.
\newblock \emph{European Journal of Mechanics A Solids}, 21:\penalty0 441--463,
  2002.

\bibitem[Hu et~al.(2016)Hu, Guo, Chen, Xie, Jing, and Peng]{hu2016}
W.~Hu, H.~Guo, Y.~Chen, R.~Xie, H.~Jing, and H.~Peng.
\newblock Experimental investigation and modeling of the rate-dependent
  deformation behavior of {PMMA} at different temperatures.
\newblock \emph{European Polymer Journal}, 85:\penalty0 313--323, 2016.

\bibitem[Federico et~al.(2018)Federico, Bouvard, Combeaud, and
  Billon]{federico2018}
C.E. Federico, J.L. Bouvard, C.~Combeaud, and N.~Billon.
\newblock Large strain/time dependent mechanical behaviour of {PMMA}s of
  different chain architectures. {A}pplication of time-temperature
  superposition principle.
\newblock \emph{Polymer}, 139:\penalty0 177--187, 2018.

\bibitem[Miehe et~al.(2009)Miehe, Goktepe, and Diez]{miehe2009}
C.~Miehe, S.~Goktepe, and J.M. Diez.
\newblock Finite viscoplasticity of amorphous glassy polymers in the
  logarithmic strain space.
\newblock \emph{International Journal of Solids and Structures}, 46:\penalty0
  181--202, 2009.

\bibitem[Cohen(1991)]{cohen1991}
A.~Cohen.
\newblock A {Pade} approximant to the inverse {Langevin} function.
\newblock \emph{Rheologica Acta}, 30:\penalty0 270--273, 1991.

\bibitem[Clayton et~al.(2024)Clayton, Murdoch, Lloyd, Magagnosc, and
  Field]{claytonZAMM2024}
J.D. Clayton, H.A. Murdoch, J.T. Lloyd, D.J. Magagnosc, and D.M. Field.
\newblock Modeling magnetic field and strain driven phase transitions and
  plasticity in ferrous metals.
\newblock \emph{Zeitschrift fur Angewandte Mathematik und Mechanik (ZAMM)},
  104:\penalty0 e202200612, 2024.

\bibitem[Miehe and Schanzel(2014)]{miehe2014}
C.~Miehe and L.-M. Schanzel.
\newblock Phase field modeling of fracture in rubbery polymers. {Part I}:
  finite elasticity coupled with brittle failure.
\newblock \emph{Journal of the Mechanics and Physics of Solids}, 65:\penalty0
  93--113, 2014.

\bibitem[Clements(2012{\natexlab{b}})]{clements2012b}
B.E. Clements.
\newblock A continuum theory of dynamically loaded polymers.
\newblock \emph{arXiv preprint}, arXiv.1207.2723, 2012{\natexlab{b}}.

\bibitem[Olabisi and Simha(1975)]{olabisi1975}
O.~Olabisi and R.~Simha.
\newblock Pressure-volume-temperature studies of amorphous and crystallizable
  polymers. {I. Experimental}.
\newblock \emph{Macromolecules}, 8:\penalty0 206--210, 1975.

\bibitem[Forbes(1977)]{forbes1977}
J.W. Forbes.
\newblock Experimental investigation of the kinetics of the shock-induced alpha
  to epsilon phase transformation in {Armco} iron.
\newblock Technical Report NSWC/WOL TR 77-137, Naval Surface Warfare Center,
  Silver Spring (MD), 1977.

\bibitem[Boettger and Wallace(1997)]{boettger1997}
J.C. Boettger and D.C. Wallace.
\newblock Metastability and dynamics of the shock-induced phase transition in
  iron.
\newblock \emph{Physical Review B}, 55:\penalty0 2840--2849, 1997.

\bibitem[Kormer(1968)]{kormer1968}
S.B. Kormer.
\newblock Optical study of the characteristics of shock-compressed condensed
  dielectrics.
\newblock \emph{Soviet Physics Uspekhi}, 11:\penalty0 229--254, 1968.

\bibitem[Jeong et~al.(2015)Jeong, Ko, Ko, and Kim]{jeong2015}
M.-S. Jeong, J.-H. Ko, Y.H. Ko, and K.J. Kim.
\newblock High-pressure elasticity of poly(methyl methacrylate) up to 31.5 gpa
  studied by {B}rillouin spectroscopy.
\newblock \emph{Current Applied Physics}, 15:\penalty0 943--946, 2015.

\bibitem[Wilkins(1980)]{wilkins1980}
M.L. Wilkins.
\newblock Use of artificial viscosity in multidimensional fluid dynamic
  calculations.
\newblock \emph{Journal of Computational Physics}, 36:\penalty0 281--303, 1980.

\bibitem[Benson(2007)]{benson2007}
D.J. Benson.
\newblock Explicit finite element methods for large deformation problems in
  solid mechanics.
\newblock In E.~Stein, R.~de~Borst, and T.J.R. Hughes, editors,
  \emph{Encyclopedia of Computational Mechanics}, volume~2, chapter~25, pages
  1--43. John Wiley and Sons, New York, 2007.

\bibitem[Morro(1980)]{morro1980b}
A.~Morro.
\newblock Shock waves in thermo-viscous fluids with hidden variables.
\newblock \emph{Archive of Mechanics}, 32:\penalty0 193--199, 1980.

\bibitem[Jordan et~al.(2016)Jordan, Casem, and Zellner]{jordan2016}
J.L. Jordan, D.~Casem, and M.~Zellner.
\newblock Shock response of polymethylmethacrylate.
\newblock \emph{Journal of Dynamic Behavior of Materials}, 2:\penalty0
  372--378, 2016.

\bibitem[Jordan et~al.(2017)Jordan, Casem, Moy, and Walter]{jordan2017}
J.L. Jordan, D.~Casem, P.~Moy, and T.~Walter.
\newblock Properties and shock response of {PMMA}.
\newblock \emph{AIP Conference Proceedings}, 1793:\penalty0 140007, 2017.

\bibitem[Lacina et~al.(2018)Lacina, Neel, and Dattelbaum]{lacina2018}
D.~Lacina, C.~Neel, and D.~Dattelbaum.
\newblock Shock response of polymethyl methacrylate ({PMMA}) measured with
  embedded electromagnetic gauges.
\newblock \emph{Journal of Applied Physics}, 123:\penalty0 185901, 2018.

\bibitem[Nunziato et~al.(1972)Nunziato, Schuler, and Walsh]{nunziato1972}
J.W. Nunziato, K.W. Schuler, and E.K. Walsh.
\newblock The bulk response of viscoelastic solids.
\newblock \emph{Transactions of the Society of Rheology}, 16:\penalty0 15--32,
  1972.

\bibitem[Nunziato and Sutherland(1973)]{nunziato1973ac}
J.W. Nunziato and H.J. Sutherland.
\newblock Acoustical determination of stress relaxation functions for polymers.
\newblock \emph{Journal of Applied Physics}, 44:\penalty0 184--187, 1973.

\bibitem[Rosenberg and Partom(1984)]{rosenberg1984}
Z.~Rosenberg and Y.~Partom.
\newblock Direct measurement of temperature in shock-loaded
  polymethylmethacrylate with very thin copper thermistors.
\newblock \emph{Journal of Applied Physics}, 56:\penalty0 1921--1926, 1984.

\bibitem[Bordzilovskii et~al.(2016)Bordzilovskii, Karakhanov, Merzhievskii, and
  Voronin]{bordz2016}
S.A. Bordzilovskii, S.M. Karakhanov, L.A. Merzhievskii, and M.S. Voronin.
\newblock Temperature measurements for shocked polymethylmethacrylate, epoxy
  resin, and polytetrafluoroethylene and their equations of state.
\newblock \emph{Journal of Applied Physics}, 120:\penalty0 135903, 2016.

\bibitem[Reinhart and Chhabildas(2006)]{reinhart2006}
W.D. Reinhart and L.C. Chhabildas.
\newblock Response to unloading and reloading of shock compressed polymethyl
  methacrylate.
\newblock \emph{AIP Conference Proceedings}, 845:\penalty0 131--134, 2006.

\bibitem[Pavlovskii(1976)]{pavlov1976}
M.N. Pavlovskii.
\newblock Measurements of the velocity of sound in shock-compressed quartzite,
  dolomite, anhydrite, sodium chloride, paraffin, plexiglas, polyethylene, and
  fluoroplast-4.
\newblock \emph{Journal of Applied Mechanics and Technical Physics},
  17:\penalty0 709--712, 1976.

\bibitem[Williams et~al.(2025{\natexlab{a}})Williams, Lloyd, Mallick, Ligda,
  and Clayton]{williams2025a}
C.L. Williams, J.T. Lloyd, D.T. Mallick, J.P. Ligda, and J.D. Clayton.
\newblock Real-time observation of twinning, detwinning, and melting in
  shock-compressed magnesium alloy {AZ31B-H24}.
\newblock \emph{Physical Review Materials}, 9:\penalty0 043603,
  2025{\natexlab{a}}.

\bibitem[Rittel(2000)]{rittel2000}
D.~Rittel.
\newblock Experimental investigation of transient thermoplastic effects in
  dynamic fracture.
\newblock \emph{International Journal of Solids and Structures}, 37:\penalty0
  2901--2913, 2000.

\bibitem[Bowen and Chen(1974)]{bowen1974}
R.M. Bowen and P.J. Chen.
\newblock Shock waves in ideal fluid mixtures with several temperatures.
\newblock \emph{Archive for Rational Mechanics and Analysis}, 53:\penalty0
  277--294, 1974.

\bibitem[Clayton(2022)]{claytonIJES2022}
J.D. Clayton.
\newblock Analysis of shock waves in a mixture theory of a thermoelastic solid
  and fluid with distinct temperatures.
\newblock \emph{International Journal of Engineering Science}, 175:\penalty0
  103675, 2022.

\bibitem[Bartkowski and Dandekar(2000)]{bartowski2000}
P.T. Bartkowski and D.P. Dandekar.
\newblock Recompression of {PMMA} following shock induced tension.
\newblock \emph{AIP Conference Proceedings}, 505:\penalty0 539--542, 2000.

\bibitem[Martinez and Williams(1980)]{martinez1980}
C.B. Martinez and M.C. Williams.
\newblock Viscosity and microstructure of polyethylene-poly (methyl
  methacrylate) melt blends: some simple interpretations.
\newblock \emph{Journal of Rheology}, 24:\penalty0 421--450, 1980.

\bibitem[Williams et~al.(2025{\natexlab{b}})Williams, Hornbuckle, Mallick,
  Parker, Clayton, and Wilkerson]{williams2025b}
C.L. Williams, B.C. Hornbuckle, D.D. Mallick, T.C. Parker, J.D. Clayton, and
  J.W. Wilkerson.
\newblock Strength-ductility synergy of a high chromium martensitic steel with
  a unique microstructure under dynamic extremes.
\newblock \emph{Acta Materialia}, 298:\penalty0 121368, 2025{\natexlab{b}}.

\bibitem[Atkins et~al.(1975)Atkins, Lee, and Caddell]{atkins1975}
A.G. Atkins, C.S. Lee, and R.M. Caddell.
\newblock Time-temperature dependent fracture toughness of {PMMA: Part 1}.
\newblock \emph{Journal of Materials Science}, 10:\penalty0 1381--1393, 1975.

\bibitem[Stephens et~al.(1972)Stephens, Heard, and Schock]{stephens1972}
D.R. Stephens, H.C. Heard, and R.N. Schock.
\newblock High-pressure mechanical properties of polymethylmethacrylate.
\newblock Technical Report R-4531, Lawrence Livermore National Laboratory,
  Livermore (CA), 1972.

\end{thebibliography}
%
%\clearpage
%\pagestyle{empty}
%\normalsize
%\input{sreply}
%
\end{document}